\documentclass[a4paper,fleqn,usenatbib]{mnras}

\usepackage[T1]{fontenc}
\usepackage{ae,aecompl}

\usepackage{graphicx}	% Including figure files
\usepackage{amsmath}	% Advanced maths commands
\usepackage{amssymb}	% Extra maths symbols
\usepackage[dvipsnames]{xcolor} 
\usepackage{subcaption}
\usepackage{pdflscape}

\newcommand{\hi}{H{\sc i}}

\newcommand{\prim}{$^{\prime}$}
\newcommand{\prin}{$^{\prime\prime}$}
\newcommand{\aprox}{${\sim}$}

\newcommand{\mJyb}{mJy\,beam$^{-1}$}
\newcommand{\kms}{~km~s$^{-1}$}

\newcommand{\degree}{$^{\circ}$}

\newcommand{\msolar}{M$_{\odot}$}

\newcommand{\mhi}{M$_{\mathrm {HI}}$}

\newcommand{\defhi}{HI deficiency}

\newcommand{\por}{$\times$}

\title[\hi~properties of clusters of galaxies]
{Environmental cluster effects and galaxy evolution:
The \hi~properties of the Abell clusters A85/A496/A2670}
\author[L\'opez-Guti\'errez et al.]{
M. M. L\'opez-Guti\'errez,$^{1}$\thanks{E-mail: mm.lopezgutierrez@ugto.mx (MMLG)}
H. Bravo-Alfaro,$^{1}$%\thanks{E-mail: hector@ugto.mx (HBA)}
J. H. van Gorkom,$^{2}$
C. A. Caretta,$^{1}$
\newauthor
F. Durret,$^{3}$
L. M. N\'{u}\~nez-Beltr\'an,$^{1}$
Y. L. Jaff\'e,$^{4}$
M. Hirschmann,$^{5,6}$ 
D. P\'erez-Mill\'an,$^{1,7}$ 
\\
$^{1}$Departamento de Astronom\'ia, Universidad de Guanajuato, 
Gto. 36000, M\'exico\\
$^{2}$Department of Astronomy, Columbia University, New York, 
NY 10027, USA\\
$^{3}$Sorbonne Universit\'e, CNRS, UMR 7095, Institut d'Astrophysique de Paris, 98bis Bd Arago, 75014, Paris, France\\
$^{4}$Inst. de F\'isica y Astronom\'ia, Fac. de Ciencias, Universidad de Valpara\'iso, Avda. Gran Breta\~na 1111, Casilla 5030, Valpara\'iso, Chile \\
$^{5}$Institut de Physique, Laboratoire d’astrophysique, \'Ecole Polytechnique F\'ed\'erale de Lausanne (EPFL), CH-1290 Versoix, Switzerland\\
$^{6}$INAF - Osservatorio Astronomico di Trieste, via G. B. Tiepolo 11, I-34143 Trieste, Italy \\
$^{7}$Instituto de Radioastronom\'ia y Astrof\'isica, 
UNAM, 58089, M\'exico
}

\date{Accepted XXX. Received YYY; in original form ZZZ}

\pubyear{2022}

\begin{document}
\label{firstpage}
\pagerange{\pageref{firstpage}--\pageref{lastpage}}
\maketitle

\begin{abstract}
We study the impact of local environment on the 
transformation of spiral galaxies in three nearby ($z <
0.08$) Abell clusters: A85/A496/A2670.  These systems 
were observed in \hi\ with the Very Large Array, 
covering a volume extending beyond the virial radius
and detecting 10, 58, 38 galaxies, respectively. 
High fractions (0.40--0.86) of bright spirals 
[log$(M_{*}/M_{\odot})=9-10$] are not detected in 
\hi. We provide further evidence of environmental 
effects consisting in significant fractions 
(0.10--0.33) of abnormal objects and a number of 
red (passive) spirals, suggesting an ongoing process
of quenching.   Ram-pressure profiles, and the
sample of the brightest spirals used as test particles 
for environmental effects, indicate that ram-pressure 
plays an important role in stripping and transforming 
late-types. Phase-space diagrams and our search for 
substructures helped to trace the dynamical stage of 
the three systems. This was used to compare the global 
cluster effects {\it vs.} pre-processing, finding that 
the former is the dominating mechanism in the studied 
clusters.  By contrasting the global distribution of 
\hi\ normal {\it vs.} \hi\ disturbed spirals in the
combined three clusters, we confirm the expected 
correlation of disturbed objects located, on average, 
at shorter projected radii.  However, individual 
clusters do not necessarily follow this trend and 
we show that A496 and A2670 present an atypical 
behavior.  In general we provide conclusive evidence 
about the dependence of the transformation of 
infalling spirals on the ensemble of cluster properties 
like mass, ICM density, dynamical stage and surrounding 
large-scale structure.
\end{abstract}


%%%%%%%%%%%%%%%%%%%%%%%%%%%%%%%%%%%%%%%%%%%%%%%%%%

%%%%%%%%%%%%%%%%% BODY OF PAPER %%%%%%%%%%%%%%%%%%

\section{Introduction}
\label{intro}

For decades, observations have shown that regions of 
high galaxy-density, such as cluster cores, are 
dominated by red E/S0 galaxies while blue spirals are 
more abundant in lower density regions.  Large amounts 
of observational evidence have confirmed the {\it 
morphology-density relation} which constitutes one of 
the clearest evidences of environment playing an 
important role in galaxy evolution \citep{Oemler74,Dressler80,Postman-Geller84,Dressler97,Goto03,Poggianti09b}.  
This ($nurture$) scenario is in full agreement with the 
$\Lambda$CDM model of hierarchical formation of structures, 
where individual galaxies are predicted, and observed 
to follow the stream from low to high-density 
environments during their life time.  This galaxy 
migration seems to be tightly correlated with
a systematic change of their physical properties 
such as morphology, gas 
content, and SF/AGN activity.  On the theoretical side
many works \citep[and references therein]{DeLucia19} 
conclude that secular mechanisms and/or conditions 
at the time of galaxy formation ($nature$) might be
more important than environment in driving galaxy 
evolution. Theoretical models are still trying to
reproduce all the observational trends linked
with environment. Nowadays it is generally
accepted that galaxies at low redshifts 
have been shaped by both, nature and nurture.
However the full understanding of the physics 
behind the galaxy evolution over cosmological 
timescales remains a very active research field 
\citep[\textit{e.g.}][]{Peng10, Boselli-Gavazzi14,Cortese21}.  
%ADD REF: CORTESE, CATINELLA /& SMITH, 2021, PASA.

%Concerning the study of the environment, the remarkable 
%absence of spirals in most of the cluster cores is 
%accompanied by the opposite growing population of 
%lenticulars and, in many cases, by the presence of 
%post-starburst galaxies which are found preferentially in 
%high density environments when observed at $z$ between 
%0.3 and 0.6 \citep[\textit{e.g.}][]{Poggianti09a}.  These facts 
%raised the question of whether a large fraction of spirals 
%is being transformed into lenticulars (S0s) during their 
%infall to galaxy clusters due to different physical 
%mechanisms \citep{Boselli-Gavazzi06,Barway07}. 

Solving the question of morphology transformation
of spirals into lenticulars, through an 
evolutionary sequence involving 
starburst/post-starburst phases 
\citep[\textit{e.g.}][]{Boselli-Gavazzi06,Barway07,
Poggianti09a,Paccagnella17,Boselli22},
requires the study along two main 
axes, the environment being one of them.  Studying
galaxies within the same cosmic epoch but living in
different environments helps to find correlations 
between galaxy properties and the physical conditions 
of their local environment
\citep{Bahe13,Sperone-Longin21}.
%(ADD REF : Bahe et al. 2013, MNRAS, 430, 3017; Sperone-Longin et al. A\&A, 2021, 647, 156).
The second axis corresponds to the cosmic evolution:
the study of galaxy properties as a function of redshift 
has shown that a short ($\leq 10^8$\,yrs)
starburst phase occurs, followed by a long term and
complex quenching process \citep{Rhee20, Cortese21}.
% ADD REF Rhee et al. 2020. 
As a consequence, the 
fraction of galaxies in the red sequence almost 
doubled between $z$ \aprox\,1 and $z =$ 0,
implying a decrease of blue galaxies 
accompanied by strong morphology evolution  
\citep{Butcher-Oemler78, Fasano00, Arnouts07}. 

Several 
consortia have been actively approaching the question based 
on large photometric and spectroscopic surveys such as 
2dF and SDSS, covering large sky areas and redshift ranges \citep{Lewis02,Balogh04,Kauffmann04,Poggianti17}. 
%
%HERE WE SHOULD ADD MORE RECENT PROJECTS LIKE THOSE ON 
%VIRGO, GASP, ETC).
%
%Combining this with simulations (Roediger09, Tonnesen-Bryan09)
%Roediger 2009, AN, 330, 888
%Tonnesen S. \& Bryan G. L. 2009, ApJ, 694, 789
%.........  pargraph gone 
%
All these efforts have found indisputable evidence for the 
impact of environment in galaxy evolution 
 \citep[and references therein]{Boselli-Gavazzi06,Boselli-Gavazzi14,Cortese21}.
%{\bf Add REF Boselli-Gav 2006, Boselli-Gav 2014]}.
However, understanding the role played by different 
physical mechanisms exerted under diverse
environment conditions
constitutes the matter of a very active debate.  
The involved physical mechanisms are classified 
in two types, hydrodynamic and gravitational. 
Hydrodynamic effects concern the stripping of 
cold/warm interstellar gas (\hi\ and H$_2$) by the 
hot intracluster medium (ICM).  The ram-pressure 
stripping \citep[RPS,][]{Gunn-Gott72} and the 
viscous stripping \citep{Nulsen82} are the most 
studied cases.   On the other side we have the 
tidal (gravitational) mechanisms occurring between 
a galaxy and the cluster potential 
\citep{Byrd-Valtonen90} or among neighbor galaxies 
\citep{Merritt83, Barnes-Hernquist96,Walker96}.
These include major mergers, accretion of low-mass
satellites, and the accumulation of fast speed 
encounters between galaxies  
\citep[the {\it galaxy harassment},][]{Moore96}.
The removal of the galaxy halo gas, known as 
galaxy starvation \citep[\textit{e.g.}][]{Larson80}, is 
predicted to occur either by hydrodynamic or gravitational
interactions.  Most of these mechanisms are predicted to 
transform a spiral galaxy into an S0, and it is known 
that more than one process might act simultaneously on 
a single galaxy.  The {\it pre-processing} of galaxies 
occuring within groups infalling towards clusters, 
seems to be particularly important \citep{Donnari21}.
Groups of galaxies are known to have lower velocity 
dispersions than clusters, allowing slower and deeper 
tidal interactions among their members.  Several 
authors \citep{Fadda08,Poggianti09b} provided 
substantial evidence that strong galaxy evolution 
is occurring in low mass systems at large distance 
from the cluster core.   However, the debate {\it 
pre-processing {\it vs.} cluster effects} remains open 
because many variables are involved, such as the 
infalling orbits, initial gas/stellar masses, 
the group/cluster properties and even the 
surrounding large scale structure 
\citep{Salerno20,Rhee20}.

In this context, studying the atomic hydrogen (\hi) of 
spiral galaxies, in combination with observations at 
other frequencies and with numerical simulations, 
has proven to be a powerful tool to study galaxy 
evolution as a function of environment.  \hi\ is 
probably the best tracer of galaxy interactions 
as it extends well beyond the stellar disk. 
%what makes the restoration force lower on the outskirts 
%of the galaxy.  
Several studies have been devoted to the nearest 
clusters such as Virgo, Hydra, A\,1367 and Coma 
\citep[\textit{e.g.}][]{Giovanelli-Haynes85, 
Cayatte91, Bravo-Alfaro00, Solanes01, Chung09, 
Scott10, Scott18, Boselli21, Wang21, Molnar22}. 
%ADD REF Wang et al. 2021, ApJ, 915, 70 
The large coverage and high sensitive \hi-surveys 
carried out with single dish telescopes, such as 
HIPASS and ALFALFA \citep{Giovanelli05,Haynes18},
%\citep{Zwaan03,Giovanelli05,Haynes18},
provided statistical information on the \hi~content of
large samples of spirals in different environments. 
All these surveys have 
shown that spirals in clusters have systematically 
less gas than their counterparts in the field.
This loss of cold gas is known to be the first 
step to the eventual quenching of star formation 
observed in cluster galaxies. 

In parallel to single-dish surveys, \hi\ synthesis 
imaging provides critical information mainly at low
redshifts. This technique unveils different types of 
disruptions like gas shrunk discs, asymmetries, 
and offsets between the gas and the stellar disks 
\citep[and references therein] {Bravo-Alfaro01, Chung09, 
Scott18, Luber22}.
\hi-maps and kinematical velocity-fields are essential 
to constrain hydrodynamical simulations of 
galaxies under different environment processesing
\citep[\textit{e.g.}][and references therein]{Vollmer12, 
Tonnesen19}.  With a few remarkable exceptions, 
\hi-synthesis imaging surveys are restricted to 
the nearby universe ($z < 0.1$) due to instrumental 
limitations and to the very long observing times 
required. As a consequence, only a few clusters have 
been fully covered by using this technique.  
The new generation of radio telescopes such 
as ASKAP and MeerKAT \citep{Koribalski20, Jonas16}, 
as well as upgraded instruments like the Westerbork 
Synthesis Radio Telescope (WSRT) and the Karl G.
Jansky Very Large Array (VLA), are nowadays pushing 
farther the redshifts of \hi-imaged galaxies. 
The BUDHIES project \citep{Jaffe13,Jaffe16,Gogate20} 
use the WSRT to characterize the HI properties in 
and around two clusters at $z=0.2$, A963 and A2192. 
The CHILES project \citep{Fernandez13,Hess19} 
%{\bf ADD REF: Fernandez+13  2013, ApJ, 770L, 29}\\
uses the VLA to image the \hi\ in a 40\,arcmin field 
out to $z=0.45$ and has made a direct detection at the 
largest distance so far, at $z=0.376$ \citep[]{Fernandez16}. 
In parallel to 
pushing the study of \hi\ to higher redshifts (the 
{\it time axis}) it is equally important to enlarge 
the sample of studied clusters lying at the same epoch 
(the {\it environment axis}).  Large-volume, \hi\ 
blind imaging surveys like the one presented in 
this work are intended to explore different clusters 
from their core to the outskirts. The \hi\ maps
and kinematics of resolved objects are studied 
as a function of the very diverse  
environments where galaxies are located.

In this paper we enlarge the sample of nearby systems 
($z <$ 0.08) imaged in \hi. Large volumes of the
Abell clusters A85, A496, A2670, were homogeneously
observed with the VLA. 
%We covered from the center to beyond one virial radius and more than 3$\sigma$ in velocity.  
We discuss their global \hi\ 
properties paying special attention to a complete
sample of bright spirals, used as test particles 
for environment effects. 
%Our \hi\ and optical images reveal a sample of objects showing perturbed morphology (in \hi\ and/or optical); 
We obtain \hi\ physical parameters of individual 
galaxies and deliver an atlas of \hi\ maps and
velocity fields.  Bright spirals not detected
in \hi\ are of particular interest. We seek 
for correlations between gas properties of 
individual galaxies within the substructures 
reported in this work as a test quantifying the
pre-processing. We present profiles of the ram 
pressure and projected phase space diagrams 
(PPS) for each cluster.
%by applying different infall velocities and anchoring forces. 
%where we plot each member galaxy at its projected distance from the cluster center and relative velocity to the cluster mean.  
Finally, we use 
the gas content and color index as a first approach 
quantifying the quenching fraction in the studied 
clusters. In forthcoming papers we will use optical
spectra and multi-band observations to complement the 
study of individual galaxies in particular those 
showing strong distortions in \hi\ and in optical.

%and the correlation of 
%gas properties with optical spectra, analyzing the 
%gas stripping mechanisms and SF/quenching histories.
%based on optical spectra and on the infalling orbits.

This paper is organized as follows: Section\,\ref{obs} 
describes the clusters under study, the observations 
and the data reduction for the \hi\ and the optical 
imaging.  In the same section we describe the 
catalogues of member objects used in this work, 
as well as the sample of the brightest spirals and 
the color-magnitude diagrams of the studied clusters.
%(Sect.\,\ref{obs_colors}).  
In Section\,\ref{results} we give the main 
observational results: we provide the \hi\ 
parameters of all detected galaxies and 
%are in Sect.\,\ref{res_detect} 
their global distribution across the observed 
clusters.
%in Sect.\,\ref{res_distrib}.  
Tables containing these data and the corresponding
\hi\ maps are given in the Appendix. 
In Sect.\,\ref{res_substructures} we study the 
dynamical state of the three clusters by applying 
two independent methods tracing substructures.
In Sect.\,\ref{disc} we calculate RPS profiles 
%and in Sect.\,\ref{disc_PPS} we trace 
and we obtain radial velocity \textit{vs.} 
cluster-centric distance plots in order to 
study the assembly history of the clusters.  In 
Sect.\,\ref{discussion} we compare the clusters
with each other and we discuss the role played 
by the cluster environment in the evolution 
of infalling galaxies.
%and which are the most likely mechanisms driving the evolution of infalling spirals. 
%The dynamical evolution and the LSS around each cluster is included in this discussion.  
We summarize our results and give our
conclusions in Sec.\,\ref{summary}.
Throughout this paper we assume $\Omega_M$\,=\,0.3, 
$\Omega_\Lambda$\,=\,0.7, and $H_0=70$\,\kms Mpc$^{-1}$.

\begin{table}
        \centering
        \caption{Properties of the studied clusters}        
        \label{tab_clusters}
        \begin{tabular}{lccc} % four columns, alignment for each
                \hline
                                         & A85 & A496 & A2670\\
                \hline
                RA (2000)     & 00 41 50 & 04 33 38 & 23 54 14\\
                Dec (2000)    &-09 18 07 &-13 15 33 &-10 25 08 \\
                $z$/$D_{c}$ (Mpc)          & 0.055/233 & 0.033/140 & 0.076/320\\
                $D_{l}$ (Mpc)         & 245.4 & 144.9 & 344.1\\
                scale 10\prim (Mpc)      & 0.69 & 0.41 & 0.95 \\
                $v_{cl}$ (km~s$^{-1}$)   & 16,607 & 9,884 & 22,823 \\
                $\sigma_{cl}$ (km~s$^{-1}$) & 1054 & 685 & 781 \\
              %  Min \hi-vel (km~s$^{-1}$)& 14,580 & 8,630 & 20,300\\
              %  Max \hi-vel (km~s$^{-1}$)& 18,440 & 11,092 & 25,400\\
                R$_{200}$ (Mpc)          & 2.54 & 1.68 & 1.86\\
  %              \rabell (\prim)        & 30.9 & 51.5 & 22.3 \\
              %  R$_{\mathrm {HI}}$ (\prim)& 50 & 55 & 40\\
                M$_{200}$ ($10^{14}$ M$_{\odot}$)& 12.6 & 3.49 & 5.07\\
                Morph (B-M)          & I & I & I-II \\
                Richness   & 130 & 134 & 224 \\
                $L_X$ ($10^{44}\, \mathrm{erg}^{-1}$)  & 9.4 & 3.8 & 2.3 \\
                LSS                  & SC (11) & isolated(*) & isolated \\
                \hline
        \multicolumn{4}{l}{Notes: }\\
        \multicolumn{4}{l}{-$D_{c}$ (comoving distance), $D_{l}$ (luminosity distance) and linear}\\
        \multicolumn{4}{l}{scale for 10\prim\ are estimated with $H_0=70$\kms Mpc$^{-1}$.}\\
        \multicolumn{4}{l}{-The central velocity ($v_{cl}$), the velocity dispersion ($\sigma_{cl}$) and $L_X$ were } \\ 
        \multicolumn{4}{l}{taken from NED, HyperLeda and BAX, respectively.}\\      
        %\multicolumn{4}{l}{-Min \hi-vel / Max \hi-vel give the velocity range covered in \hi.} \\
        \multicolumn{4}{l}{-R$_{200}$: computed following \cite{Finn05}.}\\
     %   \multicolumn{4}{l}{\cite{Finn05}.}\\
        %\multicolumn{4}{l}{-R$_{\mathrm {HI}}$: the "major axis" 
        %of the total VLA FoV. }\\
        \multicolumn{4}{l}{-M$_{200}$ was estimated following \cite{Barsanti18}.}\\
        \multicolumn{4}{l}{-LSS (large scale structure): 
        indicates that A85 is part of the }\\
        \multicolumn{4}{l}{supercluster MSCC\,039, consisting of 11 cluster members }\\
        \multicolumn{4}{l}{\citep{Chow-Martinez14}.}\\
        \multicolumn{4}{l}{(*) We 
        revisit the LSS of A496 in this work.}        
        \end{tabular}
\end{table}

\section{The observations}
\label{obs}

\subsection{The cluster sample}
\label{obs_clusters}

In this work we study the Abell clusters
A85 ($z = $\,0.055), A496 ($z =  $\,0.033), 
A2670 ($z = $\,0.076), which harbor different 
physical properties (see Table\,\ref{tab_clusters}) 
and dispose of ancillary multifrequency data.

%, including 
%X-rays, redshift catalogs, and deep optical/NIR imaging. 
A85 is more massive and more luminous in X-rays 
than the other two systems.  Despite its relaxed 
morphology it is known to be undergoing minor 
merging processes with a subcluster and groups seen 
in X-rays and in the optical \citep[]{Durret98, 
Durret05, Bravo-Alfaro09}.   More recently, this 
system has been confirmed to possess a centrally 
peaked profile in X-rays with a sloshing pattern 
\citep[]{Ichinohe15} and with some additional 
disturbed features \citep{Lagana19}.
%{\bf ADD REF Lagana, Durret, Lopez 2019; MNRAS, 484, 2807}.
The presence of jellyfish galaxies like JO201
has been associated with strong RPS in this cluster 
\citep[][and references therein]{Ramatsoku20, Luber22}.
%{\bf ADD REF Ramatsoku et al. 2020, A\&A, and references therein}.  

A496 is known to have a very relaxed morphology 
\citep{Durret00,Boue08a,Ulmer11} consistent with 
a significant coincidence between the BCG and the 
cluster centroid (position and velocity) \citep{Lopes18}.
%{\bf ADD REF Lopes et al. 2018MNRAS.478.5473}.  
Nevertheless, \cite{Lagana19}
%{\bf ADD REF Lagana, Durret, Lopez 2019; MNRAS, 484, 2807} 
reported disturbed features superposed onto a 
centrally peaked X-ray distribution.  Concerning the
LSS, A496 was reported as an isolated system 
by \cite{Chow-Martinez14}. We revisit this 
result in Sect.\,\ref{discussion}.

A2670 is the faintest of the three systems in X-rays,
however its ICM is not negligible: RPS is blamed
for strong sweep of material observed in a 
post-merger galaxy \citep{Sheen17}.
%{\bf ADD REF Sheen et al. ApJLetters, 2017, 840, L7}.
A2670 is the most irregular of the three systems 
and displays a very elongated morphology distributed 
along a NE-SW axis.  This cluster is dynamically 
young, as suggested by the diffuse X-ray emission
and the large offset in velocity of 333\kms\
(C. Caretta, priv. comm.) between the cluster and the BCG;
this clearly indicates that the massive galaxy has not 
yet settled at the bottom of the cluster potential well.

\subsection{The \hi\ data}
\label{obs_hi}

\begin{figure}
\centering
\begin{subfigure}[tbp]{0.55\textwidth}
 \includegraphics[width=\columnwidth]{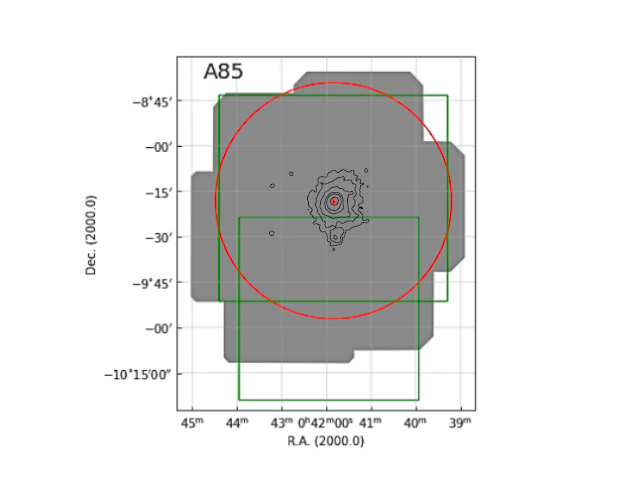}
 %\caption{}
  \end{subfigure}
  \\
 \begin{subfigure}[tbp]{0.49\textwidth}
 \includegraphics[width=\columnwidth]{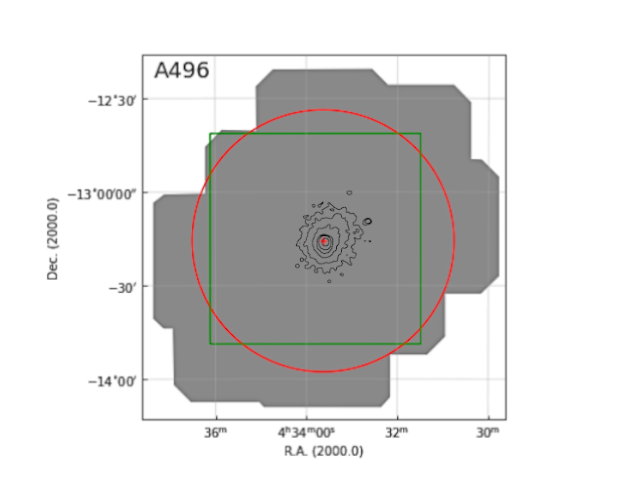}
 %\caption{}
  \end{subfigure}
  \\
   \begin{subfigure}[tbp]{0.52\textwidth}
 \includegraphics[width=\columnwidth]{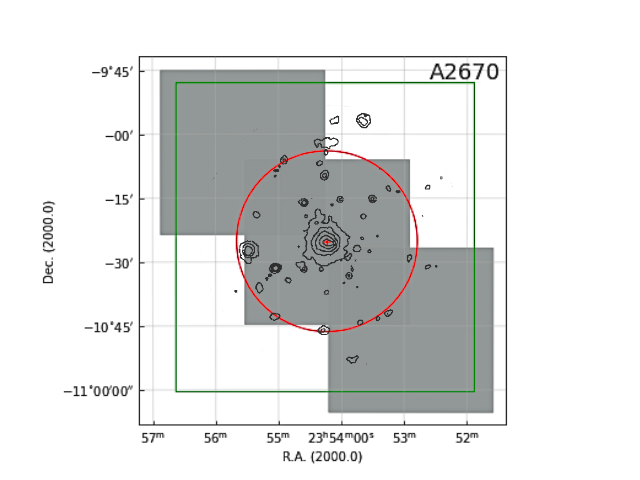}
 %\caption{}
  \end{subfigure}
  \caption{The observed fields in \hi\ (grey polygons) 
  and in optical (green squares). A85 is in the top 
  panel, A496 in the middle, and A2670 in the bottom. 
  %The \aprox2\,Mpc (1\,\rabell) radius is indicated
  The R$_{200}$ radius is indicated
  with a red circle and the black contours draw the 
  ROSAT X-ray emission. The BCG is indicated with a 
  red cross.}
  \label{fig_cubes}
\end{figure}

We take advantage of \hi-data obtained with the 
VLA\footnote{The National Radio 
Astronomy Observatory is a facility of the National 
Science Foundation operated under cooperative 
agreement by Associated Universities, Inc.} 
%{\bf add footnote for VLA: The National Radio Astronomy Observatory is a facility of the National Science Foundation operated under cooperative agreement by Associated Universities.}\\
between 1994 and 1996 for A2670, and from
2001 to 2003 for A85 and A496.  All observations 
were taken in C configuration, producing beam 
sizes of \aprox\,24\prin \por \,17\prin.  
Table\,\ref{tab_obs} gives the main observing 
parameters: the total integration time, the VLA
configuration, the beam size, the bandwidth and 
the correlator mode applied. The correlator mode
defines the number of polarizations applied. The 
correlator mode 1 uses one polarization and 
allows a larger number of velocity channels; 
mode 2 applies two polarizations and is able
to produce an improvement in sensitivity by
a factor of $\sqrt{2}$.  Hanning smoothing 
was used on the data of A85 and A496,
obtaining a set of 31 independent channels. 
For A2670 no Hanning smoothing was applied, 
obtaining a set of 63 channels (cubes NE 
and SW). The general observing strategy is 
based on a combination of several  
observing pointings at different positions 
and central velocities. This allowed to 
cover the full cluster volume beyond one 
virial radius and more than three times the 
velocity dispersion.  Fig.~\ref{fig_cubes} 
shows the total observed fields for each 
cluster and Table\,\ref{tab_cubes} gives 
the parameters of the final \hi\ data cubes. 
Throughout this paper we list the \hi\ 
heliocentric velocities using the optical 
definition.

\begin{table}
    %\centering
    \caption{\hi\ observational parameters }
    \label{tab_obs}
    \begin{tabular}{cccccc}
    \hline
 ID  &  T$_{int}$  & VLA      & beam    & Bandw &  Correl.  \\
     &  (h)       & config.  & (\prin) & MHz   &  mode     \\
 (1) & (2)         & (3)      & (4)     & (5)   &  (6)   \\
\hline   
A85  & 80         & C & 24\,\por \,17 &  6.3   &  2AC \\
A496  & 12         & C & 25\,\por \,17 &  6.3   &  2AC \\
A2670 & 90         & C & 23\,\por \,16 &  12.5  &  1A, 1D \\
  \hline 
\multicolumn{6}{l}{Column (2): The time on source, in hours. }  \\
\multicolumn{6}{l}{Column (3): The VLA configuration used. }  \\
\multicolumn{6}{l}{Column (4): The beam size in arcsec. }  \\
\multicolumn{6}{l}{Column (5): The applied bandwidth in MHz. }  \\
\multicolumn{6}{l}{Column (6): The correlator mode. }  \\
\end{tabular}  
    \end{table}

%Continuum subtracted cubes
%were CLEANed of sidelobes and then mosaiced to get 
%a large FoV format (A085 and A496) or combined
%with other cubes at different central velocities,
%in order to get a large full velocity coverage
%(see previous paragraph).  

% VLA-Arch :     AG0592; Target name empty; RA Dec;  rad 1.0d     Obs band L
%A  085; AG0592; 2001 > 2002; ToS = 80hrs;      Bndw =  6.3MHz; Chann = 32; Config C; Corr mode 2AC; 6 flds
% VLA-Arch :     Target name empty   RA 04 33 04.1  Dec -13 06 19    Obs band L
%A  496; AG0592; 2001 > 2003; ToS = 12 hrs;     Bndw =  6.3MHz; Chann = 32; Config C; Corr mode 2AC; 10 flds
% VLA-Arch :     Target name A2670;   Obs band L
%A 2670; AG0432; 1994 > 1996; ToS = 90-100hrs;  Bndw = 12.5MHz; Chann = 64; Config C; Corr mode 1A, 1D

\begin{table*}
    \centering
    \caption{The \hi-data cubes.}
    \label{tab_cubes}
    \begin{tabular}{lcccccccc}
    \hline 
Data & RA & Dec  &  FoV    & Vel. range
& $V_c$ & $\Delta\,v$ & rms &\mhi\, limit\\
cube	& (2000)   & (2000) & (\prim)    & \kms  
& \kms  & \kms  &  \mJyb & $10^8\,$\msolar \\
(1) & (2) & (3) & (4) & (5) & (6) & (7) & (8) &(9) \\
\hline
A85--LV & 00 41 51.9 & -09 18 17 & 90 & 1,366 & 15,266 & 45.5 & 0.25 & 9.0\\
A85--MV & 00 41 51.9 & -09 18 17 & 90 & 1,330 & 16,500 & 46.9 & 0.25 & 9.2\\
A85--HV & 00 41 51.9 & -09 18 17 & 90 & 1,387 & 17,744 & 46.2 & 0.25 & 9.1 \\

& & & & & & & & \\

A496--LV & 04 33 04.1 & -13 06 19 & 100 & 1,314 &  9,285 & 43.8 & 0.24 & 3.0\\
A496--HV & 04 33 04.1 & -13 06 19 & 100 & 1,324 & 10,429 & 44.2 & 0.24 & 3.0\\

 & & & & & & & & \\

A2670--C& 23 54 13.8 & -10 25 09 & 40 & 4,120 & 23,037 & 47.9 & 0.11 & 7.9\\
A2670--NE& 23 55 34.1 & -10 04 18 & 40 & 2,957 & 22,690 & 47.7 & 0.12 & 8.6\\
A2670--SW& 23 52 53.4 &  -10 46 00  & 40 & 2,957 & 22,690 & 47.7 & 0.12 & 8.6\\
\hline 
\multicolumn{9}{l}{Column (1): Field ID, where LV, MV, HV indicates low, medium and 
high velocity ranges, respectively. }  \\
\multicolumn{9}{l}{Columns (2) and (3): Central coordinates for the data cube,
after mosaicing for A85 and A496.} \\
\multicolumn{9}{l}{Column (4): Approximate diameter of the FoV, 
in arcmin. }  \\
\multicolumn{9}{l}{Column (5): Total velocity coverage, in \kms.}  \\
\multicolumn{9}{l}{Column (6); The heliocentric velocity 
of the central channel, using the optical definition. }  \\
\multicolumn{9}{l}{Column (7): Velocity resolution given by the channel width, in \kms.}  \\
\multicolumn{9}{l}{Column (8): The rms per channel in \mJyb.}  \\
\multicolumn{9}{l}{Column (9): The \hi-mass detection limit, in units 
of $10^8$\,\msolar.}  \\
\end{tabular}
\end{table*}

We applied different observing strategies 
to each cluster depending on their redshifts
and velocity dispersions.  Six VLA fields were 
pointed side by side across A85.  The 
individual fields have slight overlap in 
RA, Dec in order to avoid any gap 
in between. These fields have the same 
central velocity giving a homogeneous 
coverage across an area of \aprox100\prim\  
side after a mosaicing procedure \citep{Taylor99}.
The same six fields were observed three times, 
each one centered at a different velocity:
15,266\kms, 16,500\kms, 17,744\kms\ 
(see Table\,\ref{tab_cubes}).  The central 
velocities were chosen to slightly overlap 
producing a homogeneous coverage of some
3800\kms, \textit{i.e.} more than 3.5 times 
the velocity dispersion of A85.

A similar procedure was applied to A496. Since 
it is located at two thirds the distance of A85 
the total integration time for A496 was 
significantly reduced.  On the other hand 
this cluster is spread across a larger region 
of the sky and we had to apply ten VLA pointings 
with different RA, Dec positions in order 
to reach the one virial radius coverage. 
These fields went through a mosaicing procedure
delivering a field of view of some 100\prim.
Having a low velocity dispersion (737\kms), 
the ten fields were observed two times, each one 
centered at a different velocity: 9,285\kms\ 
and 10,429\kms. This delivered a coverage of 
\aprox\,2400\,\kms, equivalent to more than 
3.5 times the cluster velocity dispersion 
(Table\,\ref{tab_cubes}).

A  different observing strategy was applied 
to A2670.  Located at a larger distance than 
the previous clusters, A2670 needed fewer VLA 
pointing fields to cover the virial radius. 
However, due to its distance, longer observing
time was needed in order to 
reach a similar detection limit than the other 
clusters.   The one virial radius coverage 
was ensured by pointing three VLA fields 
spread along a NE-SW axis and having some 
overlap in order to avoid any gaps. Due to 
observing time limitations, the covered 
area is elongated, giving priority to the 
NE-SW axis.  Mosaicing was not applied in 
this cluster and the three data cubes were 
analyzed separately.  
The velocity coverage was ensured by applying 
a bandwidth of 12.5 MHz and correlator modes 
1A for the central cube, and 1D for the NE and
SW cubes.  This strategy delivered a central 
data cube with a coverage of 4,000\kms\ and
nearly 3,000\kms\ for the NE and SW cubes 
(Table\,\ref{tab_cubes}).

Standard VLA calibration and imaging were 
carried out by using the NRAO Astronomical 
Image Processing System (AIPS).  All final 
data cubes for A85 and A496 have 31 channels 
with velocity widths between 43.8\kms\ and 
46.9\kms. The central field A2670-C has 
88 channels while 2670-NE and 2670-SW have 63 
channels.  All data cubes for this cluster 
have a channel width close to 48\kms.  The 
imaging was made by applying \texttt{ROBUST}$=$ 
1.0 \citep{Briggs95}, which is the best 
compromise between uniform and natural 
weighting.  This maximizes the sensitivity
while keeping high spatial resolution, 
delivering a final beam size of about
24\prin \por\,17\prin\ for the 
three clusters.  The typical rms is 
\aprox\,0.24\,\mJyb\ for A85 and A496,
while \aprox0.12\,\mJyb\ was reached 
for A2670 (see Table\,\ref{tab_cubes}).
This allowed to obtain \hi-mass 
detection limits of the same order of magnitude
for the three systems.

The search for detections and the \hi\ 
analysis were carried out using AIPS.  The 
moment maps were produced 
with the AIPS task \textsc{MOMNT}, 
applied with Gaussian 
(space coordinates) and Hanning smoothing 
(velocity) to maximize the signal-to-noise ratio.  
We produced \hi-maps and velocity fields using a 
cutoff level between 1.5 and 2.0 times the rms 
measured in the corresponding data cube. As expected,
variations of noise were found at the edge of the
data cubes, in RA, Dec and velocity. We
normally discarded detections found within those 
zones.

\subsection{CFHT MegaCam optical images }
%the \hi-data cubes}
\label{obs_opt}

%Florence

We obtained optical images with the 3.6m Canada 
France Hawaii Telescope and the MegaCam instrument, 
each field having a size of around $1\times 1$~deg$^2$. 
The south tip and the infalling filament of A85 
were observed in October 2004 (program 04BF02); we 
exploit here the $u$ $g$ $r$ $i$ band images, with 
surface magnitude limits of 26.7, 27.3, 26.4, 
25.7 mag~arcsec$^{-2}$, respectively. Details can be 
found in \cite{Boue08b}.  We later retrieved archive 
images, this time centered on A85 in the $g$ and $r$ 
bands. The corresponding depths are 26.3 and 
26.1 mag~arcsec$^{-2}$. We assembled the individual 
images (five in each band) to obtain frames with 
total exposure times of 3960\,s and 8380\,s. 

A496 was observed with the same telescope and 
instrument (November 2003, program 03BF12) obtaining 
$u$ $g$ $r$ $i$ band images. The total exposure time 
is 7853~s for each image and the depths are 
27.4, 27.2, 26.2, 25.6  mag~arcsec$^{-2}$, respectively. 
Details on these data can be found in \cite{Boue08a}.

The data for A2670 were taken from the CFHT archive, 
with a total exposure time of 6160~s in the 
$u$ $g$ $r$ $z$ bands. The depths are 25.7, 26.8, 
26.3, 24.2 mag~arcsec$^{-2}$, respectively.

\subsection{The membership catalogues }
\label{redshift_cat}

We use the spectroscopic redshift compilation by 
H. Andernach (priv. comm.) to define the member 
sample of the three clusters.  This compilation 
considers as potential members all galaxies with 
published radial velocities inside a projected 
Abell radius ($R_{A} = 2.1~h_{70}^{-1}$ Mpc), 
and within a range of \aprox2500\,\kms\ from 
a preliminary estimation of the cluster velocity.  
Several observational limitations 
%affect these catalogues with a few objects , for instance, the low S/N for low surface brightness objects, missing spectra within very crowded zones, etc. This 
could prevent the redshift catalogues to be fully 
complete within the studied regions; nevertheless 
it remains the most reliable option with the
data currently available in the literature.
Taking this redshift sample, the membership 
catalogues are constructed by tracing caustic 
curves representing the escape velocity of the 
system in a projected phase-space diagram  
\citep[see,][and references therein for more details]{Serra11}. 
The curves are constructed following \citet{Chow19}.  
All galaxies within the defined limits are taken as 
cluster members.  The caustics applied here are 
intentionally {\it relaxed} (an enclosing fit) 
as the catalogues are intended to retain as many 
potential member galaxies as possible.   The 
membership catalogues contain 616, 368, 308 
objects in A85, A496, A2670, respectively.  
Due to its larger redshift A2670 is sampled through 
a larger linear radius. This bias will be taken 
into account when we compare the observed clusters 
with each other in Sect.\,\ref{discussion}.

\subsection{Catalog of the brightest spiral galaxies }
%the \hi-data cubes}
\label{obs_catalog}

\hi\ blind surveys covering large cluster
volumes allow to study the gas component 
of all the member galaxies contained within 
the observed data cubes. Even those spirals 
not detected in \hi\ are very important as 
they trace the distribution of \hi\ swept 
spirals throughout the studied cluster.
%and to study the gas depletion mechanisms at work. 
%To achieve these goals, disposing 
%of a complete sample of spirals contained 
%within the \hi\ data cubes, is a key step.
%In addition, the fraction of \hi\ non-detected 
%bright spirals constitutes a reliable indicator
%of the cluster deficiency as a whole.
With this in mind we built catalogues (RA, 
Dec, $v_{rad}$) of the brightest spirals 
contained within the VLA data cubes.
%of each studied cluster.  
%We dispose of \hi\ information for all these 
%objects disregarding if they are detected or not.
%This sample of galaxies will be used to trace 
%the degree of gas loss across the observed volumes. 
%We apply a magnitude limit with the 
%aim to obtain a degree of completeness.
%
%are complete up to a magnitude limit defined by two criteria. %First, galaxies in this sample must be bright enough to check %their morphological type through  visual inspection; and second, %these spirals must be massive  enough to be --statistically--
%
As a first step, magnitudes $g, r$ were obtained 
for all the member galaxies by applying
SExtractor\footnote{https://www.astromatic.net/software/sextractor/}
to the CFHT images referred above. A few zones  
observed by the VLA but not covered by the CFHT 
(Fig.\,\ref{fig_cubes}) were similarly analyzed
%through \textbf{$g, r$} 
%other images available, for instance from  
%SDSS,\footnote{https://skyview.gsfc.nasa.gov/current/cgi/query.pl} %and \textbf{$g, r, i$} bands
by using 
Pan-STARRS\footnote{https://panstarrs.stsci.edu/} 
frames \citep[]{Flewelling20}. 
%Absolute magnitudes were calculated by using the 
%corresponding cluster distances. 

The strategy to define the bright spiral catalog 
is as follows. We applied a limit in apparent 
magnitude of $g=$\, 19.0, 18.0, 19.5,
%$b_j$\,=\,18.5, 17.5, 19.0, 
for A85, A496, A2670, respectively.  
Considering the distance to each cluster, 
these values roughly correspond to the same 
absolute magnitude, M$_{g}$\aprox\,-18.0.  
%in case of, add as references Taylor+2011, 
%Mahajan+2018, 2018MNRAS.475..788M
%This is a conservative value for the purpose of this study; 
In {\it r}-band this limit is close to 
M$_{r}=-$18.2.   We convert these magnitudes 
to stellar mass following \cite{Mahajan18},
showing that our catalogues of bright spirals 
are complete in mass above a value of
log$ (M_{*} / M_{\odot})=9.0$.  
%Galaxies in 
%this range are expected to be detected in \hi.
%
%as it is one magnitude fainter than the small spiral M33 (M$_{\mathrm{B}}=-$19.4).  
%We will show that the absolute magnitude 
%M$_{\mathrm{B}}=-$18.4 roughly corresponds 
%to a red magnitude of M$_{\mathrm{r}}=-$19.0.
%Objects with this luminosity are expected to
%have a mass of log$ (M_{*} / M_{\odot})=9.0$ 
%\citep[see][]{Mahajan18}. 
%As we will show 
%later, our survey is capable to detect 
%galaxies two magnitudes fainter.
%
%The strategy to build our catalogs of bright spirals 
%is the following.  First we applied a SQL request 
%through HyperLEDA\footnote{http://leda.univ-lyon1.fr/}, 
%compiling all the late-type galaxies ($T \geq 1 $) 
%within the (RA, Dec) window covered by the VLA data 
%cubes.  We applied the apparent magnitude limit 
%mentioned above for each cluster.  This {\it 
%step-one} list of late-types provided by HyperLEDA 
%was complemented with the output delivered by the 
%photometric SuperCOSMOS Sky 
%Survey\footnote{http://www-wfau.roe.ac.uk/sss/}. 
%Here we selected all galaxies within the \hi\ 
%observed fields with a restriction in colour
%index as a proxy for the morphological type. 
%We applied red values of (B-R)\,=\,1.2 as a 
%typical threshold, ensuring that possible 
%red spirals in our clusters are not missed.
%This added a few objects and produced a more 
%complete list, named {\it step-two}.
%\textcolor{Purple}{\textbf{a range 0<B-R< 1, ~~
%0<B-R< 1.2, ~~ -0.2<B-R< 1.64 for A85, A496 and A2670}}
%
We did not apply a color-index threshold in 
order to include any possible red spiral. This 
method delivered bright objects whose morphology 
can be defined through visual inspection; this
was done by using combined (RGB) optical images 
from CFHT and Pan-STARRS.  All objects in the blue
cloud (see next subsection) and having stellar mass 
above log$ (M_{*} / M_{\odot})=9.0$ were considered 
as candidates to be spiral.  Our visual inspection 
confirmed spiral features for most of those objects. 
The remaining ones, before being included in 
the sample of bright spirals, should present
a combination of features based on a disky shape, 
blue ($g-r$) color, stellar mass, \hi\ content, and 
the presence of dust lanes (for the near edge-on 
ones).  As shown below, a few red spirals are 
reported in the three clusters. Their belonging to
the bright spiral sample was conditioned by
the following criteria: having a mass above  
log$ (M_{*} / M_{\odot})=9.0$, being detected 
in \hi\ or, alternatively, showing spiral features
through our visual inspection.  
%
%  Given the defined stellar mass range, these galaxies are expected to be detected in \hi\ with our survey under unperturbed conditions (see  Sect.\,\ref{results}). 
%\textbf{Margarita: I'm not sure about these two last sentences. The BSp catalogs were build to find spirals not detected in \hi}
%
This output was cross-matched with the list of 
member objects (Sect.\,\ref{redshift_cat}) in
order to ensure that these spirals are 
cluster members.
%NED\footnote{https://ned.ipac.caltech.edu/} 
%database in order to select only cluster 
%we restricted the selection to the velocity 
%A few \hi\ detections appear as non cluster 
%members. They may be galaxies outside the 
%velocity window of the member catalogues;
%the \hi\ data cubes extend slightly beyond
%such a window.  
%%Any more detailed morphological classification 
%is beyond the scope of this work.
The final catalogues of bright spirals consist 
of 65, 46, 61 member 
objects in A85, A496, A2670, respectively. 
Hereafter we denote these objects as B-Sp. 
%All these galaxies have stellar 
%masses above log$(M_{*}/M_{\odot})=9.0$, 
%with a few of them going beyond
%log$(M_{*}/M_{\odot})=10.5$ in A2670.

%
%Most of them lie in the magnitude 
%(M$_\mathrm{r}$) range  between -19.0 to
%-21.0, corresponding to stellar masses
%log$(M_{*}/M_{\odot})$ from 9.0 to 10.0.
%A few B-Sp reach M$_\mathrm{r}=$-22.0,
%or log$(M_{*}/M_{\odot})=$10.5.
%This search for the brightest late-types  
%allowed to reclassify as spirals a number 
%of objects previously reported as ellipticals 
%in NED:  ten in A85, one in A496, eight in 
%A2670 (see Tables\,\ref{tab_A2670_no_det}).
%These galaxies look red in our RGB frames but 
%our visual inspection confirmed the presence of spiral patterns. 

%  THIS MAY STILL BE OF INTEREST!!!
%
%Based on both, the bright-spiral and the redshift catalogues 
%of member
%galaxies (Andernach 2021, priv. comm.) we are able to obtain 
%a preliminary estimation of the LTG/ETG ratio in our clusters.  
%In order to be consistent, we estimate this relation 
%for member galaxies lying within the HI-cube limits 
%($\alpha, \delta, vel$) and being brighter than $M_B = -18.4$.  
%Very interestingly we obtain similar values of LTG/ETG: 
%28\%, 36\%, 27\%, for A85, A496, and A2670, respectively. 
%??? MARGARITA, NOW THAT WE
%DEFINE THE VOLUME AND THE MAGN LIMIT, DO WE STILL GET THESE
%NUMBERS?
% {\bf A85, A496 and A2670 = 56*100/193= 29\% , 32*100/90= 36\%, 50*100/171= 30\%}

\begin{figure}
\begin{subfigure}[tbp]{0.42\textwidth}
 \includegraphics[width=\columnwidth]{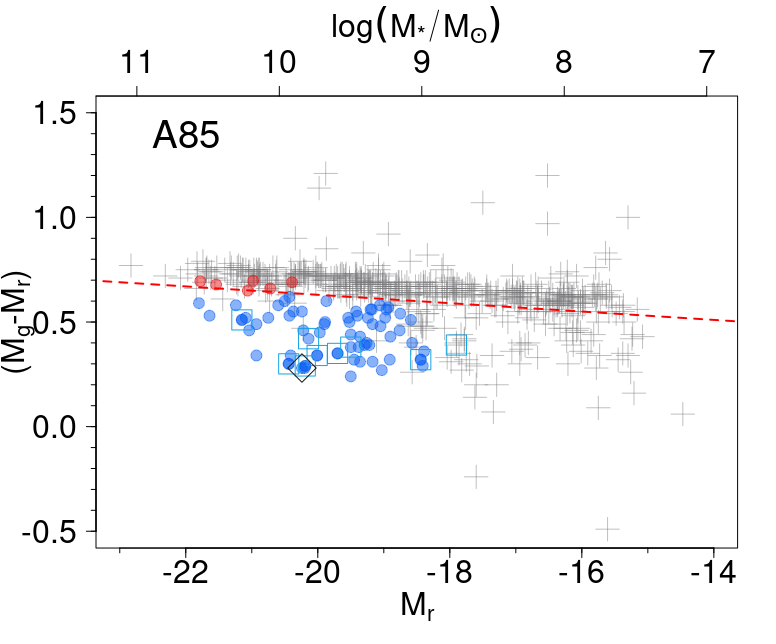}
% \caption{}
  \end{subfigure}
  \\
 \begin{subfigure}[tbp]{0.42\textwidth}
 \includegraphics[width=\columnwidth]{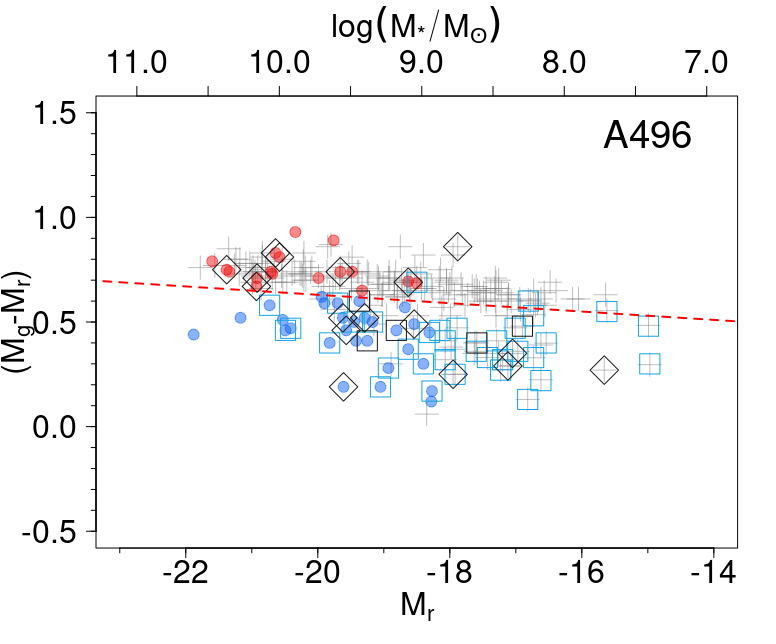}
% \caption{}
  \end{subfigure}
  \\
   \begin{subfigure}[tbp]{0.42\textwidth}
 \includegraphics[width=\columnwidth]{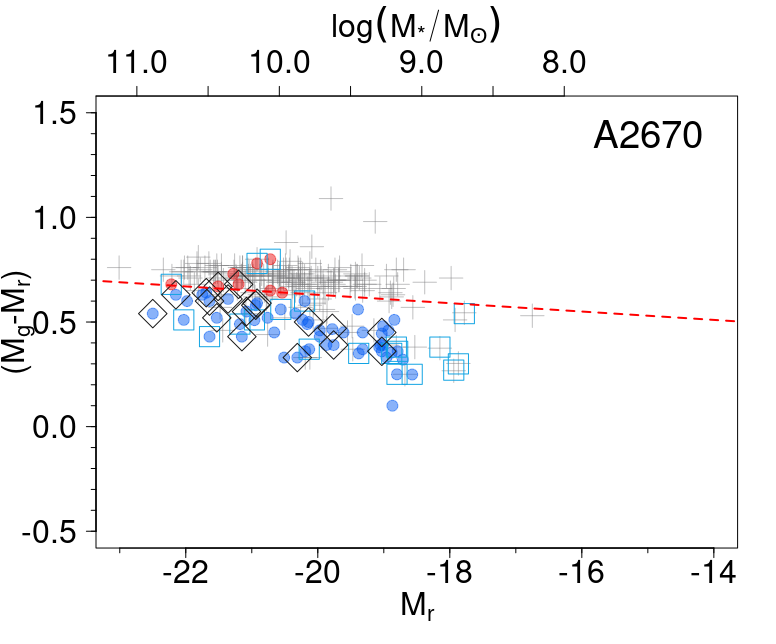}
% \caption{}
  \end{subfigure}
  \caption{Colour-magnitude diagrams showing cluster 
  members with grey crosses. The red dotted line draws 
  the border between the red sequence and the green valley
  (see text). 
  The bright spirals are shown with solid circles 
  (blue and red). Squares/diamonds indicate the \hi\ 
  normal/abnormal galaxies (see Sect.\,\ref{results}).}
  \label{fig_colour-mag}
\end{figure}

\subsection{Color-magnitude diagrams: the galaxy populations}
%the \hi-data cubes}
\label{obs_colors}

Tracing color-magnitude diagrams (CMD) in a 
cluster helps to visualize the distribution 
of different types of galaxies populating
the system.  Fig.\,\ref{fig_colour-mag} shows 
the CMD of the three clusters, giving the color 
index as a function of the absolute magnitude
M$_{r}$ (bottom axis) and stellar mass
(upper axis).  The red dotted line draws the 
border between the red sequence and the green 
valley; it is defined as:  ($g-r$) = 0.63 $-$ 
0.02\,(M$_{r}$ + 20) \citep{Masters10}.

Member galaxies in Fig.\,\ref{fig_colour-mag} 
are shown with grey crosses.  The bright 
spirals defined in the previous section 

are cluster members too; they are indicated 
with solid circles, either red 
or blue, indicating their position in the red 
sequence or in the blue cloud.  Cluster members 
in the red sequence being brighter than 
M$_{r}=-$18.2 ($i.e.$ mass above 
log$ (M_{*} / M_{\odot})=9.0$), and having no 
solid circle, correspond to early types. 

Galaxies above the same mass limit and lying 
in the blue cloud are confirmed to be part of 
the B-Sp sample (see previous subsection).
%; a few of them indicated with grey crosses (mainly in A2670) indicate galaxies lying outside the FoV of the VLA.  
The \hi\ detections 
%(see  Sect.\,\ref{results}) 
are shown with blue open 
squares (normal objects) or with black diamonds 
(\hi-abnormal objects, defined in the next 
section).  A few \hi\ detections 
%have no grey cross indicating that they are 
are not part of the 
compilation of member galaxies; either no 
redshift was available or they are outside the 
velocity window defined by the membership catalog
of Sect.\,\ref{redshift_cat}.  We added these 
objects as new cluster members.
%They have no impact in our final results.

%As we mentioned before, the fields covered
%by the VLA and the CFHT are not exactly the 
%same.  As a consequence, a few objects observed 
%in \hi\ lie outside the optical frames and we did 
%not obtain the corresponding photometry.  This 
%accounts for a minor inconsistency between the 
%number of objects reported and the corresponding 
%symbols shown in Fig.\,\ref{fig_colour-mag}. 
%Nonetheless, this fact does not affect our 
%statistical results. 

Galaxies located in the blue cloud of these
clusters are good candidates to be in a 
pre-quenching stage. Optical colors are not 
the best indicators for current star-forming 
activity due to possible differences between 
the time scales for quenching and for the 
change in color due to the stop of star 
formation. Nevertheless, not having optical 
spectra for all the objects under study, we 
consider that a blue color combined with a 
normal gas content, constitute a reasonable 
indicator for galaxies in a pre-quenching phase.
%particular for a statistical approach.
On the other side, objects above the red line
of Fig.\,\ref{fig_colour-mag} are not expected 
to have any intense star formation activity,
therefore red circles in this figure
constitute very likely passive spirals. 
%Interestingly, we
%report none in A85 while some appear in A496 
%(fifteen) and in A2670 (eight).  
A K-correction was applied to the three clusters. 
The largest K-correction values in color-index 
are slightly above 0.1~mag. As expected, these 
values correspond to red objects in A2670 (we 
used the K-correction calculator from 
GAISh\footnote{http://kcor.sai.msu.ru}).
As this correction is associated with the
galaxy distance, the y-axis of
Fig.\,\ref{fig_colour-mag} shows the 
difference in absolute magnitudes after 
the K-correction was applied.
%This accounts for the notation $(g'-r')$.
Galactic extinction is negligible for these
clusters and no correction was applied.

\begin{figure}
\begin{subfigure}[tbp]{0.44\textwidth}
 \includegraphics[width=\columnwidth]{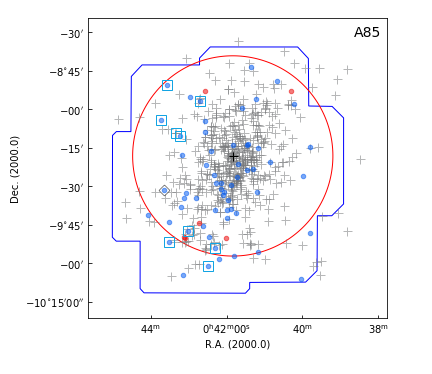}
% \caption{}
  \end{subfigure}
  \\
 \begin{subfigure}[tbp]{0.44\textwidth}
 \includegraphics[width=\columnwidth]{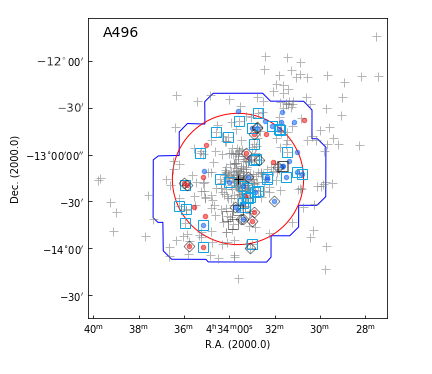}
% \caption{}
  \end{subfigure}
  \\
   \begin{subfigure}[tbp]{0.44\textwidth}
 \includegraphics[width=\columnwidth]{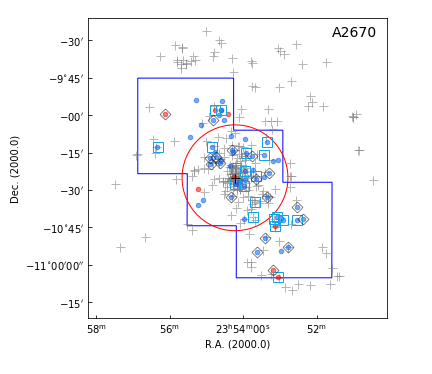}
% \caption{}
  \end{subfigure}
  \caption{The distribution of member galaxies (grey 
  crosses) in the studied clusters. \hi\ detections 
  are shown with blue squares (normal) and black diamonds 
  (abnormal).  Symbols for the bright spirals are the 
  same as in  Fig.\,\ref{fig_colour-mag}  (blue and red 
  circles).  R$_{200}$ is indicated with a red circle 
  and the blue polygon shows the VLA field of view. 
  The cluster center is indicated with a red cross.}
  \label{fig_distribution}
\end{figure}

\section{Results}
\label{results}

\subsection{The \hi\ detected galaxies }
\label{res_detect}

The catalogues of bright spirals (B-Sp) described 
above provide positions and velocities that were 
used to seek for their \hi-emission in the 
corresponding data cubes.  Most of these B-Sp 
were effectively detected in \hi\ but, in all 
three clusters, significant fractions of these 
spirals were not detected.  These objects 
constitute our sample of \hi\ non detected 
spirals. This sample is complete in mass above 
the limit log$ (M_{*} / M_{\odot})=9.0$.  The 
search for 21\,cm emission was completed by a 
thorough visual inspection of the \hi\, data cubes. 
This was done blindly, disregarding all information 
on optical positions and redshifts of the spiral 
galaxies.  Our search produced a list of emissions, 
most of them matching in position and velocity with 
a B-Sp. However, several candidate detections had 
no counterpart in the B-Sp catalog.  A subsequent 
search upon the optical-RGB frames and redshift
catalogues allowed the finding of the optical 
counterparts. As expected, these are blue, 
irregular galaxies, fainter than the limit 
defining the B-Sp sample.  
%A few of these  low-mass \hi-rich objects constitute new cluster  members.
%and they are indicated with open 
%squares and diamonds without a grey cross 
%(Figs.\,\ref{fig_colour-mag} and 
%\ref{fig_distribution}).

Our strategy to search for \hi\ emitting galaxies 
includes a conservative detection threshold above 
6 times the rms, delivering 10, 58, 38 
\hi-detections in A85, A496, A2670, respectively.
%(see Table\,\ref{tab_results}).
A few \hi\ objects are just marginally resolved 
by the VLA primary beam:  two in A85, six in A496, 
eleven in A2670.  All maps and velocity fields are 
available as online material. 
%Figs.\,\ref{fig_A85_HImaps}, \ref{fig_A496_HImaps}, \ref{fig_A2670_HImaps}.
%We display the \hi\ maps of the full sample of 
%detections disregarding the fact that they are fully 
%resolved or not. 
%
We used the \hi\ mass detection limits given in 
Table\,\ref{tab_cubes} to ensure that the sample 
of bright spirals (Sect.\,\ref{obs_catalog}) is 
potentially detected with this survey.  Obviously, 
%for the same fraction of swept gas, 
more massive spirals have higher chances to be 
detected than less massive ones. Therefore, we 
inspect the performance of our survey detecting 
galaxies with different stellar masses. We 
displayed our \hi\ mass detection threshold onto 
a \mhi\ vs M$_{*}$ plot (not shown in this work).  
We added the scaling relation obtained by 
\cite{Denes14} and we found that normal galaxies 
can be effectively detected down to 
log$(M_{*}/M_{\odot})=9.0$.  The deeper data 
cubes in A496 pushed this limit down to 
log$(M_{*}/M_{\odot})=8.7$.  This confirms that 
our full sample of B-Sp can be detected with the 
present survey under unperturbed conditions.
%having masses log$(M_{*}/M_{\odot}) \geq$9.0, are expected 

Tables\,\ref{tab_A85_det},\,\ref{tab_A496_det} 
and \ref{tab_A2670_det} give the main optical 
and \hi-parameters of the detected galaxies
in A85, A496, A2670 respectively.
The lists of non detected spirals are given 
in Tables\,\ref{tab_A85_no_det}, 
\ref{tab_A496_no_det}, \ref{tab_A2670_no_det}
for the same clusters.   Five \hi\ detections 
in A496 could not be properly measured as we
did not recover their full emission; this is 
due to limited velocity coverage of the 
corresponding data cube.  We consider these 
galaxies as \hi\ detections because the 
recovered emission has large S/N; they
are indicated with black squares in
Figs.\,\ref{fig_colour-mag} and \ref{fig_distribution}.

In Tables\,\ref{tab_A85_det},\,\ref{tab_A496_det},
\ref{tab_A2670_det}, the first five columns give 
names, coordinates and morphological type; these 
data are taken from the literature 
(NED\footnote{https://ned.ipac.caltech.edu/} and
HyperLEDA\footnote{http://leda.univ-lyon1.fr/}).  
Column 6 and 7 give optical magnitudes 
and ($g-r$) colors obtained from 
CFHT and from Pan-STARRS images.
In Column (8) we provide the stellar masses 
obtained from our M$_{\mathrm{r}}$ magnitudes. 
%and following \cite{Mahajan18}. 
Columns (9) and (10) give the optical and \hi\ 
velocities, the latter in heliocentric frame.  
We consider the value of the central channel 
displaying emission as the \hi\ velocity.  We 
give the \hi\ velocity width in Column (11) 
defined by the full range of channels containing 
emission; this constitutes a proxy for $W_{20}$. 
The \hi\ mass (Column 12) in solar masses is 
calculated by using the equation:

%\citep[][]{Haynes-Giovanelli84}: 
\begin{equation}
	\left[ \frac{M_\mathrm {HI}}{M_{\odot}} \right] =2\ldotp 36\, \times \,10^{5} \cdot \left[\frac{D_{c}}{\text{Mpc}} \right]^2 \cdot \left[ \frac{S_\mathrm {HI}}{\text{Jy km }\text{s}^{-1}} \right] 
	\label{HI_mass}
\end{equation}

\noindent
where we use the corresponding comoving distance 
($ D_{c}$) for all the members of the same system. 
The uncertainty for M$_\mathrm{HI}$ is systematically 
$\leq$ 10\%, derived from the computed error on
the total flux $S_\mathrm{HI}$.
The \hi-deficiency is estimated with the equation 
\citep[][]{Haynes-Giovanelli84}:
\begin{equation}
Def_{\mathrm {HI}}=log(M_\mathrm {HI})_{exp}-log(M_\mathrm {HI})_{obs}
	\label{ecu:defHI}
	\end{equation}

\noindent
Given that only a reduced number of our galaxies 
has reliable morphological types in the 
literature, we computed the $Def_\mathrm{HI}$ 
parameter for the two cases proposed by 
\cite{Haynes-Giovanelli84}, 
$i.e.$ for early-spirals and for late-spirals.
These values can be taken as lower and upper limits 
(Columns 13 and 14) of the deficiency parameter.   
Column (15) gives the projected distance to the 
cluster center, using the BCG position.  In Column 
(16) we provide a code indicating four different 
kinds of disruptions observed in \hi; this helps 
to define the sub-sample of \hi\ abnormal galaxies.  
The codes given in this column are the following:\\
-- {\it N} : ~galaxies with normal 
\hi\ content and distribution. \\
-- {\it Def} : indicates galaxies having lost more than 
50\% of their gas ($Def_\mathrm {HI}$\,$\gtrsim$\,0.25).  \\
-- {\it Asy} : ~denotes a resolved \hi\ asymmetry, 
or a tail, unveiled by the \hi\ maps. \\
-- {\it Pos} : ~indicates an offset $\geq$ 20\,kpc
in position, between the \hi\ and optical centroids;
this limit roughly corresponds to the diameter of 
a typical spiral and is equivalent to 19\prin, 30\prin, 
14\prin, for A85, A496, A2670 respectively.\\
-- {\it Vel} : ~denotes a velocity offset between 
the optical and \hi\ velocities;  we use three times 
the \hi\ channel width to define this offset
(\aprox 135\kms).

Galaxies having at least one of the four criteria 
mentioned above are defined as \hi\ abnormal.
We used this sub-sample of perturbed objects 
to carry out a search 
for a correlation between \hi-deficiency and 
\hi-asymmetry, finding no obvious trend. 
However this result based on gas deficient, 
yet \hi\ detected galaxies, must be taken with 
caution as these objects are caught at an early 
stripping phase.  We have no information on the 
2-D gas distribution of the even more stripped 
spirals, $i.e.$ those not detected in \hi.
Therefore a deficiency--asymmetry correlation,
developing during later gas stripping stages,
should not be discarded {\it a priori}.

%\begin{figure*}
\begin{figure}
%\begin{subfigure}[tbp]{0.66\textwidth}
\begin{subfigure}[tbp]{0.44\textwidth}
 \includegraphics[width=\columnwidth]{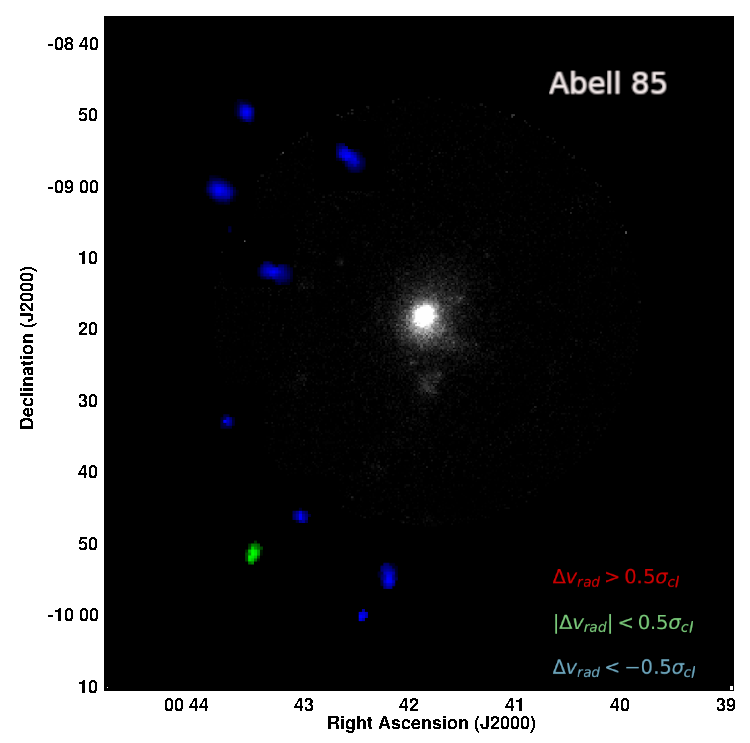}
% \caption{}
  \end{subfigure}
  \\
  %\vspace{1cm}
 %\begin{subfigure}[tbp]{0.66\textwidth}
 \begin{subfigure}[tbp]{0.44\textwidth}
 \includegraphics[width=\columnwidth]{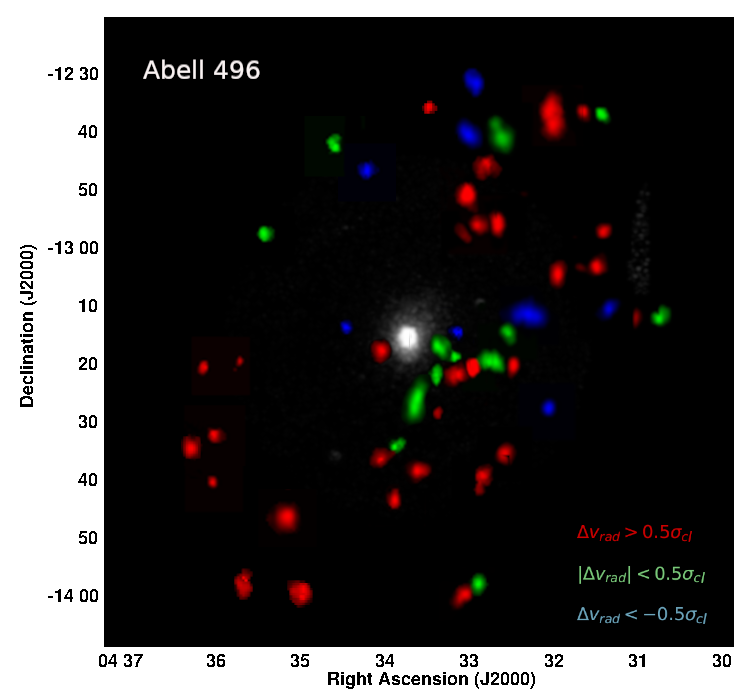}
% \caption{}
  \end{subfigure}
  \\
 % \caption{}
   %\label{fig_distribution_HImaps}
%\end{figure*}

%\begin{figure*}  \ContinuedFloat
%\centering
  % \begin{subfigure}[tbp]{0.66\textwidth}
   \begin{subfigure}[tbp]{0.44\textwidth}
 \includegraphics[width=\columnwidth]{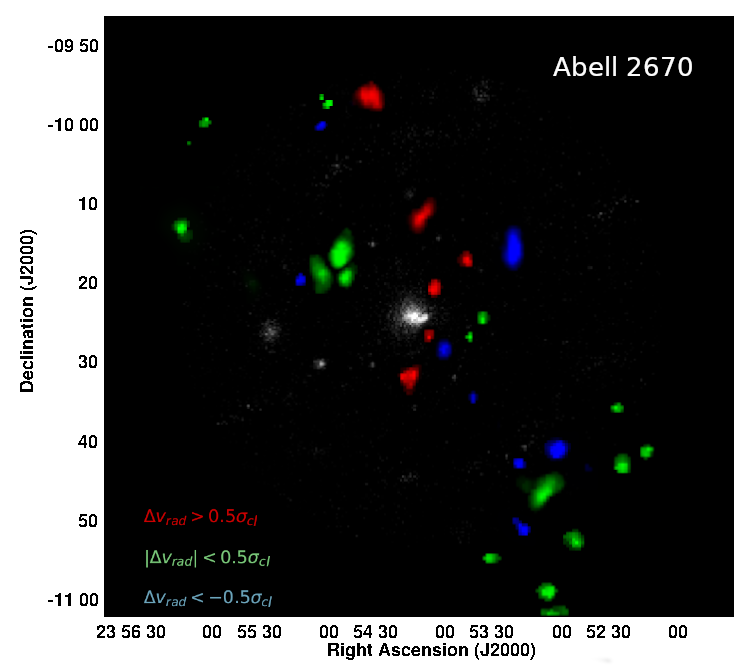}
% \caption{}
  \end{subfigure}
  \caption{The distribution of HI detected galaxies 
  in A85 (top), A496 (middle) and A2670 (bottom). 
  These objects are separated in three slices of 
  velocity relative to the cluster center 
  (see text).  The \hi\ maps are enlarged 
  by a factor of 5. The ROSAT X-ray emission is 
  shown in grey scale.}
  \label{fig_distribution_HImaps}
%\end{figure*}
\end{figure}

\begin{table}
    \centering
    \caption{Different types of galaxies in the studied clusters}
    \label{tab_results}
    \begin{tabular}{clccc}
    \hline 
& Sample & A85 & A496 & A2670\\
%	(1)  & (2)  & (3)   & (4)  \\
	\hline
(1) & Members	&	616	&	383	&	318	\\
(2) & Memb. VLA-FoV	&	603	&	336	&	254	\\
(3) & \hi-det	&	10	&	58	&	38	\\
(4) & B-Sp	&	65	&	46	&	61	\\
\multicolumn{4}{l}{}\\
(5) & B-Sp \hi-det	&	9	&	29	&	33	\\
(6) & B-Sp \hi\ non-det	&	56	&	17	&	28	\\
(7) & B-Sp in red seq	&	6	&	17	&	8	\\
\multicolumn{4}{l}{}\\
(8) & \hi-det in red seq 	&	0	&	11	&	5	\\
(9) & \hi-det abnormal	&	1	&	17	&	18	\\
(10) & \hi-det low mass	&	1	&	28	&	4	\\
\multicolumn{4}{l}{}\\
(11) & Memb. virial-region	&	140	&	203	&	108	\\
(12) & B-Sp virial-region	&	23	&	20	&	17	\\
(13) & \hi-det virial-region	&	1	&	24	&	12	\\
	\hline
%\multicolumn{84}{l}{Notes:}\\
%\multicolumn{4}{l}{--}\\
%\multicolumn{4}{l}{--}\\
\end{tabular}
\end{table}

\subsection{Global distribution of \hi\ across the studied clusters}
\label{res_distrib}

Some of our main results are displayed in 
Fig.\,\ref{fig_distribution}, 
Fig.\,\ref{fig_distribution_HImaps} and in
Table\,\ref{tab_results}.  The (RA, Dec) 
distribution of member galaxies in the three 
studied clusters is shown in Fig.\,\ref{fig_distribution}.
We use the same symbols as in Fig.\,\ref{fig_colour-mag} 
in order to emphasize the position of the bright 
spirals (blue/red circles) and the \hi-detected 
galaxies (open blue squares/black diamonds).  The 
total fields covered with the VLA are drawn with 
blue polygons and the corresponding R$_{200}$ are 
indicated with red circles.

%Considering the abnormal criteria
%described in the last section 
%given in Tables\,\ref{tab_A85_det},\,
%\ref{tab_A496_det}, \ref{tab_A2670_det} 
%we observe the following fractions of \hi\ 
%peculiar objects: 10\%,  29\%,  47\%, 
%in A85, A496, A2670 respectively.  Some 
%caution must be taken before comparing 
%one cluster with each other because 
%their different redshifts imply unequal 
%spatial resolutions {\bf MAYBE THIS 
%PARAGRAPH SHOULD BE TAKEN TO SECT. 6 AS IT MAKES 
%NO SENSE TO COMPARE THESE NUMBERS HERE}.  
%In this work we discuss the global \hi\ properties of each cluster and we postpone the study of individual galaxies (gas loss fraction,   starforming activity/quenching, etc.) to a forthcoming paper.

Fig.\,\ref{fig_distribution_HImaps} shows the 
moment-zero \hi-maps of the detected galaxies in 
each cluster.
%A85, A496, A2670 are shown in panels a, b, c, respectively.
Galaxies are set at their proper location and are 
magnified by a factor of five for clarity.  We use 
a color code in order to separate these galaxies 
in three radial velocity regimes: lower, similar
and larger than the cluster velocity.  We give 
statistical significance to the velocity slots 
by applying the same criteria as \cite{Wang21},
given as a function of the velocity dispersion 
of the cluster ($\sigma_{cl}$) given in 
Table\,\ref{tab_clusters}: 
blue color for 
$\Delta v_{rad} < -0.5\sigma_{cl}$,
green for 
$-0.5\sigma_{cl} < \Delta v_{rad} < 0.5\sigma_{cl}$, 
and red for $ 0.5\sigma_{cl} < \Delta v_{rad}$.
The X-ray emission is shown in grey scale
taking advantage of the homogeneous and 
large field map given by ROSAT. 

Table\,\ref{tab_results} reports the total number of 
objects having different observational properties 
and lying in specific cluster regions. 
Data are displayed as follows:  
Row (1) gives the full number of cluster members, 
including the new members derived from the \hi\ 
velocities.   Row (2) provides the number of galaxies 
within the VLA FoV. 
Rows (3) and (4) 
give the total number of \hi\ detections and bright 
spirals.  Rows (5), (6), (7), give the number of B-Sp 
being \hi\ detected, the B-Sp non-detected and the
B-Sp lying in the red sequence, respectively. Rows (8), 
(9), (10), give the number of galaxies detected in 
\hi\ lying in the red sequence, the \hi\ abnormal, and 
those below the limit log$(M_{*}/M_{\odot})=9.0$, 
respectively.  The last three rows, (11), (12), (13),
provide the number of galaxies projected within the 
virial region (Sect.\,\ref{disc_PPS}) being: cluster
members, bright spirals and \hi\ detections, 
respectively.  The data given in 
Table\,\ref{tab_results} are not well suited for a 
direct comparison of the clusters with each other 
because of their different redshifts and VLA coverage. 
We carry out a comparison under homogeneous conditions 
in Sect.\,\ref{discussion}. \\

\noindent
$\bullet$ {\bf The \hi\ in Abell 85}

Fig.\,\ref{fig_distribution} (upper panel) shows 
the distribution of 616 member galaxies in A85
where the solid circles indicate the 65 B-Sp.
%both those in the red sequence and those below it 
%(red and blue circles, respectively). 
The blue open squares and one black diamond 
indicate the ten \hi\ detected objects.  
%Two \hi\ detections with no solid circle, seen to the NE, correspond to optical faint objects not being part of the B-Sp catalog.
%(A85[DFL98]461 and A85[SDG98]3114); not being part of the brightest
%spiralscatalog they are indicated with empty squares.  
The cluster members (grey crosses) are rather 
homogeneously distributed across A85, while the 
B-Sp and the \hi-detected galaxies show a very
asymmetric distributions across the cluster. The 
SE zone has more bright spirals than all the other 
quadrants taken together; all the \hi-detections 
are projected as well onto the SE-NE regions (Fig.\,\ref{fig_distribution} and 
Fig.\,\ref{fig_distribution_HImaps}). The 
gas rich galaxies are projected 1.5$-$3.0\,Mpc 
away from the cluster center, while a remarkable 
absence of \hi\ detections at lower radius suggests 
the presence of a harsh environment. This 
fact is independently supported by the presence 
of strongly perturbed galaxies 
\citep[][]{Venkatapathy17, Ramatsoku20, Luber22}.

The sample of 65 B-Sp in A85 is unexpectedly 
high.  A significant fraction of these objects 
are projected around the cluster core and in the 
transition zone between the center and the SE 
(Fig.\,\ref{fig_distribution}). Six B-Sp are 
reaching the red sequence (red solid circles 
in Fig.\,\ref{fig_colour-mag}) and none of them
is detected in \hi, indicating that they have 
gone through an advanced stripping process.
%and might constitute red passive spirals. 
On the contrary, all the galaxies detected 
in \hi\ show very blue colors 
%and lie well below the red sequence 
indicating that they have not gone 
through any stripping phase. 

%The largest overdensity of B-Sp in A85 occurs in the SE. 
The asymmetry 
in the distribution of \hi-detections is even more 
striking in Fig.\,\ref{fig_distribution_HImaps}, 
showing that all --but one-- detections are in the 
low velocity domain, with a narrow $\Delta v$ of
\aprox\,15,000$-$15,200\,\kms\ (see 
Table\,\ref{tab_A85_det}).  Their low velocity 
dispersion and the large difference compared with 
the systemic velocity of A85 (\aprox\,1,500\,\kms) 
suggest that they could share a common origin. \\

\noindent
$\bullet$ {\bf  The \hi\ in Abell 496}

The middle panel of Fig.\,\ref{fig_distribution} 
shows the distribution of 383 member galaxies in 
A496, with 46 B-Sp and 58 galaxies detected in 
\hi\ (symbols are the same as in 
Fig.\,\ref{fig_colour-mag}).
A first remarkable difference of A496 when 
compared with A85 is given by the lower number 
of bright spirals (46 in A496, and 65 in A85)
but a much larger number of \hi-detections 
(58 in A496 for only 10 in A85; see 
Table\,\ref{tab_results}). 
Interestingly, \aprox50\% of the \hi\ detected 
galaxies in A496 correspond to gas-rich, low 
mass objects which are not part of the B-Sp 
sample, $i.e.$ their stellar mass is below  
log$ (M_{*} / M_{\odot})=9.0$.  This fraction
is partly explained by A496 being closer than 
the other studied clusters. However, we will show
that some peculiarities remain after considering
the observational bias due to the difference
in redshift.
%(see Sect.\,\ref{obs_catalog}).   
%The location of many of these gas rich objects (Fig.\,\ref{fig_distribution} and Fig.\,\ref{fig_distribution_HImaps}) suggest that they are only in projection onto the cluster center; low mass galaxies are expected to lose their gas more easily than their massive counterparts through ram-pressure.

%
%Therefore, the remaining 
%43 detections correspond to gas-rich, low mass objects, 
%(\textit{i.e.} 74\% in A496 for only 20\% in A85).  
%
%Fig.\,\ref{fig_colour-mag}  confirms that \hi-detections 
%in A496 are spread over a large mass range (absolute 
%magnitudes between $-$20 to $-$16) compared with A85. 
%
%most of them in the high mass limit (M$_\mathrm{B}<-$20).

The distribution of \hi-detected objects in 
A496 (Fig.\,\ref{fig_distribution})
is striking, with many \hi\ normal galaxies 
projected onto the central regions. The middle 
panel of Fig.\,\ref{fig_distribution_HImaps} 
helps to solve the paradox: the objects with 
low velocity relative to the cluster (maps in 
green, in the velocity range 
$-0.5\sigma_{cl} < \Delta v_{rad} < 0.5\sigma_{cl}$)
could be dominated by a movement on the plane
of the sky, with circular orbits still far from 
the densest ICM regions \citep{Dressler86}.  The 
sample of red maps projected close to the cluster 
core in Fig.\,\ref{fig_distribution_HImaps} have 
large components of velocity along the line of 
sight. They must be at an early stage of infalling, 
accounting for their normal gas content.  

Seventeen B-Sp in A496 are found in the red 
sequence; eleven of them still retain enough gas 
to be detected, some are classified as 
\hi\ abnormal and are indicated with black 
diamonds in Fig.\,\ref{fig_colour-mag} and 
Fig.\,\ref{fig_distribution}.  This strongly
suggests that they are under an advanced process 
of stripping. Most of the red spirals in A496 are
projected in the NW and SE quadrants; in 
Sect.\,\ref{tab_discussion} we discuss a possible 
relation between their gas properties and their 
location within the cluster.  
%Two \hi\ detected
%galaxies lying in the red sequence, and having 
%masses above log$ (M_{*} / M_{\odot})=9.0$, do
%not belong to the B-Sp sample.  This is due to
%their radial velocity which is within the VLA data 
%cube but outside the redshift membership catalog
%(Fig.\,\ref{fig_colour-mag}).  
\\

\noindent
$\bullet$ {\bf  The \hi\ in Abell 2670}

The bottom panel of Fig.\,\ref{fig_distribution} 
shows the distribution of 315 member galaxies 
in A2670, emphasising the position of 61 bright 
spirals and 38 \hi-detections (same symbols 
as previous figures).  
%A2670 is known to be dynamically younger than the other two systems based, in part, on the diffuse X-ray emission.
%(Fig.\,\ref{fig_cubes}).  
%We report a lower number of \hi\ detections compared with A496. 
From the 61 B-Sp reported in this cluster, 
28 are not detected in \hi, suggesting an active 
sweeping process.  Fig.\,\ref{fig_colour-mag} 
shows 8 B-Sp in the red sequence, some of them 
retaining enough gas to be detected in \hi.
%two are \hi\ abnormal and three display normal \hi. 
%where a few are seen through low inclination angles and can be reddened by dust. However, very few of these red spirals are detected in \hi, indicating that they might constitute true passive spirals.

Fig.\,\ref{fig_distribution} unveils a complex 
distribution of the \hi\ detections in A2670
which are mostly projected onto the cluster 
core and to the SW.    Most of the \hi\ 
non-detected bright spirals in A2670 (solid 
blue/red circles without square or diamond in 
Fig.\,\ref{fig_distribution}), are located 
near the core and to the NE. 
Fig.\,\ref{fig_distribution_HImaps} 
%helps to better understand the accretion history of spirals in A2670.  This figure 
shows a trend of abundant \hi\ rich galaxies in 
the SW while the NE beyond 1\,R$_{200}$ is very 
poor in \hi.  The center shows a balance between 
shrunk and extended \hi\ maps but we remind 
that strong projection effects might be present
in this zone. 

%In the following sections we discuss the full \hi\ 
%distribution throughout the studied clusters on 
%the basis of a substructure analysis, 
%the ram-pressure profiles and PPS diagrams.

\section{Cluster substructure analysis}
\label{res_substructures}

In order to get a better understanding of the 
assembly history of the studied clusters we
%investigate the importance of preprocessing within 
%subgroups versus the effect of the cluster environment.  
carried out a search for substructures based on 
two independent methods: the tests proposed by
\citet{Dressler-Shectman88} and by 
\citet{Serna-Gerbal96}.  
For the two tests we applied the compilation of 
member galaxies described in Sect.\,\ref{obs}.  
%(Table\,\ref{tab_results}). 
%As mentioned before, these catalogues have 616, 
%368, 308 objects in A85, A496, A2670, respectively.  
In general we have found large coincidences 
%of results obtained 
with the two methods.  
%Only two groups, 
%one in A85 and one in A496, were unveiled by the 
%$\Delta$-test but are not seen by the S-G method. The 
%opposite occurred with a group detected in A2670 
%by the S-G method which was not seen by the 
%$\Delta$-test. 
The detected substructures are 
shown in Fig.\,\ref{fig_subs_A2670} and they 
will be used, in combination with the positions 
of \hi\ abnormal galaxies, as a test for 
pre-processing (Sect.\,\ref{discussion}). 
%Next, we describe and compare 
%the results obtained with the two methods.

\subsection{Substructures: the $\Delta$-test}
\label{res_delta}

We applied an improved version of the well known
$\Delta$-test \citep{Dressler-Shectman88},
which is based on the parameter-$\delta$, defined by:

\begin{equation}
\delta^2 = \frac{N_{nn}+1}{\sigma_{cl}^2} [(v_{local} - v_{cl})^2 + 
(\sigma_{local} - \sigma_{cl})^2]
\end{equation}

\noindent 
where $v_{cl}$ and $\sigma_{cl}$ are the cluster global 
parameters, and $v_{local}$ and $\sigma_{local}$ are the 
local parameters calculated for every galaxy 
relative to its $N_{nn}$ nearest neighbours.  In this 
work we assume $N_{nn}$ = 10; modifying this parameter 
does not change significantly our results 
\citep[][]{Bravo-Alfaro09}.  

\begin{figure}
 \begin{subfigure}[tbp]{0.42\textwidth}
  \includegraphics[width=\columnwidth]{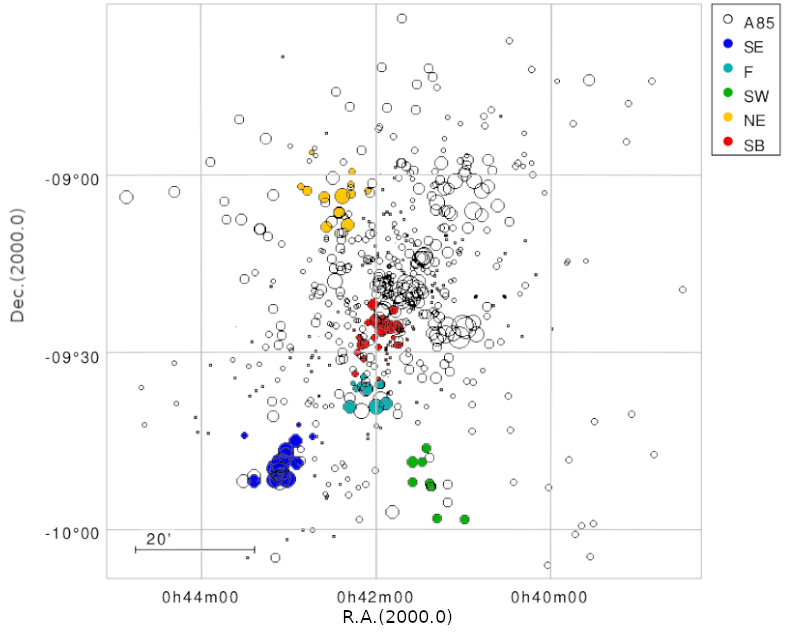}
%  \caption{}
 \end{subfigure}
 \\
 \begin{subfigure}[tbp]{0.42\textwidth}
  \includegraphics[width=\columnwidth]{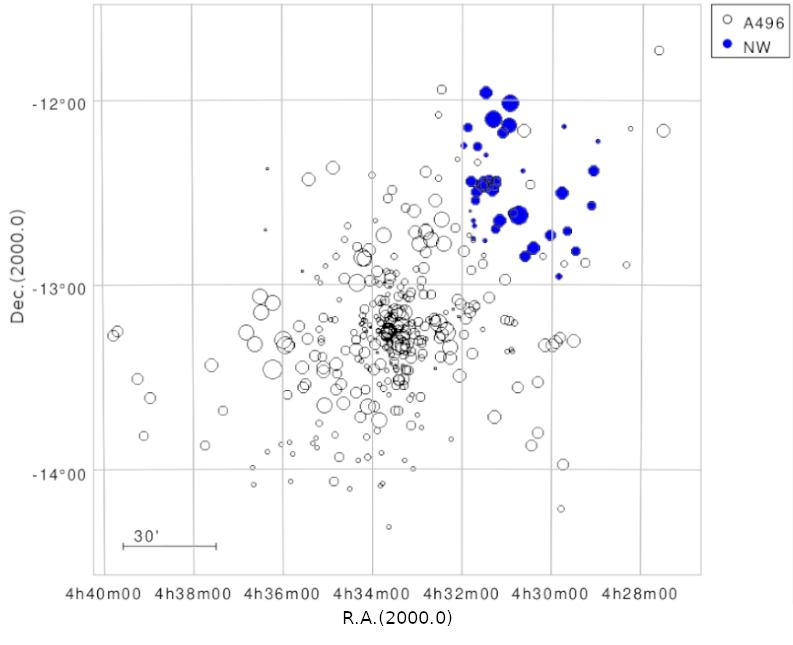}
%  \caption{}
 \end{subfigure}
 \\
  \begin{subfigure}[tbp]{0.42\textwidth}
  \includegraphics[width=\columnwidth]{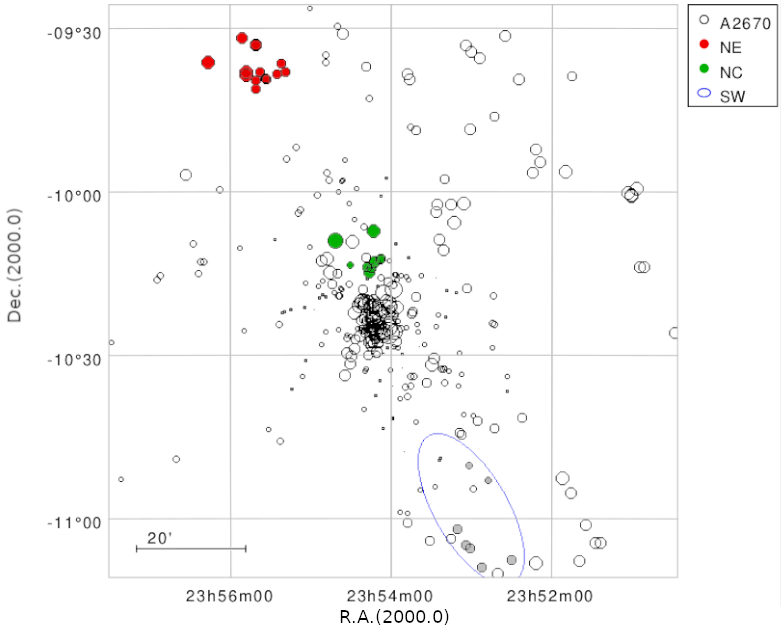}
%  \caption{}
 \end{subfigure}
 \caption{Substructures detected with the $\Delta$-test. Sizes
 of the circles grow with the probability to belong to a 
 substructure.  Colored circles are part of the extracted 
 groups. (Top): In A85 all but one substructure (the NE) 
 are detected as well with the Serna-Gerbal method. 
 (Middle): The $\Delta$-test for A496 unveiled one large
  substructure to the NW. The Serna-Gerbal test found no 
  substructures in this cluster.  (Bottom): The $\Delta$-test 
  in A2670 detected two substructures, the NC and NE. 
  The Serna-Gerbal test unveiled these two groups plus 
  a third group in the SW, indicated with a blue ellipse.}
   \label{fig_subs_A2670}
\end{figure}

The $\Delta$-test detects a substructure as a 
concentration of galaxies showing high values of 
the parameter-$\delta$; this has proven to be a 
reliable tracer of substructures.  However, this 
method loses some 
accuracy in certain cases, for instance, when two 
groups at different distances from the observer
are projected along the same LOS.   In this case 
the $\Delta$-test will produce one single 
overestimation of the parameter-$\delta$, detecting 
only one structure.  This is critical in the absence 
of a dominating large structure, like in 
bimodal/multimodal systems.  In order to improve 
this method we introduce
dividing surfaces on a 3D distribution of RA, 
Dec and $v_{loc}$.  This separates the sub-groups 
of galaxies that share the same kinematic 
properties within their particular caustics, 
but somehow detached from the kinematics of the 
whole cluster. That is, we do not take into 
account only the parameter-$\delta$, but also 
the $v_{loc}$ and the phase-space distribution
%(clustercentric distance $\times$ relative line-of-sight
%velocity) 
of galaxies that are candidates to be members 
of the substructures (more details in Caretta 
et al., 2022, in prep.).  Values of the 
parameter-$\delta$  are illustrated in 
Fig.\,\ref{fig_subs_A2670} by black circles; 
these circles grow with 
the probability of a galaxy to belong to a group.
Individual values of the parameter-$\delta$ will
be available upon request.  \\

\noindent
$\bullet$ {\bf $\Delta$-test results in A85}
%\subsubsection{D-S substructures in A85}

We revisit the analysis of substructures of 
A85 \citep{Bravo-Alfaro09} by
taking advantage that the number of redshifts 
for member galaxies has grown significantly in 
the last years \citep[\textit{e.g.},][]{Boue08b}.
For the full sample of cluster members we 
report a velocity dispersion of 
$\sigma_v = 1,054$\kms.  
%With the enlarged sample of redshifts we detect 
We report the following 
substructures.\\
%(Fig.\,\ref{fig_subs_A2670}), 
%some of them previously reported by 
%\cite{Bravo-Alfaro09}:\\

\noindent
- The main cluster body, with 532 galaxies, has a central 
velocity $v$\,=\,16,650\,\kms, and $\sigma_v = 1,092$\kms. \\
- The South-East (SE) substructure, with 15 members, has 
$v$\,=\,15,157\,\kms, and $\sigma_v = 298$\kms. \\
- The {\it Filament} group {\it F}, with 10 objects, 
$v$\,=\,15,840\,\kms, and $\sigma_v = 372$\kms; this 
and the SE substructures did not change significantly,
compared with previous works, as only few new 
redshifts are provided for these regions.  \\
- The Southern Blob (SB) is better defined than in
previous articles; it has 38 members, 
$v$\,=\,16,947\,\kms, and $\sigma_v = 282$\kms.\\
- An interesting difference compared with previous 
works is the C2 substructure  
\citep[see Fig. 4 of][]{Bravo-Alfaro09} which 
has lost its signal because the dominant object, 
the jellyfish galaxy KAZ-364 (also known as JO201), 
lies slightly outside of the caustic contours.  This 
galaxy is outside the velocity range of our \hi\ cubes, 
so it is not included in the present analysis. Despite 
a velocity of \aprox3,000\kms, relative to the cluster, 
\cite{Bellhouse17} reported this object as a fast, 
infalling cluster member being part of a small group.\\
- We unveiled two new low-mass substructures 
($ < $ 2\% of the total number of member galaxies);
the first is at the South-West with 9 galaxies, 
$v$\,=\,16,236\,\kms, and $\sigma_v = 183$\kms.\\
- Another new substructure lies 
to the NE with 12 galaxies, having $v$\,=\,16,345\,\kms, 
and $\sigma_v = 300$\kms.\\

\noindent
$\bullet$ {\bf $\Delta$-test results in A496}
%\subsubsection{D-S substructures in A496}

%We applied the improved $\Delta$-test to the complete 
%sample of 368 members which has a $\sigma_v = 685$\kms\
%(Table\,\ref{tab_clusters}). 
We detected a main cluster body with 330 galaxies and a 
$\sigma_v = 674$\kms.  We clearly unveiled one single 
substructure located \aprox 1\degree\ (2.5\,Mpc) NW
from the cluster center, with 38 members (above 10\% of
the total mass), $v$\,=\,9,338\,\kms, and $\sigma_v = 457$\kms\  
(see Fig.\,\ref{fig_subs_A2670}).    Our \hi\ survey
covers only a small portion of that group, delivering 
only limited information on the cold gas associated 
with this structure. \\

%
%Unlike what we find with the next method (below), the results
%here are very convincing that A496 is substructured. \\

\noindent
$\bullet$ {\bf $\Delta$-test results in A2670}
%\subsubsection{D-S substructures in A2670}

We obtained the following results for A2670: \\
%having a $\sigma_v = 781$\kms\ (Table\,\ref{tab_clusters}).   \\
- The main cluster body emerges with 287 members, 
$v$\,=\,22,806\,\kms, and $\sigma_v = 805$\kms.  \\
- We unveiled two low-mass structures, the first one 
projected close to the center (NC, \aprox 10\prim\ to 
the north)  with 7 members, $v$\,=\,22,883\,\kms,
and $\sigma_v = 20$\kms.  This velocity suggests 
a merger with the main cluster body.  \\ 
- The NE group lies \aprox50\prim\ 
(4.5\,Mpc), NE from the cluster center, 
with 14 members, $v$\,=\,23,115\,\kms, and 
$\sigma_v = 424$\kms\ (Fig.\,\ref{fig_subs_A2670}).
This structure is just north from the VLA FoV 
(Fig.\,\ref{fig_distribution}) and we do not 
dispose of \hi\ information for these galaxies.  \\

\subsection{Substructures: the Serna-Gerbal test}
\label{res_serna}

In parallel to the $\Delta$-test described above, we
applied the Serna-Gerbal (S-G) method 
\citep{Serna-Gerbal96} to the three clusters.  
This test introduces a fundamental, 
independent element to detect substructures: the 
relative potential energies of each pair of galaxies 
by using individual galaxy brightness as a proxy 
for the stellar mass. We use the $r$-band magnitude for
this goal.  In practice the S--G test requires  
an input catalogue of positions, velocities, magnitudes 
(converted to masses assuming a value of M/L), and a 
defined number $n$ of galaxies needed to build a 
group.  We used $n=10$ as a representative value.  
The S--G test works hierarchically, 
seeking for lower level substructures within a 
primary (more massive) detected group.  This 
delivers an output list of galaxies belonging to 
each substructure and provides the mass of the 
corresponding group. Absolute masses of individual
structures are not highly accurate, but the relative 
masses of the groups, normalized to the total member
sample, are very reliable. \\

\noindent
$\bullet$ {\bf S-G results in A85}
%\subsubsection{S-G substructures in A85}
%label{res_serna_A085}

%5starting with the full catalogue of member galaxies, 
%having magnitudes and redshifts between 
%13,796\kms\, and 19,132\kms. 
The S--G program computed for A85 a mass 
of M$_\mathrm{tot}=6.51 
\times 10^{14}$~M$_\odot$ for the entire 
cluster, and a velocity dispersion of
$\sigma_{tot}=1,037$\kms.  The S--G test 
detected the main cluster body (identified as 
\#1) with 587 galaxies, a relative mass 
M$_\mathrm{1}=0.792\,$M$_\mathrm{tot}$, and 
$\sigma_{1}=949$\kms.  A secondary group (\#2)
is detached from the main cluster body, lying 
south from the center and having 21 galaxies, 
M$_\mathrm{2}=0.079\,$M$_\mathrm{tot}$, and 
$\sigma_{2}=184$\kms.  This structure is 
coincident with the SB group detected by 
the $\Delta$-test.  

In the next step we took Structure \#1 as the
starting point, and we unveiled two further 
substructures; the first one (identified as 
\#1.1) has 144 galaxies with 
M$_\mathrm{1.1}=0.218\,$M$_\mathrm{tot}$, and 
$\sigma_{1.1}=527$\kms.  Some galaxies from this
group account for the SE subcluster, detected 
as well with the $\Delta$-test.  The second 
substructure (\#1.2) has 14 galaxies, a relative 
mass M$_\mathrm{1.2}=0.01\,$M$_\mathrm{tot}$, 
and $\sigma_{1.2}=180$\kms.  Finally, the group 
\#1.1 harbors two additional 
substructures:  the first (\#1.1.1) has 74 galaxies 
with M$_\mathrm{1.1.1}=0.125\,$M$_\mathrm{tot}$, 
and $\sigma_{1.1.1}=481$\kms. This group coincides
in position with the minor SW structure obtained 
with the $\Delta$-test.  Another dozen galaxies from 
the structure \#1.1.1 emerges as a group lying between
the SB and the {\it Filament}, unveiling a more complex
region, south of the core, than what is shown by 
the $\Delta$-test.  A final subgroup  (\#1.1.2)
appears with 10 galaxies, a relative mass  of
M$_\mathrm{1.1.2}=0.009\,$M$_\mathrm{tot}$ and 
$\sigma_{1.1.2}=173$\kms.  This substructure matches 
with the {\it Filament} group detected with the 
$\Delta$-test.  With the exception of the NE 
group, the S--G test successfully confirmed 
the substructures found with the $\Delta$-test
in A85. \\
%On the other hand, the S--G unveiled a more
%complex distribution of galaxies south of the cluster 
%core, than the D--S test. 
%in previous works \citep[\textit{e.g.}][]{Durret98,Bravo-Alfaro09}.\\
%We also tried applying the same method to spiral galaxies only. 
%Out of a total sample of 83 galaxies, we find two structures of 45 and 28 galaxies, each containing only about  $5 \times 10^{-5}$ of the total cluster mass, with respective values of $\sigma_v$ of 471\kms\, and 385\kms. Each of these structures can again be divided into two structures of comparable masses and velocity dispersion, suggesting that spiral galaxies in A85 are indeed quite substructured.  \\

\noindent
$\bullet$ {\bf S-G results in A496}

We analyzed the full sample of member galaxies 
%with magnitudes and 
%redshifts in the range $8,176<v<11,831$\kms. 
but we found no major substructures in A496. Even 
modifying the parameter {\it n} (for instance with 
{\it n=5}) we did not find significantly 
different results. The S--G method finds that A496 is a 
relaxed cluster with negligible substructuring, 
in agreement with previous studies 
\citep[\textit{e.g.}][] {Boue08a}. \\

\noindent
$\bullet$ {\bf S-G results in A2670}

%We started from the full catalog of member galaxies
%dispose of 311 galaxies in A2670 with magnitudes 
%and velocities in the range $21,125<v<24,778$\kms.
In the first step, the S--G test detected two groups 
in A2670. The largest structure (\#1) has 123 galaxies 
with a $\sigma_v$ comparable to that of the entire 
cluster, having  M$_\mathrm{1}=0.490\,$M$_\mathrm{tot}$.
The second group (\#2) has 10 galaxies, a mass 
M$_\mathrm{2}=0.016\,$M$_\mathrm{tot}$, and
$\sigma_{\#2}=215$\kms.  Group (\#2) is placed 
\aprox40\prim\ SW from the cluster center and 
was not detected by the $\Delta$-test 
(see Fig.\,\ref{fig_subs_A2670}). Very 
interestingly, this substructure embraces the zone 
where several \hi\ rich galaxies are reported (see 
bottom panel of Fig.\,\ref{fig_distribution_HImaps}). 
These objects have similar physical properties 
(position, velocity, gas richness), giving further 
support for this group to be real.

In a second run, the large structure (\#1) is 
decomposed into two substructures  
(\#1.1 and \#1.2) having 83 and 11 
galaxies respectively. The former contains 
M$_\mathrm{1.1}=0.270\,$M$_\mathrm{tot}$,
and a velocity dispersion comparable to that of 
the cluster.   The group \#1.2 contains 
M$_\mathrm{1.2}=0.180\,$M$_\mathrm{tot}$, and
$\sigma_{1.2}=433$\kms.
%suggesting it contains rather massive galaxies.
%  WHY IS THIS FLORENCE?
The fact that the substructures \#1 and \#1.1 
present high $\sigma_v$, comparable to that 
of the whole cluster, reduces the statistical
significance for these groups, as shown by
\cite{Serna-Gerbal96}. 
However, part of substructure \#1.1 extends
from the centre to the north, coinciding with 
the {\it NC} group found with the $\Delta$-test.
The same occurs with substructure \#1.2, 
coinciding with the {\it NE} group, reinforcing 
the existence of these two groups 
(Fig.\,\ref{fig_subs_A2670}). \\

The two methods, the $\Delta$-test and the 
Serna-Gerbal, confirm the very complex substructure
system of A85, contrasting with the very low level 
shown by A496.  The two tests show A2670 as an 
intermediate case with a few, not very massive 
substructures.

%\section{Discussion}
%\label{disc}

\section{Ram pressure and Phase-space diagrams}
\label{disc}

With the aim of studying the effects exerted by the 
ICM on the cold gas content of individual galaxies, we 
compute the ram-pressure stripping (RPS) profiles for 
the three studied clusters and compare with different 
values of the anchoring force
(see the review by \cite{Boselli22} for further 
details.)
Next, we build and analyze the PPS diagrams for each 
cluster displaying the position of all member galaxies.
%trying to distinguish infalling from virialized galaxies.
%We trace therein the RPS profiles, the escape velocity, and the virialized zone.  This helps to distinguish infalling from virialized galaxies and provides useful information on the assembling degree of the studied clusters.

\subsection{Ram pressure stripping in A85, A496, A2670}
\label{disc_rps}

%For decades this simple model has been successfully tested by many simulations and it is widely used to estimate RPS \citep[][and references therein]{Cortese21}. This equation is known to be more accurate describing galaxies under face-on movement against the ICM but some authors \citep[\textit{e.g.}][]{Roediger06} have found that only edge-on movements deviate from this modelling. 
%In order to quantify the cluster environment effects 
%on individual galaxies 

%\begin{equation} \label{eq_Cavaliere}
%    \rho(r) = \rho_{0} [1 + (r/r_{\mathrm c} ) ^{2} ] ^{-3\beta/2} 
%\end{equation}

\begin{table}
        \centering
        \caption{Parameters of the ICM model}        
        \label{tab_beta}
        \begin{tabular}{lccc} % four columns, alignment for each
                \hline
                         & A85 & A496 & A2670\\
                \hline
                $r_c$ (kpc)& 82 & 30 & 174\\
                $\rho_0$ (cm$^{-3}$) & 0.0257 & 0.0407 & 0.00383\\
                $\beta$ &0.532 &0.484 &0.700\\
                \hline
        \multicolumn{4}{l}{Notes: Parameters for A85 and A496 are}\\
%        \multicolumn{4}{l}{Parameters for A85 and A496 } \\
        \multicolumn{4}{l}{taken from \cite{Chen07}. We followed }\\
        \multicolumn{4}{l}{\cite{Cirimele97} for A2670.}\\
        \end{tabular}
\end{table}

\begin{figure}
\begin{subfigure}[tbp]{0.40\textwidth}
 \includegraphics[width=\columnwidth]{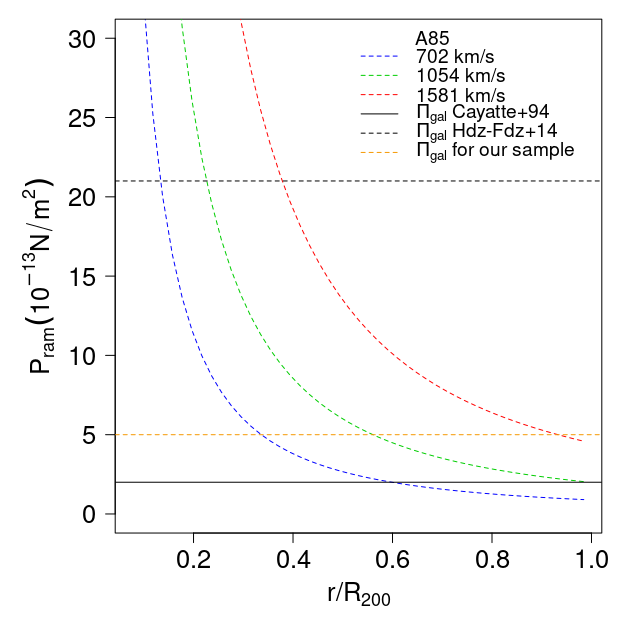}
% \caption{}
  \end{subfigure}
  \\
 \begin{subfigure}[tbp]{0.40\textwidth}
 \includegraphics[width=\columnwidth]{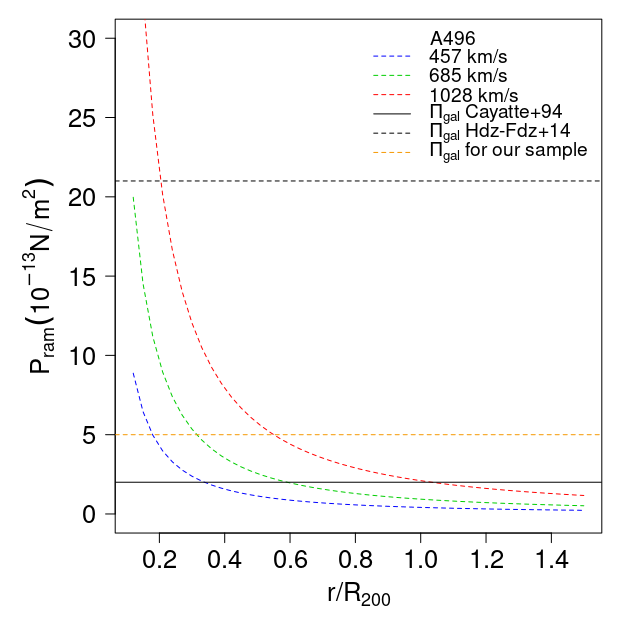}
%\caption{}
  \end{subfigure}
  \\
   \begin{subfigure}[tbp]{0.40\textwidth}
 \includegraphics[width=\columnwidth]{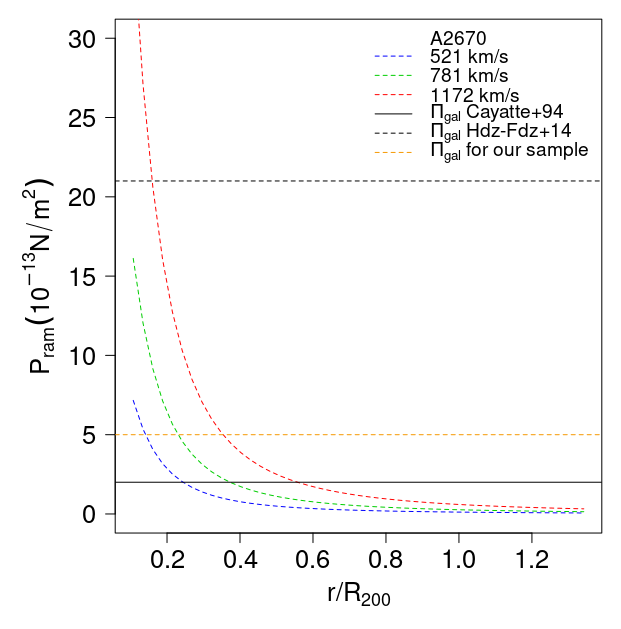}
 %\caption{}
  \end{subfigure}
  \caption{The RPS as a function of the cluster-centric 
  distance.
  %; A85 is on top, A496 in the middle and A2670 
  %in the bottom panel. 
  Three representative velocities 
  relative to the ICM are displayed, taking the cluster 
  velocity dispersion as a reference: 0.66{$\sigma_{cl}$} 
  in blue;  1.0{$\sigma_{cl}$} in green, 1.5{$\sigma_{cl}$} 
  in red.  Three values of the anchoring force are shown in order to
  evaluate the RPS effects (see text). }
  \label{fig_rps}
\end{figure}

In order to estimate the RPS radial profiles 
we follow the theoretical definition proposed 
by \cite{Gunn-Gott72}, 
$ RPS = \rho_\mathrm{ICM} \times v_\mathrm{rel} ^2$.
%
%\begin{equation} \label{eq_Pram}
%    P_\mathrm{ram} = \rho_\mathrm{ICM} \times v_\mathrm{rel} ^2 
%     RPS = \rho_\mathrm{ICM} \times v_\mathrm{rel} ^2 
%\end{equation}
Here $\rho_\mathrm{ICM}$ is the local
density of the intracluster medium, and
$v_\mathrm{rel}$ is the galaxy velocity
relative to the ICM.
We estimate the ICM density 
($\rho_\mathrm{ICM}$) as a function of the cluster radius
by applying the hydrostatic-isothermal $\beta-$model 
of \cite{Cavaliere-Fusco-Femiano76},
$ \rho(r) = \rho_{0} [1 + (r/r_{\mathrm c} ) ^{2} ] ^{-3\beta/2}. $
%It must be stressed that this model considers a smooth, spherical ICM distribution while some substructuring is likely to occur, as demonstrated by \cite{Tonnesen07}.   
%The possible 
%effects of this non-smooth ICM on the studied 
%clusters are considered in Sect.\,\ref{discussion}.  
Table\,\ref{tab_beta} gives the parameters of the 
{$\beta$}-model for the studied clusters:  the core 
radius $r_c$, the central density $\rho_0$ and the 
$\beta$ value used to calculate the density as a 
function of the cluster radius.  
%A caveat about the method based on equations \ref{eq_Pram} and \ref{eq_Cavaliere} comes from two unknown quantities, the galaxy velocity along the plane of the sky and the real 3D cluster-centric distance.  The typical proxy for a 3D case consists in applying statistical correction factors to both, the projected radius and to the radial velocity: $r \sim (2 / \pi) r_{3D} $, and $v_{3D} \sim (\sqrt{3}) v_{rad} $.  After estimating the two cases we keep the 2D approach because it facilitates the analysis of Sect.\,\ref{disc_PPS} and because RPS values do not deviate significantly from the 3D case which increases the RPS value very slightly.  If the 2D approach underestimates the real galaxy velocity, it also overestimates the local density, with the two correction factors nearly cancelling each other. In the end, the use of the 2D case has no major impact on our statistical results.  
The corresponding RPS 
profiles are shown in Figure\,\ref{fig_rps} 
where we display three representative 
velocities which are multiple values of 
the velocity dispersion of each cluster:  
0.66{$\sigma_{cl}$} in blue, 1.0{$\sigma_{cl}$} 
in green, and 1.5{$\sigma_{cl}$} in red. This 
adds physical meaning to the comparison of 
the RPS intensity among different clusters.

In order to quantify the effects produced by the 
RPS on individual objects it must be compared 
with the anchoring force 
per unit surface produced by the galaxy 
gravitational potential, given by 
$\Pi_{gal} = 2\pi \,G \,\Sigma_{\mathrm s} \,\Sigma_{\mathrm g} $.
%
%Eq.\,\ref{eq_Fr}:
%
%\begin{equation}\label{eq_Fr}
% \Pi_{gal} = 2\pi \,G \,\Sigma_{\mathrm s} \,\Sigma_{\mathrm g} 
%\end{equation}
%\noindent
Here, ${\Sigma_s}$ and ${\Sigma_g}$ are the 
surface densities of the stellar and gaseous 
discs, respectively.  RPS must overcome the 
anchoring force in order to succeed the gas
stripping.  Several values of $\Pi_{gal}$ have 
been proposed, 
going from 2.0\por10$^{-13}$\,N\,m$^{-2}$ for 
spirals in Virgo \citep{Cayatte94}, to 
2.1\por10$^{-12}$\,N\,m$^{-2}$ for a Milky 
Way-like galaxy \citep{Hernandez14}. One 
step further was made by \cite{Jaffe18} who 
modelled the anchoring force as a function of
the galaxy radius, the stellar mass and the 
gas fraction of a sample of spirals.  We follow 
these authors in order to seek for values of 
$\Pi_{gal}$ being representative for our 
sample of B-Sp.

From Fig.\,\ref{fig_colour-mag} we see that
most of the B-Sp are in the range of mass 
between 
log$ (M_{*} / M_{\odot})=9.0$ and 
log$ (M_{*} / M_{\odot})=10.5$.
%\citep{Mahajan18}.
No significant difference in mass is expected between 
our sample and that studied by \cite{Jaffe18}, so we 
use their Figure 5 \citep[and Fig. 4 in][]{ Jaffe19},
which provide the anchoring force at a radius of 
twice the disc scale length.  As we are interested 
in the RPS effects exerted at the external edge 
of spirals, we take this value as a proxy for the 
anchoring force acting at the outskirts of the 
galaxy.  This strategy delivers $\Pi_{gal}$ in 
the range from 10$^{-12}$ to 10$^{-13}$\,N\,m$^{-2}$.  
We take the middle value 5.0\por10$^{-13}$\,N\,m$^{-2}$  
as a reference for our sample of B-Sp.
%A conservative error bar is given by the interval 
%10$^{-12}$-10$^{-13}$\,N\,m$^{-2}$.  
As a comparison, the 
$\Pi_{gal}$ value reported by \cite{Cayatte94} 
predicts very strong RPS effects, even in the 
low velocity and at large cluster-centric 
domains (see Fig.\,\ref{fig_rps}). The opposite 
is given by the value 
of $\Pi_{gal}$ proposed by \cite{Hernandez14}.
%which predicts that RPS would have visible effects
%only at extremely high velocities and at very small 
%cluster-centric radii.  
We indicate the three 
values of $\Pi_{gal}$ in Figure\,\ref{fig_rps}.  
It is clear that the reference value
$\Pi_{gal}=$\,5.0\por10$^{-13}$\,N\,m$^{-2}$ is 
more realistic for the bright spirals under
study so we use it along the forthcoming analysis. 
%In Sect.\,\ref{discussion} we give further  arguments for our reference value fitting  with our observations. 

Some uncertainties are introduced though. 
Particularly important is the dispersion in the
%are the error bars in the absolute magnitude--stellar mass relation and those 
relation between the anchoring force
and the stellar mass, with a strong dependence 
on the galaxy radius. In this respect \cite{Jaffe18} 
have shown that the anchoring force is strongly 
dependent on the degree of mass concentration; 
this implies that low mass objects with a high 
central density could have a similar $\Pi_{gal}$
at their center than a more massive/extended 
galaxy at its edge.  Despite these 
uncertainties we will show (Sect.\,\ref{discussion}) 
that our reference value of $\Pi_{gal}$ match 
reasonably well with our observations on a 
statistical basis.  
%A detailed calculation of $\Pi_{gal}$ profiles for individual galaxy masses, radii and the amount of stripped gas will be tackled in a forthcoming paper. The implications of the RPS profiles of Figure\,\ref{fig_rps} are discussed in Sect.\,\ref{discussion}.

We repeat this analysis for the low-mass galaxies,
$i.e.$ those in the range of stellar mass 
log$(M_{*} / M_{\odot})$ between 8.0--9.0,
%
%In addition to the bright spirals we reported 
%a sample of low mass \hi\ rich objects seen in 
%the three clusters, more conspicuously in A496 due 
%to its lower redshift (see Fig.\,\ref{fig_colour-mag} 
%and Table\,\ref{tab_results}). 
%Most of these
%objects have stellar mass 
%log$(M_{*} / M_{\odot})$ in the range 8.0--9.0.   
%By applying the same procedure we did for the 
%sample of B-Sp we 
obtaining an upper limit for the anchoring force of
$\Pi_{gal}=$\,10$^{-13}$\,N\,m$^{-2}$. Galaxies 
in this low regime of $\Pi_{gal}$ are
predicted to be easily stripped of their 
cold gas when they approach the virial radius.
We further discuss about these objects in
Sect.\,\ref{discussion}.

\begin{figure*}
\centering
\begin{subfigure}[htbp]{0.8\textwidth}
 \includegraphics[width=\columnwidth]{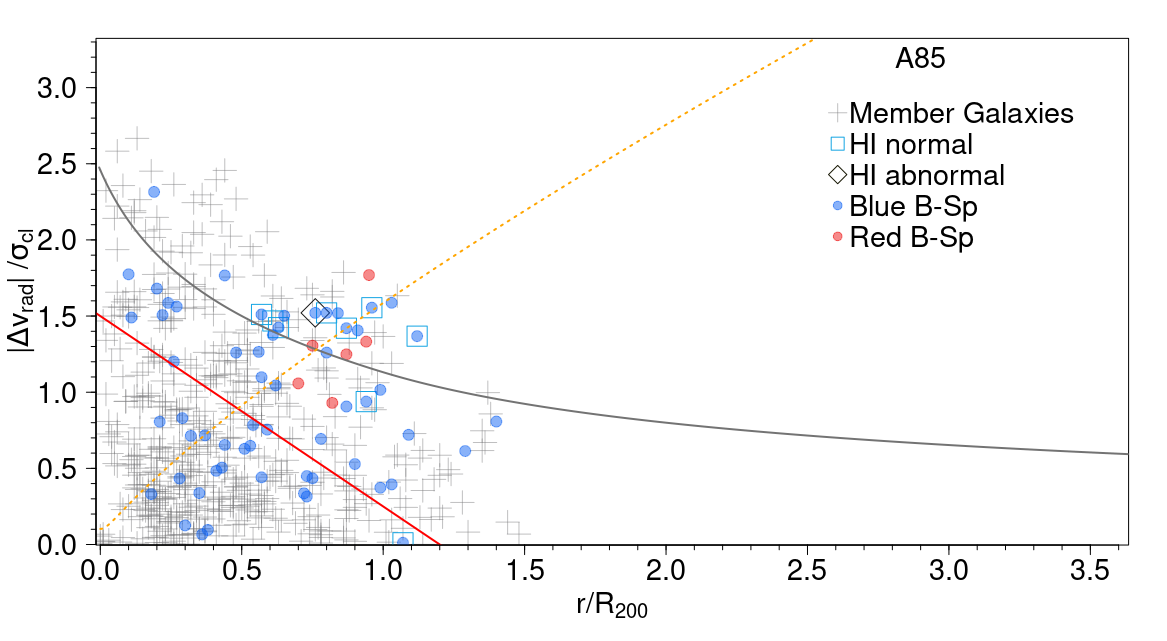}
% \caption{}
  \end{subfigure}
  \\
  %\vspace{1cm}
 \begin{subfigure}[tbp]{0.8\textwidth}
 \includegraphics[width=\columnwidth]{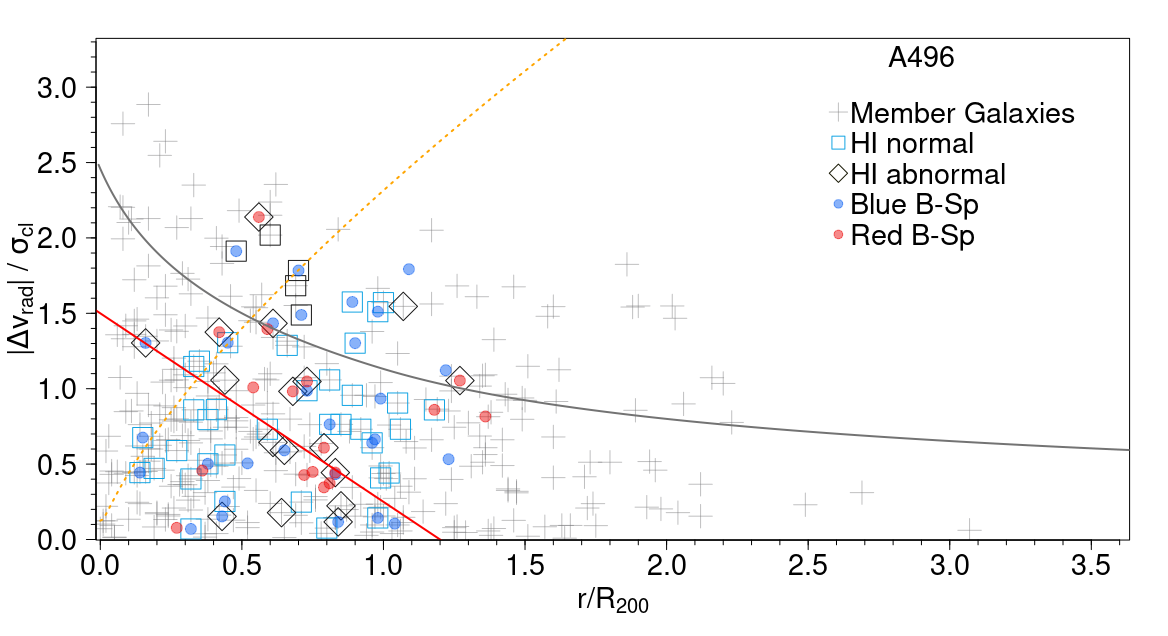}
% \caption{}
  \end{subfigure}
\\
   \begin{subfigure}[tbp]{0.8\textwidth}
 \includegraphics[width=\columnwidth]{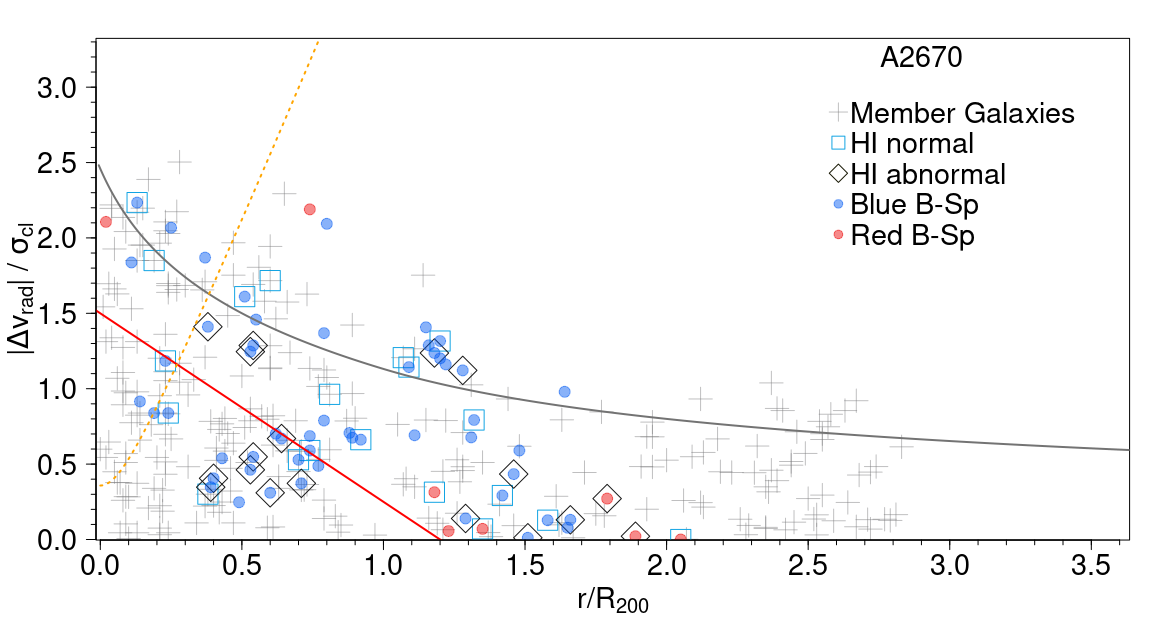}
% \caption{}
  \end{subfigure}
  \caption{PPS diagrams for A85 (top), A496 (middle) and A2670 
  (bottom).  Symbols are the same as in Figs.\,\ref{fig_colour-mag}
  %Member galaxies are shown with grey crosses;
  %solid circles indicate the B-Sp, both blue and red. The 
  %\hi\ detected galaxies are shown with blue open squares 
  %(normal) and black diamonds (abnormal). 
  and \ref{fig_distribution}. The solid red line defines 
  the virialized region (see text) and the solid grey 
  line indicates the escape velocity. The dotted orange line 
  shows the region where P$_{\mathrm{ram}}\,=\,\Pi_{gal}$ 
  (see Sect.\,\ref{disc_rps}).}
  %and the dotted green line 
  %indicates \hi\ detection threshold .
  \label{fig_phase-space}
\end{figure*}

\subsection{Phase-space diagrams and cluster 
accretion history}
\label{disc_PPS}

The projected phase-space (PPS) diagram of a 
cluster helps to distinguish virialized from 
infalling galaxies  \citep[{\textit e.g.}][]
{Kent-Gunn82,Geller13,Oman13, Jaffe15, Rhee17, 
Cortese21}.   
%PPS diagrams display galaxy peculiar velocities, relative to the cluster systemic velocity, as a function of the projected clustercentric distance. 
We present a version of PPS where the parameters 
$\Delta$v and the radius are normalized by 
the velocity dispersion $\sigma_{cl}$, and 
by R$_{200}$, respectively. 
%This plot approaches a 3D view of the system on the basis of the 2D available data. 
%Due to projection effects these diagrams deliver 
%lower limits to the real cluster-centric 
%distance and to the 3D galaxy velocity 
%relative to the cluster velocity.
On the y-axis we display the absolute 
value of the peculiar velocity ($|\Delta v_{rad}| / \sigma_{cl} $). 
%normalized to the velocity dispersion of the cluster 
On the x-axis the projected position relative 
to the cluster center is normalized by R$_{200}$. 

Fig.\,\ref{fig_phase-space} shows the PPS diagrams 
for A85, A496 and A2670 with the same symbols as 
in previous figures. 
%with member galaxies indicated 
%with grey crosses.  We emphasize the bright spirals 
%with solid circles, blue or red, depending on their
%position in the CMD.  The \hi\ detections are shown
%with blue open squares and black open diamonds,
%depending if they are \hi\ normal or abnormal 
%(see Sect.\,\ref{results}).
The escape velocity of each cluster is shown with 
a solid grey line; these are more conservative than 
the caustics used to define the member sample
and helps to obtain a better contrast between
virialized and infalling galaxies
\cite[\textit{e.g.}][]{Jaffe15, Wang21}.
%accounting for the cluster members projected 
%beyond the grey curve.  
This escape velocity was estimated on the 
basis of a NFW halo with a concentration of 
$c = 6$; other values of this parameter do 
not change the escape velocity significantly.  
The orange dotted line draws the boundary 
where RPS equals the anchoring force, 
$\Pi_{gal}=$\,5.0\por10$^{-13}$\,N\,m$^{-2}$.

%We remind that RPS is estimated from a 
%$\beta$-model, so this border is to be taken 
%with cautious. 

%The green dotted line indicates the \hi\ detection limit of this survey, given by six times the rms per channel (see Sect.\,\ref{results}). 

%We confirm that other values typically used for the 
%concentration parameter, for instance $c = 4$, does 
%not produce significantly different results.
%Several zones can be separated in the PPS diagrams 
%corresponding to different physical regions 
%in the cluster. First, 

At the lower left corner of the PPS diagram we find 
the galaxies projected at short cluster-centric 
distances and having low (radial) velocities 
relative to the cluster.  These objects are 
likely settled onto the cluster potential after 
completing --at least-- one pericentric passage.
This part of the diagram is identified as the 
{\it virialized zone} and we define its boundaries 
as follows.  On the x-axis, we followed authors  
\citep[\textit{e.g.}][]{Jaffe15, Kopylova-Kopylov18}
who propose that the virial radius lies in
the range 1.2--1.4\,R$_{200}$.
%the virial radius $R_{vir}$ depends on 
%both, cosmology and redshift \citep{Bryan-Norman98}.
%({\bf REF  Bryan \& Norman 1998, ApJ 495, 80}).  
%At low {\it z},
%$R_{vir}$ is closer to $R_{100}$ than $R_{200}$
%and we find that values of $R_{vir}$ between
%1.2 and 1.4\,$R_{200}$ are a good compromise 
%\citep[see for instance][]{Jaffe15, Kopylova-Kopylov18}.
On the y-axis the criteria to define the virial 
zone is based on 
statistics, by considering a Gaussian distribution 
of velocities. We use the value of 1.5\,{$\sigma$} 
as a more restrictive limit (compared with 
3\,{$\sigma$}) in order to avoid contamination 
by foreground/background galaxies. Considering 
these boundaries we define the virialized zone 
as the triangle limited by the solid red line in
Fig.\,\ref{fig_phase-space}. This triangle zone
is more conservative than the corresponding 
squared region and has been applied by authors 
like \cite{Jaffe15} and \cite{Wang21}.  On 
the opposite corner of the PPS diagrams, 
galaxies infalling for the first time lie 
in the large radii/velocity domain.  After a 
few $10^9$\,yrs, equivalent to the crossing 
time for a typical cluster, infalling galaxies 
will lose kinetic energy due to dynamical 
friction and violent relaxation, and they will 
settle deep into the cluster gravitational 
potential. 
%At the end of this process, galaxies will 
%gather at the virialized zone.  This is 
%expected to occur after some pericentric 
%crossings.  

The distribution of different types of
galaxies in the PPS diagrams offer 
important hints on the dynamical stage
of the three clusters. 
%Though, a direct comparison of the full
%observed fields of the three systems must 
%be done with caution. Due to differences 
%in redshift and VLA coverage A2670 has a 
%membership catalog and observations 
%reaching much larger radius compared with 
%A85 and A496.  To overcome this bias we 
%carry out a direct comparison
%(Sect.\,\ref{discussion}) by considering 
%only to those zones homogeneously covered 
%with the VLA.
%\textbf{Margarita: I think that you want to say: 
%restricted to 1\,R$_{200}$, instead of VLA.}
%
%As a consequence, some of the B-Sp and \hi\ detections in A2670 are projected up to 2.5\,$R_{200}$,  while the corresponding samples of A85 and A496 are limited to a radius about 1.5\,$R_{200}$.  We take this bias into account through the following discussion and some aspects of the three systems are only compared within the 1.5\,$R_{200}$ region. 
%
The top panel of Fig.\,\ref{fig_phase-space} 
shows A85 behaving as a typical evolved system 
with a high concentration of member galaxies 
within the virialized zone. The escape velocity 
curve in this plot is lower than the one reported 
by \cite{Bellhouse17}; this is due to the
higher mass used for A85 by these authors 
%(M$_{200}=$1.58\times 10$^{15}$\msolar).
$\text{M}_{200}=1.58 \times 10^{15} \text{M}_{\odot}$
We see a clear trend where, with one single
exception, all B-Sp lying below the escape 
velocity curve are not detected in \hi, suggesting
that these galaxies had gone through an 
advanced process of gas stripping. 
%In the next section we discuss about the red B-Sp lying beyond the escape velocity in the same plot.  
The gas rich objects in A85 lie in the high 
radius, high velocity domain, indicating that 
they approach the cluster for the first time. 
Despite some similarities between A85 and other
systems like A693 ($e.g.$ mass, morphology and 
dynamical stage), their distribution of galaxies 
across their PPS differs significantly \citep{Jaffe15}.
%In Sect.\,\ref{discussion} 
%we will further discuss about the intriguing 
%distribution of blue/red bright spirals 
%which are not detected in \hi.

%The amount of B-Sp located in the large distance/velocity domain suggests a continuous flow of spirals from the outskirts into the central regions.  Many of the gas deficient spirals (solid blue dots) projected along the escape velocity curve, and inwards, are located within the active zone between the main cluster body and the SE subcluster (see Sect.\,\ref{res_distrib} and Sect.\,\ref{discussion}).  

Similar to A85, A496 harbors a crowded 
virialized zone  
(medium panel of Fig.\,\ref{fig_phase-space}).
This is in agreement with the advanced level 
of relaxation reported by different authors 
and with the low degree of substructuring 
in A496 reported in this work
(Sect.\,\ref{res_substructures}).  The 
striking number of \hi\ rich galaxies 
lying within the virialized zone can be 
explained by projection effects as
described in Sect.\,\ref{res_distrib}.
%
%split in two groups; Fig.\,\ref{fig_distribution_HImaps} shows a number of galaxies illuminated in red which must be under strong projection effects, having a large velocity component along the line of sight. A second group of galaxies are shownin green in the same figure; they have low (radial) relative velocities and are mostly moving along the plane of the sky. 
%
The PPS diagram of A496 shows a slight trend 
of \hi\ abnormal galaxies (black diamonds) 
and red spirals (red solid circles) being more 
abundant towards the virialized zone. 
%We  discuss this issue in more detail in the next section.  

A2670 is dynamically young accounting for 
the complex distribution of galaxies in the 
PPS diagram and a virialized zone significantly 
less concentrated than in the other two systems.
In this respect A2670 looks like other young 
clusters like Hydra \citep{Wang21}.
%(Fig.\,\ref{fig_phase-space} bottom panel).  
A number of \hi\ detections is projected onto 
the virialized zone, certainly explained by 
projection effects.
%(red maps in Fig\,\ref{fig_distribution_HImaps}). Others must be dominated by circular orbits along the plane of the sky (green maps in the same figure).  
Other \hi\ galaxies are located in 
the high radius ($\geq 1.5\,r$/R$_{200}$) domain, 
matching with the substructure reported onto 
the SW in Sect.\,\ref{res_substructures}.  
Several of these detections correspond to 
abnormal \hi\ galaxies (black diamonds) and 
to red spirals which are good candidates to 
be under pre-processing.  \\

%{\bf TO MERGE WITH SECT. 6}
%In general we expect to find more \hi\ 
%detections far away from the virialized 
%region. This is true for A85, but A496 
%and A2670 harbor many exceptions.  In the
%next section we will show that the 
%statistical difference in location between 
%the \hi\ normal and abnormal B-Sp 
%in A496 and A2670 is not as significant 
%as it occurs in other nearby clusters,
%like Virgo, A1367 and Coma, and in higher 
%redshift systems like A963 
%\citep[\textit{e.g.}][]{Jaffe16}. 

\section{Discussion}
\label{discussion}

%\noindent
\subsection{The \hi\ and the effects of environment }

The results described so far provide undeniable 
evidence on the important role played by 
environment in the transformation of spirals in 
the three studied clusters.  The global cluster 
\hi\ deficiency is an important parameter that 
correlates with the properties of galaxies in 
most of the nearby clusters.   In this work we 
approach the cluster deficiency as the ratio of 
high \hi\ deficient B-Sp (\textit{i.e.} those
not detected in \hi) normalized by the 
total number of B-Sp found within each system.
%(see Sect.\,\ref{obs_catalog} and Table\,\ref{tab_discussion}). 
Cluster deficiencies
calculated in this way are: 0.86, 0.37, 0.46, 
for A85, A496, A2670, respectively.  A85 appears
as an extreme case of \hi\ deficiency and, since 
it is the most massive of the three studied 
systems, this result matches the model prediction 
of more massive halos producing larger gas loss 
\citep[\textit{e.g.}][]{Donnari21}.  
%Table\,\ref{tab_discussion} helps to make a
%deeper comparison among the studied clusters.
%
%In order to overcome observational biases, the
%results displayed in this table are restricted 
%to the largest physical radius being covered by 
%the VLA in the three systems. We take the 
%conservative value of 1\,R$_{200}$ to carry
%out this comparison.
%Furthermore, 
%if we consider the spirals in the red
%sequence (Fig.\,\ref{fig_colour-mag}), we see
%that some of these objects in A496 and A2670 still
%retain enough gas to be detected in \hi. 
%This is not the case in A85 where no red spirals are detected in \hi\ (see Table\,\ref{tab_discussion}); the environment in this cluster is apparently rougher than in the other two systems due to stronger ram-pressure. 

Next we discuss our results in the frame of 
four milestone results and predictions on 
galaxy evolution in clusters which are based 
on observations and on theoretical models: (1) 
The dependency of individual galaxy properties 
(like gas content and SF quenching) on the 
cluster-centric distance. 
(2) The commonly accepted role played by RPS 
in galaxy transformation, at least in the inner 
cluster regions.  (3) The pre-processing of
galaxies at large cluster-centric radii is 
thought to be responsible for the transformation
of large fractions of galaxies. (4) Models of 
galaxy evolution predict the transformation 
(quenching and gas sweeping) of large fractions 
of satellite galaxies previous to their infalling
\citep[we refer to the following works and 
references therein,][]{Peng10, Rhee17,Jung18,
Roberts19,Cortese21,Donnari21,Healy21,Boselli22}.
%
%In the low redshift universe the effects on galaxies 
%infalling to the high mass halos of clusters, e.g. on 
%the gas content, color and quenching, are known to 
%strongly correlates (inversely) with the cluster-centric 
%radius.  In general, the environment seems to have 
%weaker influence on galaxies in the high 
%stellar-mass range, like $10^{10}$\msolar\ and 
%beyond. {\bf ADD REF ROBERTS+2019; ApJ 873, 42}.  
%

We explore how the three studied clusters fit 
within these predictions taking into account 
the results described in the previous sections.  
With this aim we use our sample of B-Sp 
%(9.0~$\leq ~$log$ (M_{*} / M_{\odot})\leq\,$10.0)
%(log$ (M_{*} / M_{\odot})\leq\,$10.0) 
as test particles for environment effects. As 
a first approach we consider that a normal \hi\ 
content {\it plus} a blue color constitute an 
indicator of a non-quenched galaxy.  As
mentioned previously, optical colours are not 
the most accurate tracers for quenching but 
they are often used as an alternative, 
distinguishing active from non-active galaxies,
for instance in statistical studies as the 
present one. 
%\citep[e.g.][]{Tacchella19}.
%
%With the aim of better understanding the origin 
%of the observed galaxy transformations,
%we carry out a first stage comparison of our 
%observational results with predictions made by
%models of galaxy evolution in clusters.  We 
In the following discussion we take our 
sample of B-Sp as analogues of the  
{\it satellite} galaxies defined in 
cosmological simulations. We observe that 
none of the B-Sp dominates any of the groups 
reported in Sect.\,\ref{res_substructures}, 
giving support to this approach. Current models 
\citep[\textit{e.g.}][]{Xie20}  predict that 
evolution of low mass galaxies, those having 
log$ (M_{*} / M_{\odot}) <$ 10.3  (matching 
the range of our B-Sp) is dominated by 
environmental effects.  Above this threshold 
AGN feedback is expected to drive the galaxy 
quenching.
%Very few of these B-Sp reach the value of log$ (M_{*} / M_{\odot})=\,$10.  Even if the mass alone does not constitute a criteria on the satellite status, 

%Above this threshold AGN feedback 
%is expected to drive the galaxy quenching.  

%Now we discuss the main results obtained for each cluster followed by a composite view of the three clusters where we explore the behaviour of the sample of bright spirals in such composite cluster.  \\

%
%This constitutes a lower limit for the cluster deficiency as some of the detected B-Sp, three in A496 and eight in A2670, are reported as \hi-deficient (see Table\,\ref{tab_A496_det} and \ref{tab_A2670_det}). This would rise the \hi\ deficiency values to 0.63 and 0.70 for these two clusters.  Even if, the three studied clusters coincide in the \hi\ deficiency, 
%

%The three clusters behave quite differently in other aspects such as, (a) the cluster-centric distribution of gas-rich/gas-poor spirals, (b) the balance between {\it cluster vs. pre-processing} effects, and (c) the degree of processing of the infalling low-mass satellites.  

%\noindent
%$\bullet$ A85  

\subsection{Global properties of the studied clusters}

%We describe the major physical properies derived from our observations and 
Here we compare some physical properties of
the studied clusters with each other. In order 
to avoid part of the observational bias we focus
on the inner regions, more specifically on 
the largest physical radius being observed 
with the VLA in the three systems. 
%This roughly 
%coincides with the virialized radius and,
%(see Fig.\,\ref{fig_distribution}) at 1.3\,R$_{200}$. 
%in order to be more conservative, 
Based on this criteria we take the conservative 
value of 1\,R$_{200}$ as comparison radius.  
Obviously, some bias remain due to different 
cluster distances which has an effect on the 
population of \hi\ detected galaxies. For 
instance, typical \hi\ masses reported in 
A85 and A2670 approach 8\,\por\,10$^{9}$\,
\msolar, while in A496 this value is 
\aprox2\,\por\,10$^{9}$\,\msolar. As a 
consequence, in this last cluster we report
more \hi\ galaxies in the low mass regime 
(Sect.\,\ref{results}). Nonetheless, some 
peculiarities described below remain even 
after considering this observational bias. 
Table\,\ref{tab_discussion} resumes the data 
corresponding to 1\,R$_{200}$ for the three 
clusters.  \\

%If we consider the number of member objects within this inner zone, A85 shows -by far- the largest number of member galaxies of the three systems under study, despite A496 being the closest one. 
%
%Table\,\ref{tab_discussion} shows other intriguing results, for instance, the high number of B-Sp in A85 but the absence of red spirals in this system.  A496 harbors a striking number of low mass \hi\ detections compared with the other two clusters, while A2670 shows a much higher fraction of \hi\ abnormal galaxies than A85 and A496.  We will analyze these issues in a forthcoming paper.

\noindent
$\bullet$ {\bf A85}:  This stands out as a very 
\hi\ deficient system.  The \hi\ detections 
represent a fraction of 0.016 of the member galaxies 
within 1\,R$_{200}$  (Table\,\ref{tab_discussion}). 
The majority of the B-Sp not detected in \hi\ 
(Fig.\,\ref{fig_phase-space}) gather 
towards the virial zone of the PPS diagram.  
%Fig.\,\ref{fig_distribution} shows 
%a  remarkable grouping of stripped spirals 
%along the frontier between A85 and the SE subcluster.  
%Despite the asymmetric distribution of B-Sp 
%and \hi\ detected galaxies 
%(Sect.\,\ref{res_distrib}), 
This cluster shows the expected cluster-centric 
distribution of gas-rich/gas-poor spirals. The 
strongest ICM effects in A85, compared with A496 
and A2670, are supported by two results given in 
Table\,\ref{tab_discussion} within 1\,R$_{200}$: 
(a) the low fraction of \hi\ detected galaxies 
relative to the member objects 
(0.01, 0.19, 0.11, in A85, A496, A2670 respectively); 
and (b) the fraction of B-Sp not detected in \hi, relative 
to the B-Sp reported within the same region (0.88, 
0.31, 0.51, in A85, A496, A2670 respectively).  
These numbers do not correlate with redshift so 
we discard any observational bias as responsible 
for this result.

%The number of swept spirals 
%confirms an active transformation from blue, star-forming galaxies, to gas stripped, quenched objects. Such galaxy transformation 
These effects are well explained by RPS.
We showed (Sect.\,\ref{disc_rps}) 
that this mechanism might exceed the 
anchoring force up to a radius close
to 0.6\,$r$/R$_{200}$ for typical ($1\sigma$)  
velocities, and up to 0.9\,$r$/R$_{200}$ 
for faster infallers ($1.5\sigma$; see 
Fig.\,\ref{fig_rps}). 
%This would imply that the environment in A85 plays a major role transforming the incoming spirals.  
As a matter of fact, the central regions of 
A85 could be an extreme case even compared 
with Coma, where a number of gas deficient 
spirals are still detected within 1\,Mpc, 
with similar \hi\ mass limit detection than 
the present work \citep{Bravo-Alfaro00, 
Healy21, Molnar22}.  In A85 the only 
\hi\  projected close to the center 
is the high-speed infalling jellyfish JO201, 
which shows signs of strong ram pressure 
stripping \citep{Ramatsoku20, Luber22}.

In parallel to RPS, we inspect the role 
played by pre-processing and by the shocked 
interface between A85 and the SE-group, in
the transformation of infalling spirals.
%, in the transformation of late types in this cluster. 
%This assertion is based on the substructure analysis reported 
%in Sect.\,\ref{res_substructures}.
%and on the \hi\ properties of galaxies belonging to those groups.  
A large majority of the B-Sp which are
not detected in  \hi\  are located in the 
SE quadrant, where the shocked interface is 
found. The complexity of this region is 
consistent with the presence of two groups 
(the South Blob and the Filament groups) 
in addition to the SE subcluster itself
(Fig.\,\ref{fig_subs_A2670}).  Another 
jellyfish galaxy (JO200) is located
in this zone being so deeply stripped that
it was not detected in \hi\ down to a mass 
limit of 9\,\por\,$10^8$\,\msolar\  
(A85[DFL98]286 in Table\,\ref{tab_A85_no_det}).  
\cite{Venkatapathy17} already proposed this
galaxy to be in a more advanced stage of
transformation than JO201.  

A remaining number of \hi\ non-detected B-Sp 
galaxies lies in the NW quadrant of A85 
(Fig.\,\ref{fig_distribution}). 
These objects are seen slightly beyond the 
escape velocity curve of the PPS of
Fig.\,\ref{fig_phase-space}. No substructures 
are reported in that zone; therefore we 
favor the backsplash hypothesis \citep{Gill05,
Hirschmann14}.
%Their stellar masses log$(M_{*} / M_{\odot})$ 
%are in the interval 9.5--10.0.
This suggests that these galaxies had entered A85 
from the SE and now we see them projected to the 
NW.  Timescales for RPS quenching (\aprox1\,Gyr) 
and for cluster crossing (\aprox3\,Gyr) are 
in good agreement. Considering that morphology 
transformation, disregarding the responsible
mechanism, takes longer timescales 
(\aprox10\,Gyr), backsplash is a plausible 
explanation for the spiral morphology 
and their location within the cluster 
\citep[]{Boselli-Gavazzi06, Boselli-Gavazzi14,
Cortese21}.

As a first approach to quantify pre-processing
in A85 we obtained the 
number of \hi\ perturbed galaxies, and those 
non-detected, within each of the groups
reported in Sect.\,\ref{res_substructures}. 
We obtained the following numbers: 
SB:3, Filament:~4, SE:~4, SW:~0, NE:~1.  This 
suggests that pre-processing could be at work 
within the groups located SE of A85. However 
it is difficult to separate 
pre-processing from the effects of the interface 
shocking region. The later covers very well the 
zone where the B-Sp not detected in \hi\ are found.

%An individual study of this sample of possible backsplash objects is of great interest and will tackled in a forthcoming paper.

The sample of blue, \hi\ normal galaxies in 
A85 projected between 1.5\,Mpc and 3.0\,Mpc 
east of the cluster center raises some 
unanswered questions. They are confirmed as 
cluster members and, being in the stellar 
mass range log$(M_{*} / M_{\odot})$ 
between 9.0 and 10.0, these objects are
predicted to have high probability to be 
quenched before their complete infall onto 
a massive cluster like A85 
\citep[\textit{e.g.}][]{Bahe19}. For instance, 
\cite{Xie20} estimates that \aprox60\% of 
galaxies would be quenched in the mass range 
given above and halos (clusters) with masses 
above log$(M_{halo} / M_{\odot})=$\,14.0,
which is the case for this cluster. 
%(see Table\,\ref{tab_clusters}).  
%A statistical approach of this issue will be tackled in a forthcoming paper.

In any case the gas rich galaxies in A85 seem 
to be part of a continuous flow of newcomers 
occurring during several Gyr.  This scenario 
is supported by the fact that A85 is part of 
the supercluster MSCC-039 which includes eleven 
systems \citep{Chow-Martinez14}. The large scale 
structure joining A85 with its neighbors, in 
particular A117A, projected at \aprox12\,Mpc 
SE of A85 (C. Caretta 2021, priv. comm.), could 
be responsible for the feeding of fresh 
\hi\ galaxies and with the presence
of the SE subcluster.  \\

\noindent
$\bullet$ {\bf A496}:  This cluster constitutes 
a puzzle not only compared with the two other 
systems under study but also relative to the 
low-z clusters previously imaged in \hi.  On 
one hand A496 is confirmed as a relaxed system 
and, in agreement with this, it is very 
concentrated.  The PPS of 
Fig.\,\ref{fig_phase-space} shows that more 
than 50\% of member galaxies lie within the 
virialized zone.
%compared with 23\% in A85, and 
%34\% in A2670 (see Table\,\ref{tab_results}).
However, A496 presents some features in 
apparent contradiction with an evolved 
system. First, the high number of \hi\ 
detections (58) constitutes 17\% of the 
member galaxies within the VLA FoV (this 
is below 2\% in A85; see 
Table\,\ref{tab_results}).  Even if we 
compensate for A496 being closer than the other
two systems the trend clearly persists. Another 
paradox is the absence of the expected 
cluster-centric distribution of \hi\ galaxies.  
This is observed in clusters like Coma and 
A85, comparable to A496 in dynamical stage.  
Figs.\,\ref{fig_distribution} 
and \ref{fig_distribution_HImaps}
%and a plot  of \hi\ deficiency {\it vs.} radius (not shown in this work), account 
clearly show the lack of correlation. 
As mentioned before, particularly strong
projection effects should explain the \hi-rich 
galaxies projected onto the cluster core.  
This scenario fits with active ram-pressure 
stripping in this cluster. Our estimations 
predict that RPS surpass the anchoring 
force at \aprox0.4\,$r$/R$_{200}$ for typical 
$1\sigma$ velocities, and up to 
0.6\,$r$/R$_{200}$ for velocities of 
$1.5\sigma$ (Fig.\,\ref{fig_rps}). This 
implies that \hi\ normal spirals are hardly 
expected near the cluster core.

%In this scenario, A496 appears as a less evolved system compared with A85; in the later, the majority of LTGs have already lost their gas envelope while in A496 a large population of LTGs have not yet reached the densest cluster environment, explaining the high number of \hi\ detections, including many low-mass galaxies. 

Taking the full sample of B-Sp as test particles 
for environment effects we find that 37\% 
of them have lost a large fraction of gas and 
were not detected in \hi\ 
(Table\,\ref{tab_results}). 
%We remind that the detection threshold for this cluster is as low as 3\,\por\,$10^8$\,\msolar\ (see Table\,\ref{tab_cubes}).  
Furthermore, among the \hi\ detected B-Sp a 
number of them are reported as abnormal 
(Table\,\ref{tab_A496_det}).  
%Another important sign of transformation of spirals is based on
If we consider the color index, one third of the 
B-Sp in A496 are located in the red sequence. 
%(Fig.\,\ref{fig_colour-mag}). 
All these results clearly indicate an important 
degree of spiral transformation in this cluster.

Another intriguing result in A496 is the 
large number of \hi\ detected galaxies with
%(open squares/diamonds with no solid circles in Fig.\,\ref{fig_phase-space}). 
masses below log$ (M_{*} / M_{\odot})$=9.0, 
some of them having \hi\ mass 
$\geq$ 3\por10$^{9}$\,\msolar\ (see 
Table\,\ref{tab_A496_det}).
%They represent nearly half of all the detected galaxies in the cluster.  
In similar clusters, like Hydra, the average 
\hi\ mass for the same range of stellar mass
is significantly lower, below 10$^{9}$\,\msolar\
\citep[]{Wang21}.  
Even considering the closer distance of A496
(the nearer the cluster, larger the fraction 
of low-mass \hi\ detected objects) we expected 
similar galaxies to be found in A85 and A2670. 
Taking into account our \hi\ detection 
limits and by scaling the corresponding \hi\ 
flux to the distance of A85 and A2670, our 
survey would be capable to detect such 
gas-rich galaxies with stellar masses around 
log$ (M_{*} / M_{\odot})$=8.5.
%\citep{Denes14}.  
However these objects are absent, more 
conspicuously in A85.
%and are spread across diverse zones of the cluster.
%, from the virialized 
%region to the limits of the escape velocity 
%curve (see Fig.\,\ref{fig_phase-space}). 

The low-mass, gas-rich and very blue galaxies 
in A496, are very likely in a pre-quenching 
phase. These objects deserve 
a more detailed study; current models 
predict strong pre-processing and a 
significant quenching fraction, going up 
to 80\% for log$ (M_{*} / M_{\odot})=$\,9.0, 
at $z=0$ \citep[]{Donnari21}. In a forthcoming
paper we will try to determine, statistically, 
how close to a massive halo these low-mass 
galaxies last before extinguishing. This  
could help to restrict theoretical models. 
%  STILL TO USE IN THE CONCLUSIONS
%and combining the red passive spirals to estimate the time from the cluster accretion to the final quenching (i.e. the quenching time-scale) 

Beside the projection effects, the full 
distribution of \hi\ galaxies in A496
(Fig.\,\ref{fig_distribution}) deserves 
a closer look. The NE quadrant is almost 
empty of \hi\ objects. In the south and
south-east regions, all the \hi\ detections 
belong to the high velocity domain (red maps 
in Fig.\,\ref{fig_distribution_HImaps}),
suggesting a common origin.  Finally the 
north-west is the richest zone of both, \hi\ 
detections and member galaxies. This could be
associated with the only substructure found 
in this cluster (Sect.\,\ref{res_substructures}).  
Concerning pre-processing, the very low degree 
of substructures suggest that this mechanism 
is not playing a major role transforming 
infalling galaxies in A496.  Instead, the 
cluster environment seems to dominate the 
processing of spirals.

\begin{figure}
    \centering
    \includegraphics[width=\columnwidth]{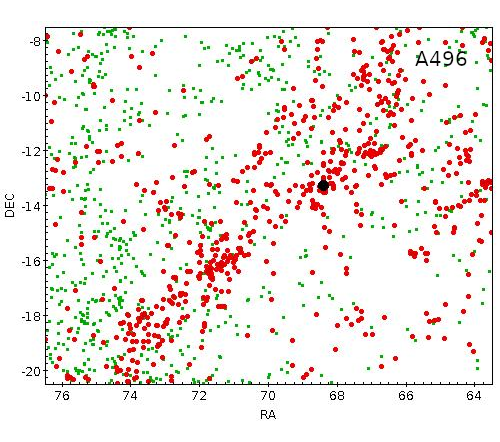}
    \caption{6dF galaxies (green dots) found within a 
    {\it square} of 12\degree\ of side around A496 
    (black diamond).  Red dots indicate the 498 galaxies 
    having velocities within the 3$\sigma$ interval around A496. }
    \label{Fig_LSS_Flo}
\end{figure}

\begin{table*}
    \centering
    \caption{Different types of galaxies within 1\,R$_{200}$:
    comparing the inner regions of A85, A496, A2670.
}
    \label{tab_discussion}
    %\begin{tabular}{llcccccc}
   \begin{tabular}{
  p{\dimexpr.18\linewidth-2\tabcolsep-1.3333\arrayrulewidth}% column 1
  p{\dimexpr.15\linewidth-2\tabcolsep-1.3333\arrayrulewidth}% column 2
  p{\dimexpr.08\linewidth-2\tabcolsep-1.3333\arrayrulewidth}% column 3
  p{\dimexpr.08\linewidth-2\tabcolsep-1.3333\arrayrulewidth}% column 4
  p{\dimexpr.08\linewidth-2\tabcolsep-1.3333\arrayrulewidth}% column 5
  p{\dimexpr.08\linewidth-2\tabcolsep-1.3333\arrayrulewidth}% column 6
  p{\dimexpr.08\linewidth-2\tabcolsep-1.3333\arrayrulewidth}% column 7
  p{\dimexpr.08\linewidth-2\tabcolsep-1.3333\arrayrulewidth}% column 8
  }
    \hline 
 & & \multicolumn{2}{c}{A85} &  \multicolumn{2}{c}{A496}&  \multicolumn{2}{c}{A2670}\\
Galaxies  &	Normalized by:	&	\# 	&	(fract.) &	\# 	&	(fract.)	&	\# 	&	(fract.) \\
	(1)  & (2)  & (3)   & (4)  & (5)  & (6)  & (7)  &  (8)  \\
	\hline
	Members	&		&	574	&		&	271	&		&	195	&		\\
 \multicolumn{8}{l}{}\\
\hi-det	&	Members &	~8	&	0.01	&	52	&	0.19	&	21	&	0.11	\\
B-Sp	&	Members &	58	&	0.10	&	39	&	0.14	&	35	&	0.18	\\
 \multicolumn{8}{l}{}\\
B-Sp \hi-det	 &	B-Sp &	~7  &	0.12 &	27	&	0.69	&	17	&	0.49	\\
B-Sp \hi\ non-det&	B-Sp &	51 &	0.88 &	12	&	0.31	&	18	&	0.51	\\
B-Sp in red seq	&	B-Sp &	~6  &	0.10 &	14	&	0.36	&	~2	&	0.06	\\
 \multicolumn{8}{l}{}\\
\hi-det in red seq 	& \hi-det 	& ~0 &	0.00 &	~7 & 0.13	& ~0	&	0.00	\\
\hi-det abnormal	& \hi-det 	& ~1 &	0.13 &	15 & 0.29	& 10	&	0.48	\\
Low mass \hi-det	& \hi-det 	& ~1 &	0.13 &	21 & 0.40	& ~3	&	0.14	\\
%Members	&		&	616	&		&	368	&		&	308	&		\\
%Members VLA	&		&	603	&		&	323	&		&    244	&    \\
%\hi-det & Members VLA &	10	&	(0.02)	&	58	&	(0.18)	& 38 &	(0.16)	\\
%B-Sp &	Members VLA	&	59	&	(0.10)	&	43	&	(0.13)	& 55 & (0.22)	\\
%\multicolumn{8}{l}{}\\ 
%B-Sp \hi-det &	\hi-det	& 9	&	(0.90)	&	26	&	(0.45)	&29	&	(0.76)	\\
%B-Sp \hi\ non-det  & B-Sp &	50	& (0.85) &  17  &	(0.39)	&26	&	(0.47)	\\
%B-Sp in red seq & B-Sp 	& 0	& (0.00)    &	15	& (0.35)  & 9	& (0.16)   \\
% \multicolumn{8}{l}{}\\ 
%\hi-det in red seq	& \hi-det &	0 &	(0.00)	& 11 &	(0.19)	& 6	&(0.16)	\\
%\hi-det abnormal    & \hi-det & 1 & (0.10) &  17 &  (0.29)  & 17 & (0.45) \\
%Low mass \hi-det  &	\hi-det	&	1 &	(0.10) &	43	& (0.74) & 5 & (0.13)	\\
% \multicolumn{8}{l}{}\\ 
%B-Sp in red seq &	VLA-FoV	&	25	&	(0.04)	&	15	&	(0.05)	&	21	&	(0.09)	\\
%\multicolumn{8}{l}{}\\ 
%Virial-reg	&	Members &	140	&	(0.23)	&	190	&	(0.52)	&	105	&	(0.34)	\\
%B-Sp  in virial-reg	& B-Sp	&	11	&	(0.20)	&	17	&	(0.53)	&	9	&	(0.20)\\
%\hi-det in virial-reg & \hi-det	& 0	&	(0.00)	&	24	&	(0.41)	&	8	&	(0.21)	\\
%%RPS-reg	&	full members&	48	&	(0.08)	&	26	&	(0.07)	&	26	&	(0.08)	\\
%%RPS-reg + Esc-vel  & full members & 16	&	(0.03)	&	18	& (0.05) &	13	& (0.04)	\\
\hline
\multicolumn{8}{l}{Notes: }\\ 
\multicolumn{8}{l}{-- All objects mentioned in Columns 1 and 2 are
restricted to 1\,R$_{200}$.}\\ 
\multicolumn{8}{l}{-- The numbers of galaxies mentioned in Column 1 are 
given in columns 3 (A85), 5 (A496), 7 (A2670).}\\
\multicolumn{8}{l}{-- Column 2 indicates the sample used to normalize the 
number of galaxies referred in Column 1.}\\ 
\multicolumn{8}{l}{-- The fractions {\it "Column (1) / Column (2)"} 
are given in Columns 4 (A85), 6 (A496), 8 (A2670).}\\
%\multicolumn{8}{l}{-- Members VLA: indicates the number of member galaxies within the volume defined by the VLA data cubes.}\\ 
%\multicolumn{8}{l}{-- HI-det / HI non-det: galaxies detected / not 
%detected in HI. } \\
%\multicolumn{8}{l}{-- B-Sp: the number of bright spirals reported in Sect.\,\ref{obs_catalog}.}\\
%\multicolumn{8}{l}{-- \hi-abnormal: indicates objects showing 
%the \hi\ perturbations described in Sect.\,\ref{res_detect}. }\\
%\multicolumn{8}{l}{-- B-Sp phot: the bright spirals within the optical frames for which photometry was estimated in Sect.\,\ref{obs_opt}}\\
%\multicolumn{8}{l}{-- Red seq: indicates above the red sequence defined in Fig.\,\ref{fig_colour-mag}}. \\
%\multicolumn{8}{l}{-- Virial-reg: indicates the virialized region 
%shown in the PPS of Fig.\,\ref{fig_phase-space}. }\\
%\multicolumn{8}{l}{-- RPS-reg: indicates the region left of the RPS curve in the PPS.}\\
%\multicolumn{8}{l}{-- Esc-vel indicates the region below the escape velocity curve.}
\end{tabular}
\end{table*}

We explored the surrounding large scale structure 
of A496 in order to better understand the 
replenishment of gas-rich galaxies. We analyzed 
6dF Survey data within an area of 12 degrees 
around A496; the center of this cluster is
indicated with a black solid circle 
in Fig.\,\ref{Fig_LSS_Flo}. We applied a galaxy 
request restricted to the redshift range of 
0.0255 -- 0.0407 (7552 -- 11962 \kms), 
corresponding to the 3$\sigma$ interval of 
A496.  The output of 498 galaxies is displayed 
in the same figure.  This plot shows in 
green the positions of all the galaxies found in 
6dF while the objects in red correspond to the 
selected redshift interval. Important galaxy 
concentrations are clearly seen, drawing a 
filament in the NW-SE direction, having 
redshifts of 0.037 and 0.038 (the redshift 
of A496 is 0.033).  The nearest structure to 
A496 lies to the NW and could be at the origin 
of the large number of gas-rich objects in the 
NW zone of A496 and the substructure reported 
in the same region (Fig.\,\ref{fig_subs_A2670}).  
Other observational evidences support this 
filament of large scale structure, like the 
X-ray cluster MCXC J0445.1-1551, SE from A496. 
This system was not identified originally by 
Abell and is found at RA=71.5, Dec=-16, 
centered on the cD galaxy NGC 1650.  These 
filaments account for the observed replenishment 
of gas rich galaxies, in particular from the NW. \\

\noindent
$\bullet$ {\bf A2670}:  This system is dynamically 
younger than the other studied clusters so the 
high fraction of B-Sp within the VLA FoV 
(24\%, see Table\,\ref{tab_results}) 
is not unexpected. The fraction of B-Sp within 
1\,R$_{200}$ is larger than in the other two systems 
(Table\,\ref{tab_discussion}).  However, opposite 
to other young clusters like Virgo \citep{Chung09}, 
Hydra \citep{Wang21} and Abell 963 \citep{Jaffe15}, 
the correlation between \hi\ deficiency and 
cluster-centric radius is not seen in A2670.
%(plot not shown in this work).  
This is due to the large number of \hi\ 
rich galaxies projected onto the cluster 
core (Figs.\,\ref{fig_distribution} and 
\ref{fig_distribution_HImaps}) which are 
located into the virialized zone of the 
cluster (Fig.\,\ref{fig_phase-space}).  
Our estimations of RPS in A2670 predict 
significant environment effects, even if 
less intense than in A496 and A85.  In 
A2670, RPS is expected to dominate within
$r$/R$_{200}\leq$\,0.3, for relative velocities 
of $\sim 1.0\sigma$, and within a radius of 
$r$/R$_{200}\leq$\,0.4, for higher velocities 
$\sim 1.5\sigma$ (Fig.\,\ref{fig_rps}). 
Therefore most of the \hi\ detections projected
in the very central region of A2670 are 
explained by projection effects, either being 
fast infallers along the line of sight or 
objects with non-radial orbits.

We use the B-Sp properties to investigate the 
strength of the environment effects in A2670. 
Fig.\,\ref{fig_colour-mag} shows that a number 
of B-Sp have reached the red sequence; five 
of them retain enough gas to be 
detected in \hi.  Interestingly, almost half of 
the \hi\ detected galaxies are classified as 
abnormal (Table\,\ref{tab_results}) confirming 
a rather harsh cluster environment.  The fraction
of \hi\ abnormal objects within 1\,R$_{200}$ is 
much larger in A2670 (0.48) than the other two 
clusters. We quantify the importance of pre-processing 
in A2670 by contrasting the positions of the 
\hi\ perturbed B-Sp with the substructures 
reported in Sect.\,\ref{res_substructures}. 
We found only two of such galaxies in the 
north-center (NC) group and four in the SW,
including 3 \hi\ abnormal and one non-detected 
object.  The SW region seems to be the only 
zone of this cluster where pre-processing is 
at work.  
%The NE group do not harbor any B-Sp galaxy.  

The elongated morphology of A2670, following 
a NE-SW axis, is confirmed 
by the distribution of member galaxies in
Fig.\,\ref{fig_distribution}.  The global 
positions of \hi\ detections help to trace 
the accretion history of A2670.  
%Despite the fact that our \hi\ survey covers an elongated region along the same NE-SW axis, 
When we focus beyond 1.0\,R$_{200}$, a strong 
asymmetry is seen between the very \hi\ rich 
south-west and the \hi\ poor north-east region 
(Fig.\,\ref{fig_distribution_HImaps}).  This
strongly suggest the former as the preferred 
direction of infalling spiral galaxies.  
%
%The later zone  harbors only a few detections, most of them consisting of shrunk \hi\ disks  and a few red spirals.  
%Some of these objects 
%could be of backsplash origin. This is based
%on the evidence of the SW as the preferred 
%direction of infalling galaxies.  
%
%In addition, we observe a
%region devoid of \hi\ galaxies  
%(see Fig.\,\ref{fig_distribution_HImaps}). 
%This gap is ring shaped with inner/outer 
%radii of 15\prim / 25\prim, equivalent to 
%0.8/1.3\,$R_{200}$. This is visible as well 
%in the PPS diagram of Fig.\,\ref{fig_phase-space}.
%Concerning the \hi\ rich, very low-mass galaxies 
%(masses log$ (M_{*} / M_{\odot})<9.0$), they
%constitute only 10\% of all the detected objects 
%in A2670 while they represent 48\% in A496.

%Due to their low anchoring force some of these galaxies are expected to be gas-stripped at their current infalling stage.
%A forthcoming paper will study these objects
%in more detail.

\begin{figure*}\centering
    \begin{subfigure}{0.4\linewidth}
    \includegraphics[width=\linewidth]{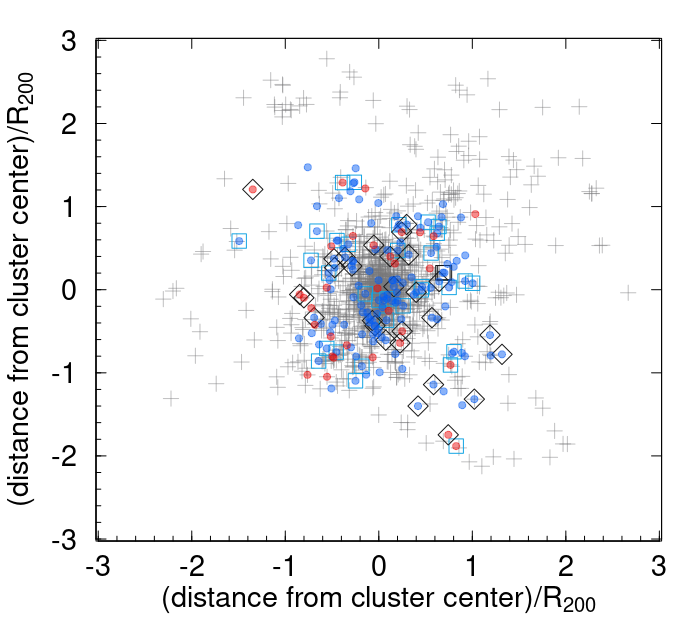}  
    \end{subfigure}
    \begin{subfigure}{0.55\linewidth}
    \includegraphics[width=\linewidth]{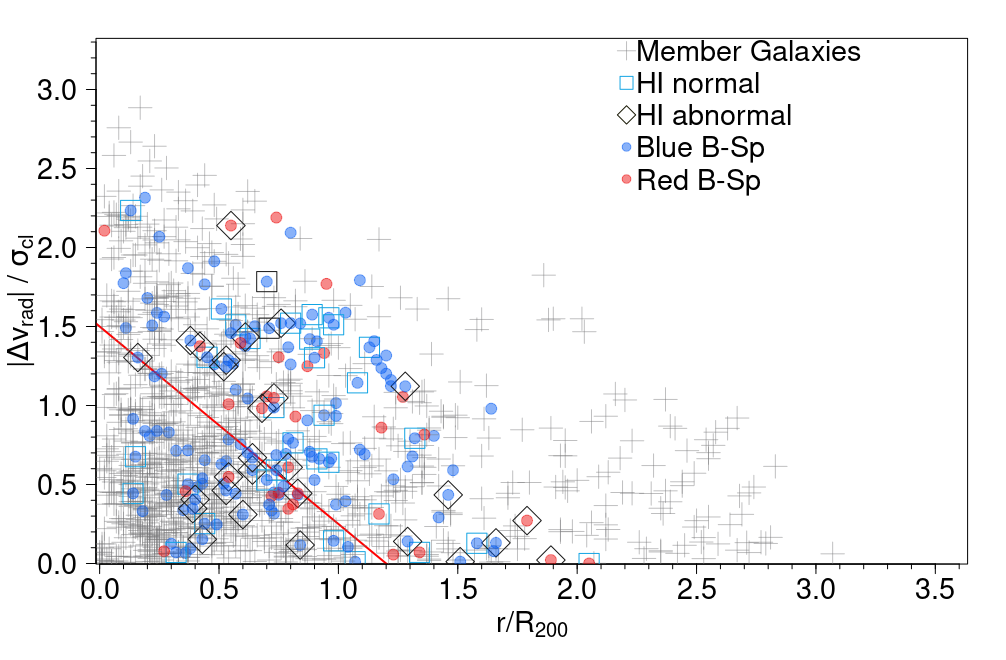}  
    \end{subfigure}
    \caption{Composite plots of three clusters showing the 
    member galaxies (grey crosses), highlighting the bright 
    spirals; blue/red spirals are shown with solid circles; \hi\ detections are given with squares and 
    diamonds, indicating normal and abnormal objects, respectively. 
    %Solid circles without square/diamond correspond to non-detected spirals.
    Left panel: the projected position on the sky, normalizing 
    cluster-centric distances by R$_{200}$. Right panel: positions 
    of galaxies in the phase space diagram. The red line delimits the 
    virialized zone as defined in Fig.\,\ref{fig_phase-space}.}
    \label{fig_composite}
\end{figure*}

%\subsection{Individual clusters: comparison within 1\,R$_{200}$ }
%\label{disc_comparison}

\subsection{The evolutionary stage of the bright spirals: a three-cluster composite view }\label{disc_3clusters}

In order to explore the global behavior of the
large sample of bright spirals (B-Sp) collected in
this work we create a composite cluster constituted 
by the three studied systems.  The left panel of 
Fig.\,\ref{fig_composite} shows the distribution, 
on the plane of the sky, of all member objects 
and all the bright spirals (we use the same 
symbols than in previous plots).
%in order to highlight the blue/red spirals (blue/red solid circles), and \hi\ normal/abnormal galaxies (empty squares/diamonds). 
The right panel of Fig.\,\ref{fig_composite} 
shows the PPS diagram of the composite cluster. 
In order to properly display the three systems 
in a single plot we normalize the galaxy 
cluster-centric distance by the 
corresponding R$_{200}$.  The relative 
velocities are normalized by the corresponding
cluster velocity dispersion $\sigma_{cl}$ 
\citep[see][]{Luber22}.  
%We display the absolute value of this relative velocity as we did in Fig\,\ref{fig_phase-space}. 

%Fig.\,\ref{fig_composite} provides a clear view of 
%the location of the bright spiral sample across the 
%composite cluster. 
We study the behaviour of the B-Sp across the 
composite cluster by looking for trends between
their distribution and their \hi\ properties.
We split the full catalog of B-Sp in two 
sub-samples named: {\it normal} B-Sp
and {\it disturbed} B-Sp. The former is 
constituted by those \hi\ detected B-Sp,
displaying normal distribution and gas content.
The sub-sample of {\it disturbed} B-Sp includes 
those being \hi\ abnormal {\it plus} those \hi\ 
non-detected.  Defined in this way, the disturbed 
sub-sample includes galaxies undergoing strong 
degrees of transformation.  Next we estimate 
the mean projected distances for each sub-sample 
and we compare them. We use units of $r$/R$_{200}$, 
obtaining the following results: 

\noindent
normal B-Sp: ~~~$ \mu_{1}$=0.63;\\ disturbed B-Sp:   $ \mu_{2}$=0.56.

This confirms the expected scenario where \hi\ 
disturbed spirals are located, on average, closer 
to the cluster center than \hi\ normal ones. The 
difference between means is, however, not found 
to be significant enough ($1 - p \lesssim 90\%$) 
by the Welch's t-test. As we next show, this is 
most likely due to the heterogeneous trends seen 
in the individual clusters.  
%This confirms the expected scenario where disturbed 
%spirals are located, in average, closer to the 
%cluster center than the normal ones. However, the
%difference is not conclusive.  We quantify 
%the difference between the means 
%$ \mu_{1}$ and $\mu_{2}$ is conclusive, we examine
%the level of significance by applying
%the statistical Student's t-test and explore 
%the null hypothesis $H_{0}:\mu_{1} = \mu_{2}$.
%where the sub-index "1" corresponds to the normal sub-sample.  
%We find that the difference in average
%projected distance of the normal and the 
%disturbed sub-samples is confirmed with a low
%confidence level (p-value of 0.117). This means 
%that we might reject the null hypothesis $H_{0}$ 
%with a confidence level of only \aprox90\%.
%The correlation of more processed spirals laying 
%closer to their parent halo is not very strong 
%because of the heterogeneous trends seen in the 
%individual clusters. 
In order to quantify such
trends we contrasted the average distances of
the same sub-samples within each system: \\

\noindent
A85: ~~~normal B-Sp: $ \mu_{1}$=0.86 {\it vs.} disturbed B-Sp: 
$ \mu_{2}$=0.59.\\
~A496: ~~normal B-Sp: $ \mu_{1}$=0.66 {\it vs.} disturbed B-Sp: 
$ \mu_{2}$=0.75. \\
A2670: normal B-Sp: $ \mu_{1}$=0.96 {\it vs.} disturbed B-Sp: 
$ \mu_{2}$=0.85\\

This confirms, in A85, the very strong trend where 
disturbed B-Sp lay at lower cluster-centric radius
(confidence value $>$\,99\%). Oppositely, A496 has a very 
atypical behaviour with an inverse trend of normal B-Sp 
located, in average, at lower radius than their disturbed 
counterparts. Despite the projection effects cited
above, such a trend has never been observed in other 
clusters.  A2670 is an intermediate case where 
%even 
%if the normal B-Sp are located at larger radius than 
%the disturbed ones, the level of confidence for the 
the difference between $ \mu_{1}$ and $ \mu_{2}$
has low statistical significance.
%can not be rejected as the {\it p-}value is very low, 
%of \aprox85\%.   
Other statistical methods produce similar results,
for instance the Two-sample test for the mean, 
considering the one-tail case of the z-Test.

\section{Summary and Conclusions}
\label{summary}

This paper is devoted to study galaxy 
evolution and environment effects occurring 
at the core and outskirts of three nearby 
Abell clusters: A85/A496/A2670.  We analyzed 
the transformation of spirals as a function 
of their  local environment by studying 
different clusters lying at roughly the same 
cosmological epoch.  This work is 
based on a blind, VLA \hi-imaging survey, with 
data cubes covering a volume that extends beyond 
one virial radius and more than three times the 
velocity dispersion of each cluster. 
%We combine the \hi\ data with CFHT multi-band optical images and a substructure analysis. 
%
%
%In addition, \hi\ synthesis imaging at low redshifts resolved spatially most of the detected objects providing fundamental information on the physical mechanisms affecting individual galaxies and permitting a direct comparison of \hi\ morphology and kinematics with numerical simulations of physical mechanisms such as RPS or tidal interactions. As a first result we deliver a catalog of \hi\ parameters and images (\hi\ maps and velocity  fields) for the detected galaxies: 10, 58, and 38, in A85, A496, and A2670 respectively.
%
%
%We report two additional samples of galaxies which deserved further attention: (a) the sub-sample of B-Sp that have migrated into the red sequence ($i.e.$ the passive spirals), and (b) the sample of blue, low-mass ($log (M_{*} / M_{\odot}) \leq$ 8.5), \hi\ rich galaxies.  The later are used for a first approach testing the degree of quenching in the clusters outskirts.  
Our main results and conclusions can be 
summarized as follows.\\

\noindent
$\bullet$ We report \hi\ parameters
for 10, 58, and 38 galaxies in A85, A496, and A2670,
respectively, as well as an atlas of the 
corresponding \hi\ maps and velocity fields.
We defined a complete sample of bright spirals 
%up to M$_{\mathrm{B}}=-$18.4 
with masses above log$(M_{*} / M_{\odot})=9.0$
%M$_{\mathrm{r}}=-$19.0), and 
and we used them 
as test particles for environment effects.  
This sample helped as well to obtain a catalog of 
the spirals not detected in \hi\ and to trace 
the accretion history of the parent clusters. \\

\noindent
$\bullet$ We report further evidence for cluster 
environment playing an active role in transforming
spirals, both in their \hi\ content and in their 
color.  We measured the global \hi\ deficiency 
of the three clusters, finding a very high value 
(85\%) for A85, and less conspicuous deficiencies 
for A496 and A2670.  The expected cluster-centric 
distribution of \hi\ rich/poor galaxies is only 
observed in A85. The other two systems show a 
very complex distribution of their \hi, 
with a high fraction of gas-rich objects lying 
within the virialized zone. Strong projection 
effects must be responsible.\\

%\noindent
%$\bullet$ A496 appears as a less evolved system 
%compared with A85; in the latter, most of the spirals 
%have already lost their gas envelope, while 
%the other two systems show large numbers of spirals 
%at early infalling stages.  We report more 
%\hi\ detections in A496 and A2670, compared with 
%A85. A fraction of the \hi\ galaxies 
%are classified as \hi\ abnormal objects. \\

\noindent
$\bullet$  We report a number of bright spirals that 
have moved into the red sequence in the three clusters. 
The majority of these objects are \hi\ disturbed. 
This sample of red spirals 
constitutes additional evidence for systematic 
transformation from normal, blue spirals, into 
gas poor, passive objects.\\

\noindent
$\bullet$ We traced ram-pressure stripping 
profiles in order to measure the strength 
of environment effects due to the intracluster 
medium. 
%We estimated an {\it ad hoc} value of 
%the anchoring force fitting at the outskirts 
%of the sample of bright spirals in the studied clusters.  
We find that RPS accounts for the 
disrupted/stripped galaxies in the central 
regions of the three studied clusters. The 
inner 1.5\,Mpc (roughly one virial radius) 
of A85 appears to be particularly harsh, 
only compared with clusters like Coma. \\

%\noindent
%$\bullet$ In addition to the \hi\ abnormal
%galaxies in the central regions of the studied 
%clusters, which are likely affected by RPS, we 
%report a sample of peculiar galaxies lying away 
%from the densest ICM zone. The global distribution 
%of \hi\ rich/poor spirals strongly suggests that 
%these objects constitute backsplash galaxies. 
%They have low gas content (some are not 
%even detected in \hi) and/or belong to the 
%red spiral sample.  Their locations match 
%the time-scales for gas stripping and cluster 
%crossing.  \\

\noindent
$\bullet$ Our substructure analysis confirms 
A85 as a complex system with several minor 
groups and a major SE subcluster merging 
with the main cluster body. We report two 
new groups in A85, one to the SW and a second 
to the NE. In A496 we unveiled a large group 
NW of the cluster center. We revisited the 
surrounding large scale structure of A496 
and we detected a filament running NW-SE. 
A2670 is an intermediate case between the 
two other studied clusters, having three 
minor groups running along the NE-SW axis.  
The distributions of \hi\ rich/poor
galaxies in the three clusters reveal the 
following (preferred) directions of galaxy 
infalling: the SE for A85, NW for A496, 
and SW for A2670.\\

\noindent
$\bullet$  We explored the relevance of
pre-processing by 
measuring the number of \hi\ disturbed 
galaxies within each of the reported 
substructures.  In A85 we found an important
number of such disturbed objects, SE from the
cluster center.  They could be produced by a 
shocked interface due to a cluster-subcluster 
merger process and/or by pre-processing. The 
other two clusters show no evidence for strong 
pre-processing with the exception of the SW 
group of A2670. Our results suggest that the 
effects of the global cluster environment are 
equally (or even more) important than pre-processing 
in the transformation of spiral galaxies. \\
%In A85 
%we reported a few \hi\ non-detected galaxies
%belonging to the SE subcluster and to the 
%Filament group.
%indicating some pre-procesing in this region.  
%However, a much larger number 
%of \hi\ non-detected spirals are spread across 

%This suggests that the 
%corresponding shocking region could be the
%most important processing mechanism in this 
%zone of the cluster.  

%\noindent
%$\bullet$  Our preliminary search for a 
%correlation between \hi-deficiency and 
%\hi-asymmetry showed no trend. However, 
%since we have no clue on the gas distribution 
%of the \hi\ non-detected objects, we stress 
%that a  {\it deficiency--asymmetry} correlation,
%developing at more advanced stages of gas 
%stripping, cannot be discarded {\ a priori}.  \\

\noindent
%$\bullet$  In addition to the bright spirals
%we detected a number of gas rich, blue dwarfs, 
%with masses log$ (M_{*} / M_{\odot})$ in the 
%range 8.0--9.0.  Many of them are likely only 
%in projection close to the center, in particular 
%for A496. The systemic distances between these 
%objects and their parent massive halos could help 
%restricting the quenching conditions predicted by 
%current models of galaxy evolution, improving the 
%match between observational results and the models. \\

%We used the sample of \hi\ detected B-Sp, and the blue, \hi\ rich, low-mass ($log (M_{*} / M_{\odot}) \leq$ 8.5) galaxies, to quantify the degree of quenching at different cluster-centric radii for the three clusters.  This was done to compare our large volume observations with the model predictions of low-mass satellite galaxies expected to be largely quenched by pre-processing, previous to the infall.  All the galaxies considered in our analysis fit the criteria (low-mass ($log (M_{*} / M_{\odot}) 
%< $ 11.0), satellite and SF) typically applied in the modelling of galaxy evolution approaching  massive halos.  We have found, in all three studied clusters, a significant number of non-quenched galaxies at different cluster-centric radii, in contradiction with model predictions.  This shows that matching the observational results with theoretical model predictions still remains a major challenge. 

\noindent
$\bullet$  The systematic transformation of spirals
in the studied clusters cannot be associated to 
one single parameter of the parent system. 
%Instead,
%we have found that all the cluster properties must 
%be considered in order to better understand
%the evolution of infalling galaxies.  
Therefore, 
in addition to the cluster mass and $L_{X}$, the 
dynamical stage and the feeding of new spirals 
from the large scale structure help to explain the 
observed properties of the spiral population and 
their degree of transformation. 

%In particular, we 
%conclude that the role played by 
%pre-processing is not the same in different studied clusters. 
%The presence of these and other phenomena depend on all the physical properties of the studied clusters.  

%We postpone for the next papers of this series the following topics: (i) The role played by tidal, gravitational mechanisms in the Sp -- S0 transformation. (ii) A deeper study of the red-passive spirals reported in this work, and (iii) of those objects showing strong perturbations in \hi\ and in optical, some of them looking like jellyfish galaxies.  (iv) To study the properties of the possible backsplash galaxies reported in this work, which could provide fundamental information on the journey of infalling spirals across massive clusters.

%\newpage

%\vspace{2cm}

\section*{Acknowledgements}

The authors thank the anonymous referee for her/his 
helpful comments which helped to improve this paper.
M.M.L.G., H.B.A. and C.A.C acknowledge funding from 
CONACyT and DAIP. H.B.A. also thanks {\it Institut 
d'Astrophysique de Paris} for its support during his 
working visits.  We acknowledge Y. Venkatapathy for 
his help enlarging the redshift catalogues used in 
this work.  F.D. acknowledges continuous support from 
CNES since 2002.  Y.J. acknowledges financial support 
from FONDECYT Iniciaci\'on 2018 No. 11180558 and ANID 
BASAL Project FB210003. \\

\noindent
{\bf Data availability.}\\
Data available on request.

%%%%%%%%%%%%%%%%%%%%%%%%%%%%%%%%%%%%%%%%%%%%%%%%%%

%\section*{References}


%\bibliographystyle{mnras}


%%%%%%%%%%%%%%%%%%%%%%%%%%%%%%%%%%%%%%%%%%%%%%%%%%

%%%%%%%%%%%%%%%%% APPENDICES %%%%%%%%%%%%%%%%%%%%%

\newpage

\appendix

%\section*{Appendix}
\onecolumn
%\\
\section{\hi-detected galaxies in A85, A496, A2670.}
\label{app_hi_det}

%\newpage
%\onecolumn
%\begin{landscape}
\begin{table}
    \centering
    \caption{The \hi-detected galaxies in Abell\,85 }
    \label{tab_A85_det}
    \resizebox{\textwidth}{!}{%
    \begin{tabular}{clcccccrcccrcccc}
    \hline 
ID &Name & RA & Dec & Morph. & $g$ & color & M$_{*}$ & Opt-vel  & \hi-vel & $\Delta v$ & \mhi & \multicolumn{2}{c}{Def$_{\mathrm {HI}}$} & Dist. & \hi \\
    & & (2000)   & (2000)   & type & mag & ($g-r$)& &\kms &  \kms & \kms  & $10^9$\msolar  &   Sa-Sb & Sc-Irr & Mpc & pert. \\
(1)  & (2)      & (3)      & (4)   & (5)  & (6)    & (7) & (8) & (9) &(10) & (11)& (12)&(13) & (14) & (15) & (16)\\
\hline
%1&079 &00 40 31.7&-09 13 20&LV&15220.4&91.0&1.89&0.22&0.34&1.37\\
%2&139 &00 41 14.2&-08 55 54&LV&15129.4&91.0&0.87&0.09&0.17&1.65\\
1&323 &00 42 18.7&-09 54 14& S?$^*$& 17.86 & 0.376 & 9.50 &15618  &15608&137&8.2&-0.27&	-0.13&	2.39 & {\it N} \\
2&347 &00 42 29.5&-10 01 07&S?$^*$ & 17.62 & 0.352 & 9.59 & 15165 &15175&91&2.5&-0.03&	0.08&	2.86&{\it N}\\
3&374&00 42 41.5&-08 56 49& S?$^*$& 16.39 & 0.512 & 10.26 & 15106 &15107&318&13.8&-0.29&	-0.14&	1.60&{\it N}\\
4&986362$^b$&00 43 01.7&-09 47 34 &-  &17.27  & 0.418 & 9.79  &15110&15107&227&6.5&-0.39&	-0.27&	2.23&{\it N}\\
5&461&00 43 14.3&-09 10 21&- & 18.84 & 0.318 & 9.01 & 15015 &15016&227&10.2&-0.85&	-0.76&	1.44&{\it N}\\
6&3114$^a$&00 43 19.5&-09 09 13& -& 19.47 & 0.390 & 8.76 & 15060 &15061&136&3.7&-0.46&	-0.37&	1.54&{\it N}\\
7&486&00 43 31.2&-09 51 48&S & 16.81 & 0.288 & 9.82 & 16619 &16615&230&7.2&-0.09&	0.05&	2.72&{\it N}\\
8&491&00 43 34.0&-08 50 37&S?$^*$& 17.05 & 0.297 & 9.94 & 14968 &14993&182&11.6&-0.56&	-0.44&	2.43&{\it N}\\
9&496&00 43 38.7&-09 31 21&S?$^*$ & 16.99 & 0.281 & 9.84 & 15004 &14993&182&4.3&0.04&	0.17&	1.94&{\it Pos}\\
10&502&00 43 43.9&-09 04 23&- & 17.30 & 0.340 & 9.73 & 15004 &15038&182&11.8&-0.33&	-0.19&	2.03&{\it N}\\
\hline 
\multicolumn{16}{l}{Column (1): Sequence ID number.}\\
\multicolumn{16}{l}{Column (2): [DFL98] galaxy name  \citep{Durret98}; ($^a$) [SDG98] galaxy name \citep{Slezak98}; 
($^b$) PGC names \citep{Paturel89}.}\\
\multicolumn{16}{l}{Columns (3) and (4): RA, Dec coordinates from NED. } \\
\multicolumn{16}{l}{Column (5): The morphological type from NED, except ($^*$) taken from HyperLEDA.}  \\
\multicolumn{16}{l}{Columns (6) and (7): The $g$ magnitude and ($g-r$) 
color index estimated in this work from CFHT/PanSTARRS images.}\\ 
\multicolumn{16}{l}{Column (8): Stellar mass given by log(M$_{*}$/\msolar).}\\ 
\multicolumn{16}{l}{Column (9): Optical velocity from our
redshift membership catalog (see Sect.\,\ref{obs}).}\\
%except ($^*$) taken from optical spectra (Lopez-Gutierrez, M. 2019, com. priv).}   \\
\multicolumn{16}{l}{Column (10): Heliocentric \hi\, 
velocity. The uncertainty is taken as one-half the 
velocity resolution, $i.e.$ \aprox23\,\kms\ (see Table\,\ref{tab_cubes}).}  \\
\multicolumn{16}{l}{Column (11): The galaxy velocity width, the uncertainty is \aprox46\,\kms\, (see Table\,\ref{tab_cubes}).}  \\
\multicolumn{16}{l}{Column (12): The \hi-Mass following \cite{Haynes-Giovanelli84}; uncertainties are lower than 10\%; see the text for more details.}  \\
\multicolumn{16}{l}{Columns (13) and (14): The upper and lower limits for \defhi, following \cite{Haynes-Giovanelli84}.}\\
\multicolumn{16}{l}{Column (15): The projected cluster-centric distance considering the position of the corresponding BCG.}  \\
\multicolumn{16}{l}{Column (16): Qualitative code indicating 
if the galaxy is normal in \hi\, (N), otherwise it is perturbed. See the text 
for more details.} \\
%{(N) normal, (deff) gas defficient,
%(assym) strong asymmetries,
%(pos-off) offset in position, (vel-off) offset in velocity.
%See text for more details.}
\end{tabular}
}%end resizebox
\end{table}
%\end{landscape}

\onecolumn
%\begin{landscape}
\begin{table}
    \centering
    \caption{The \hi-detected galaxies in Abell\,496}
    \label{tab_A496_det}
    \resizebox{\textwidth}{!}{%
    \begin{tabular}{clcccllrcccrcccc}
\hline 
ID &Name & RA & Dec & Morph. & $g$ & color & M$_{*}$ & Opt-vel  & \hi-vel & $\Delta v$ & \mhi & \multicolumn{2}{c}{Def$_{\mathrm {HI}}$} & Dist. & \hi \\
    & & (2000)   & (2000)   & type & mag & ($g-r$)& &\kms &  \kms & \kms      &$10^9$\msolar  &   Sa-Sb & Sc-Irr & Mpc & pert. \\
(1)  & (2)      & (3)      & (4)   & (5)  & (6)    & (7) & (8) & (9) &(10) & (11)& (12)&(13) & (14) & (15)&(16) \\
\hline
1&603&04 30 49.8&-13 12 28& -& 16.92 & 0.193 & 9.29 & 9786&9724 &88&1.7&-0.19	&	-0.11	&	1.63&{\it N}\\
2&672&04 31 03.4&-13 11 21& Sc& 17.69 & 0.302 & 8.99 & 10776 &10783 &177&2.3&-0.14	&	-0.04	&	1.51&{\it N}\\
3&B042910.53-130437.3&04 31 30.0&-12 58 15&-&18.83 & 0.272	&	8.44 &- &10407	&44&1.9&-1.00	&	-0.97	&	1.42&{\it N}\\
4&853&04 31 32.0&-13 14 34&- & 17.14 & 0.278	&	9.23 & 9208 &9264&219&2.6&-0.22	&	-0.12	&	1.22&{\it N}\\
5&903$^{***}$&04 31 40.3&-13 07 11&- & 17.48 & 0.455	&	9.18 &10904 &-%10959.5
&-&$>$2.3&-	&	-	&	1.19&\\
6&926$^{***}$&04 31 43.1&-13 07 10& -& 17.08 & 0.596	&	9.44 & 11106 &-%11092.4
&-&$>$1.0&-	&	-	&	1.16&\\
7&958&04 31 47.7&-12 44 23& -& 19.38 & 0.224	&	8.16 & -&10186	&132&0.8&-0.16	&	-0.10	&	1.64&{\it N}\\
8&982&04 31 50.3&-12 43 54& -&17.68 & 0.174	&	8.93 & 10919&10871	&89&1.0&-0.14	&	-0.06	&	1.64&{\it N}\\
9&989&04 31 51.8&-13 07 43&- & 18.96 & 0.288	&	8.39 &10006 &10275	&221&3.3&-0.38	&	-0.28	&	1.08&{\it Vel}\\
10&1054&04 32 02.8&-13 29 48& I& 16.73 & 0.464	&	9.53 & 10289 &9329	&175&1.5&-0.03	&	0.07	&	1.08&{\it Vel, Pos}\\
11&1099&04 32 09.6&-12 41 42& Sc& 15.70 & 0.584	&	10.07 &10323 &10319	&309&7.3&-0.07	&	0.08	&	1.60&{\it N}\\
12&1169&04 32 20.4&-13 14 11&Sb & 16.39 & 0.401	&	9.65 & 8990 &8979	&262&8.9&-0.52	&	-0.41	&	0.76&{\it N}\\
13&1179&04 32 21.8&-13 15 48&- & 17.55 & 0.371	&	9.10 &10057 &10076	&176&1.4&-0.15	&	-0.06	&	0.74&{\it N}\\
14&1275&04 32 41.8&-13 03 21& -& 19.10 & 0.349	&	8.36 &10608 &10473	&88&1.0&-0.17	&	-0.10	&	0.73&{\it Vel}\\
15&04.56-13.38$^{**}$&04 32 45.3&-13 23 08& -& 21.12& 0.296 &	7.39 & -& 10694	&88 &0.6&-0.11	&	-0.05	&	0.59&{\it N} \\
16&1303&04 32 45.5&-12 33 39&- & 19.68 & 0.399	&	8.12 &- &9264 &44&3.5&-1.20	&	-1.16	&	1.75&{\it N}\\
17&1315&04 32 48.1&-12 42 52&- & 16.36 & 0.191	&	9.55 &9954 &9615	&132&3.6&-0.17	&	-0.06	&	1.39&{\it Vel, Asy}\\
18&1324&04 32 49.0&-12 42 19&- &18.09 & 0.254	&	8.78 & 10037& 9615	&132&1.1&-0.16	&	-0.09	&	1.41&{\it Vel, Asy}\\
19&1338&04 32 51.1&-13 03 07&- & 18.92 & 0.320	&	8.43 & 10474 &10473	&88&1.1&-0.33	&	-0.27	&	0.68&{\it N}\\
20&1354&04 32 54.2&-13 23 50& -& 17.16 & 0.498	&	9.35 & 9932 &9922	&220&3.6&-0.62	&	-0.54	&	0.53&{\it N}\\
21&1363&04 32 54.7&-12 53 25& -& 18.11 & 0.250	&	8.77 & -&10385	&88&2.5&-0.49	&	-0.41	&	0.98&{\it N}\\
22&04.55-13.06$^{**}$&04 32 55.4&-13 03 33&- & 21.33& 0.484 & 7.40 &- &10429	&88&1.0&-0.78	&	-0.75	&	0.64&{\it N}\\
23&198$^c$&04 32 56.1&-13 36 38& Sc & 15.65 & 0.709	&	10.16 & 11349 &10561	&177&2.5&-0.16	&	-0.06	&	0.92&{\it Vel}\\
24&3760&04 32 56.8&-12 46 41& SBbc & 15.23 & 0.745	&	10.37 &10602 &10606	&354&5.3&0.14	&	0.29	&	1.22&{\it Asy}\\
25&1386&04 32 59.8&-13 42 33& -& 16.95 & 0.739	&	9.57 &10557 &10473	&265&3.1&-0.10	&	0.01	&	1.13&{\it Asy}\\
26&1389&04 33 00.1&-13 57 28& -& 19.65 & 0.535	&	8.21 & -&10186	&44&1.4&-0.71	&	-0.66	&	1.70&{\it N}\\
27&1399&04 33 02.0&-12 43 00& -& 15.79 & 0.457	&	9.96 & 9361&9373	&263&5.4&-0.59	&	-0.49	&	1.35&{\it N}\\
28&1418&04 33 05.3&-13 26 55& -& 18.42 & 0.472	&	8.75 &- &10473	&89&1.2&-0.12	&	-0.03	&	0.55&{\it N}\\
29&1419&04 33 05.9&-13 59 54& -& 18.84 & 0.860	&	8.75 & 10943&10971	&266&1.6&0.36	&	0.47	&	1.79&{\it Def}\\
30&1457&04 33 10.9&-13 14 10& -& 17.79 & 0.495	&	9.05 & 8991&8914	&131&1.8&-0.32	&	-0.24	&	0.27&{\it Pos}\\
31&1461&04 33 11.4&-13 28 02& -& 18.14 & 0.461	&	8.87 & 10670&10650	&177&2.8&-0.54	&	-0.46	&	0.55&{\it N}\\
32&1474&04 33 13.7&-13 27 56& -& 18.94 & 0.415	&	8.47 &10159 &10120	&88&1.1&-0.31	&	-0.25	&	0.54&{\it N}\\
33&1482&04 33 15.1&-13 21 19& -& 18.04 & 0.317	&	8.83 & 10207&10186	&132&0.8&0.03	&	0.11	&	0.31&{\it N}\\
34&225$^c$&04 33 17.4&-12 59 01&Sb & 15.61 & 0.674	&	10.16 & 10826 &10915	&266&0.6&0.35	&	0.44	&	0.70&{\it Def}\\
35&1512&04 33 19.6&-13 33 39&- & 19.20 & 0.358	&	8.32 & 10268&10253	&88&0.7&-0.17	&	-0.10	&	0.74&{\it N}\\
36&1547&04 33 23.3&-13 20 21& Sc & 16.89 & 0.504	&	9.48 & 10188 &10208	&265&3.8&-0.27	&	-0.17	&	0.23&{\it N}\\
37&1549&04 33 23.9&-13 31 03& S & 16.74 & 0.590	&	9.59 & 10228 &10231	&397&10.4&-0.75	&	-0.64	&	0.63&{\it N}\\
38&1565&04 33 25.4&-13 41 09& Sc & 16.75 & 0.524	&	9.55 & 10866 &10827	&354&3.1&-0.12	&	0.00	&	1.02&{\it Pos}\\
39&1584&04 33 29.0&-13 41 15& -& 20.40 & 0.273	&	7.72 &9443 &9527	&44&0.4&0.31	&	0.39	&	1.02&{\it Def}\\
40&1624&04 33 34.6&-12 38 34& -& 19.40 & 0.329	&	8.21 & -&10539	&133&1.4&-0.22	&	-0.14	&	1.48&{\it N}\\
41&1656$^{***}$&04 33 38.5&-13 35 46&I & 16.98 & 0.411	&	9.38 &11194  &-%11092.4
&-&$>$0.9&-	&	-	&	0.80&\\
42&1709&04 33 44.7&-13 33 50& I & 17.07 & 0.523	&	9.40 & 9779 &9724	&176&1.2&0.23	&	0.33	&	0.72&{\it Def, Pos}\\
43&1747$^{***}$&04 33 50.9&-13 44 18& -& 18.62 & 0.400	&	8.61 & 11037&-%11092.4
&-&$>$0.8&-	&	-	&	1.15&\\
44&1834&04 34 02.0&-13 17 49& Sbc& 15.89 & 0.470	&	9.92 & 10347 &10319	&132&2.3&-0.01	&	0.10	&	0.25&{\it N}\\
45&1849&04 34 04.1&-12 48 52 &- & 18.56 & 0.356	&	8.62 & 9002 &9023	&88&1.1&-0.02	&	0.07	&	1.10&{\it N}\\
46&1860$^{***}$&04 34 06.3&-13 39 44&- & 19.40 & 0.476	&	8.29 & 11267 &-%11092.4
&-&$>$1.8&-	&	-	&	1.00&\\
47&1967&04 34 23.9&-13 15 32&- & 18.16 & 0.405	&	8.83 & 9480 &9483	&132&0.7&0.08	&	0.15	&	0.45&{\it N}\\
48&2036&04 34 36.8&-12 45 30& -& 19.08 & 0.128	&	8.25 & 9936 &9944	&88&1.5&-0.49	&	-0.43	&	1.33&{\it N}\\
49&2246&04 35 10.5&-13 59 34& -& 18.01 & 0.685	&	9.03 & - &10451 &132&3.0&-0.56	&	-0.48	&	1.96&{\it N}\\
50&2252&04 35 11.0&-13 45 18& -& 17.97 & 0.446	&	8.94 & 10963 &11004	&87&2.9&-0.32	&	-0.23	&	1.49&{\it N}\\
51&B043258.99-130504.4&04 35 18.3&-12 58 58&- & 20.78 & 0.554	&	7.70 &- &10054	&44&1.3&-0.47	&	-0.41	&	1.18&{\it N}\\
52&2447&04 35 45.5&-13 58 37&- & 17.86 & 0.694	&	9.09 & - &10606	&265&4.3&-0.34	&	-0.24	&	2.11&{\it Asy}\\
53&2485&04 35 52.0&-13 19 47& Sab & 16.10 & 0.810	&	10.00 &10301 &10363	&309&1.4&0.43	&	0.55	&	1.31&{\it Def, Pos}\\
54&2503&04 35 54.5&-13 34 49& S0-a& 19.29 & 0.472	&	8.34 & - &10385	&88&0.8&-0.53	&	-0.49	&	1.53&{\it N}\\
55&2516&04 35 56.7&-13 19 49&- & 18.70 & 0.331	&	8.54 & 10609 &10584	&133&1.0&-0.11	&	-0.03	&	1.36&{\it N}\\
56&2525&04 35 57.8&-13 43 56& -& 19.65 & 0.604	&	8.25 &- &10385	&177&1.3&-0.15	&	-0.07	&	1.76&{\it N}\\
57&2534&04 35 59.5&-13 18 07& Sab& 16.06 & 0.830	&	10.03 & 10188 &10098	&220&0.9&0.55	&	0.67	&	1.38&{\it Def}\\
58&04.60-13.55$^{**}$&04 36 14.2&-13 33 20& -& - & - & - & - &10960	&44&0.8&-0.76	&	-0.73	&	1.66&{\it N}\\
\hline 
\multicolumn{16}{l}{Column (1): Sequence ID number.}\\
\multicolumn{16}{l}{Column (2): [SDG99]-SRC galaxy name \citep{Slezak99}; ($^c$) [DFL99] galaxy name \citep{Durret99}; ($^{**}$) The names were constructed using RA, Dec in hours and decimal degrees,}\\
\multicolumn{16}{l}{they are objects not cataloged in NED, and the coordinates shown for these objects are those of the HI emission.  ($^{***}$) The emission extends 
%}\\
%\multicolumn{16}{l}{
beyond the observed range.}\\
\multicolumn{16}{l}{Columns (3) and (4): RA, Dec coordinates from NED. } \\
\multicolumn{16}{l}{Column (5): The morphological type from NED.}  \\
\multicolumn{16}{l}{Columns (6) and (7): The $g$ magnitude and ($g-r$) 
color index estimated in this work from CFHT/PanSTARRS images.}\\ 
\multicolumn{16}{l}{Column (8): Stellar mass given by log(M$_{*}$/\msolar)}\\ 
\multicolumn{16}{l}{Column (9): Optical velocity from our
redshift membership catalog (see Sect.\,\ref{obs}).}   \\
\multicolumn{16}{l}{Column (10): Heliocentric \hi\, 
velocity. The uncertainty in taken as one-half the 
velocity resolution, $i.e.$ \aprox22\, \kms\, (see Table\,\ref{tab_cubes}).}  \\
\multicolumn{16}{l}{Column (11): The galaxy velocity width, the uncertainty is \aprox44\,\kms (see Table\,\ref{tab_cubes}).}  \\
\multicolumn{16}{l}{Column (12): The \hi-Mass following \cite{Haynes-Giovanelli84}; uncertainties are lower than 10\%; see the text for more details.}  \\
\multicolumn{16}{l}{Columns (13) and (14): The upper and lower limits for \defhi, following \cite{Haynes-Giovanelli84}.}\\
\multicolumn{16}{l}{Column (15): The projected cluster-centric distance considering the position of the corresponding BCG.}  \\
\multicolumn{16}{l}{Column (16): Qualitative code indicating 
the \hi\, appearance, either normal or perturbed. See the text 
for more details.} \\
\end{tabular}
}%end resizebox
\end{table}
%\end{landscape}

%\begin{landscape}
 \begin{table}
    \centering
    \caption{The \hi-detected galaxies in Abell\,2670 }
    \label{tab_A2670_det}
     \resizebox{\textwidth}{!}{%
    \begin{tabular}{clcccllrcccrcccc}
    \hline 
ID &Name & RA & Dec & Morph. & $g$ & color  & M$_{*}$ & Opt-vel  & \hi-vel & $\Delta v$ & \mhi & \multicolumn{2}{c}{Def$_{\mathrm {HI}}$} & Dist. & \hi \\
    & & (2000)   & (2000)   & type & mag & ($g-r$) & &\kms &  \kms & \kms      &$10^9$\msolar  &   Sa-Sb & Sc-Irr & Mpc & pert. \\
(1)  & (2)      & (3)      & (4)   & (5)  & (6)    & (7) & (8) & (9) &(10) & (11)& (12)&(13) & (14) & (15) & (16)\\
\hline
1&J235222.31-104134.7&23 52 22.3&-10 41 35&- &19.03 & 0.364	&	9.28 & 22814 &22666	&	238	&	5.4	&-0.47	&	-0.36	&	2.81&{\it Vel, Pos}\\
2&J235232.39-104154.1&23 52 32.4&-10 41 54 &- &19.35 & 0.249	&	9.07 & - &22595	&	95	&	4.4	&-0.57	&	-0.48	&	2.65&{\it N}\\
3&J235233.13-103640.8&23 52 33.1&-10 36 41 & -& 18.34 & 0.388	&	9.62 & 22724 &22690	&	191	&	4.8	&-0.35	&	-0.24	&	2.40&{\it Pos}\\
4&J235247.36-105259.1&23 52 47.3&-10 52 59 & E?$^*$& 16.78 & 0.640	&	10.51 & 22721 &22714	&	143	&	6.6	&0.19	&	0.35	&	3.08&{\it Def, Pos}\\
5&3325027$^b$&23 52 55.8&-10 42 04&- & 19.16 & 0.354	&	9.21 & 21795 &21929	&	95	&	2.6	&-0.20	&	-0.10	&	2.24&{\it N}\\
6&J235303.74-110454.9&23 53 03.7&-11 04 55& S?$^*$& 16.30 & 0.679	&	10.76 & 22823 &22738	&	191	&	7.0	&	-0.01	&	0.14	&	3.82&{\it N}\\
7&J235305.67-104055.6&23 53 05.7&-10 40 56& Sbc & 16.51 & 0.434	&	10.49 & 21930 &21953	&	427	&	14.4	&-0.31	&	-0.17	&	2.02&{\it N}\\
8&J235308.87-104417.3&23 53 08.9&-10 44 17& E? & 17.96 & 0.802	&	10.06 & 22578 &22547	&	667	&	25.6	&-0.60	&	-0.46	&	2.19&{\it N}\\
9&J235310.49-104142.3&23 53 10.5&-10 41 42&- & 20.00 & 0.267	&	8.78 &-  &21858	&	142	&	3.6	&-0.79	&	-0.72	&	2.00&{\it N}\\
10&J235310.87-110203.0&23 53 10.9&-11 02 03& S?$^*$& 17.33 & 0.684	&	10.29 & 22806 &22785	&	96	&	3.4	&	0.12	&	0.25	&	3.52&{\it Def}\\
11&J235319.07-102303.2&23 53 19.1&-10 23 03& Sc& 17.07 & 0.614	&	10.36 & 22300 &22305	&	236	&	4.2	&0.13	&	0.27	&	1.19&{\it Def}\\
12&J235320.90-101047.7&23 53 20.9&-10 10 48& -& 18.67 & 0.352	&	9.44 & 22306 &22423	&	283	&	2.7	&-0.06	&	0.05	&	1.70&{\it N}\\
13&2dFGRS S888Z177&23 53 20.9&-10 32 39&- & 19.14 & 0.445	&	9.28 & 23114 &23298	&	142	&	3.0	&-0.29	&	-0.20	&	1.31&{\it Vel}\\
14&J235324.23-104916.6&23 53 24.2&-10 49 17&- & 17.70 & 0.330	&	9.88 & 21947 &21976	&	190	&	7.1	&-0.67	&	-0.58	&	2.38&{\it Pos, Asy}\\
15&J235326.38-101549.0&23 53 26.4&-10 15 49&E?$^*$ & 18.22 & 0.603	&	9.82 & 22410 &22399	&	425	&	12.6	&-0.11	&	0.04	&	1.31&{\it N}\\
16&J235338.22-105443.9&23 53 38.2&-10 54 44& S?$^*$& 17.47 & 0.589	&	10.16 & 22484 &22476	&	143	&	6.7	&-0.22	&	-0.09	&	2.71&{\it Pos}\\
17&J235339.45-102430.1&23 53 39.4&-10 24 30&- &20.11 & 0.302	&	8.74 & -&22588	&	142	&	2.6	&-0.44	&	-0.35	&	0.72&{\it N}\\
18&J235339.87-102547.6&23 53 39.9&-10 25 48& Sa & 17.26 & 0.552	&	10.23 &22552  &22612	&	189	&	2.6	&0.16	&	0.28	&	0.73&{\it Def, Pos}\\
19&J235341.02-103453.9&23 53 41.0&-10 34 54&- & 19.93 & 0.375	&	8.87 & - &21482	&	188	&	3.2	&-0.57	&	-0.49	&	1.11&{\it N}\\
20&J235344.08-104041.2&23 53 44.1&-10 40 41&- & 20.54 & 0.544	&	8.70 & - &23575	&	48	&	2.1	&-0.51	&	-0.45	&	1.51&{\it N}\\
21&J235346.66-101614.8&23 53 46.7&-10 16 15& Sab& 16.31 &0.629	&	10.73 & 23796 &23964	&	143	&	3.7	&0.36	&	0.52	&	0.97&{\it Def, Vel, Pos}\\
22&J235356.57-101510.4&23 53 56.6&-10 15 10& Sd & 17.07 & 0.495	&	10.28 & 21565 & 21529	&	94	&	2.2	&-0.51	&	-0.44	&	0.94&{\it N}\\
23&141225$^b$&23 53 57.1&-10 22 29&Sa & 17.31 & 0.509	&	10.17 & 21898 &21834	&	423	&	10.6	&-0.46	&	-0.33	&	0.43&{\it N}\\
24&141233$^b$&23 53 58.5&-10 28 27& -& 19.26 & 0.360 &	9.17 & 22168 &22140	&	283	&	5.0	&-0.56	&	-0.47	&	0.44&{\it N}\\
25&J235411.73-102115.1&23 54 11.7&-10 21 15&- & 19.26 & 0.375	&	9.18 &24268 &24321	&	95	&	2.3	&-0.30	&	-0.21	&	0.35&{\it N}\\
26&J235413.59-102748.2&23 54 13.6&-10 27 48& E? & 17.79 & 0.563	&	9.99 & 24568 &24536	&	143	&	1.7	&0.03	&	0.14	&	0.24&{\it N}\\
27&J235418.24-101349.5&23 54 18.2&-10 13 50& Sc & 17.43 & 0.580	&	10.17 & 23251 &23227	&	474	&	11.0	&0.08	&	0.25	&	1.00&{\it Def}\\
28&J235419.51-103300.3&23 54 19.5&-10 33 00& Sa & 16.77 & 0.601	&	10.49 & 23925 &23964	&	429	&	11.5	&-0.25	&	-0.11	&	0.71&{\it Asy, Pos}\\
29&J235435.77-095748.6&23 54 35.8&-09 57 49& S?$^*$& 17.94 & 0.370	&	9.79 &23442 &23407	&	144	&	9.2	&-0.30	&	-0.17	&	2.46&{\it N}\\
30&J235438.15-101907.6&23 54 38.1&-10 19 08& Sab & 16.76 & 0.523	&	10.44 & 22506 &22517	&	378	&	10.0	&-0.19	&	-0.05	&	0.75&{\it Asy}\\
31&J235444.17-101655.9&23 54 44.2&-10 16 56& Scd & 17.00 & 0.426	&	10.26 & 23184 &23120	&	383	&	7.3	&-0.01	&	0.14	&	0.98&{\it Pos}\\
32&J235446.43-095752.9&23 54 46.4&-09 57 53& - & 17.73 & 0.780	&	10.16 & 22768 &22666	&	334	&	4.6	&-0.05	&	0.07	&	2.50&{\it N}\\
33&J235449.71-100151.0&23 54 49.7&-10 01 51& - & 18.43 & 0.466	&	9.63 & - &21857	&	237	&	2.7	&0.36	&	0.50	&	2.20&{\it Def, Pos}\\
34&J235451.55-101238.9&23 54 51.5&-10 12 39& - & 19.13 & 0.253	&	9.17 & 23284 &23216	&	191	&	3.2	&-0.27	&	-0.17	&	1.37&{\it N}\\
35&J235453.62-101931.6&23 54 53.6&-10 19 32& Im & 18.13 & 0.505	&	9.79 & 21818 &21787	&	235	&	4.5	&-0.24	&	-0.12	&	0.99&{\it Pos}\\
36&J235454.17-101717.0&23 54 54.2&-10 17 17& Sc& 15.83 & 0.544	&	10.89 & 23065 &23168	&	478	&	10.9	&-0.01	&	0.15	&	1.12&{\it Asy}\\
37&J235607.62-095938.1&23 56 07.6&-09 59 38& E$^*$& 17.00 & 0.670	&	10.43 & 23035 &23168	&	96	&	2.8	&0.31	&	0.44	&	3.34&{\it Def, Pos}\\
38&J235619.93-101246.2&23 56 19.9&-10 12 46& S?$^*$& 16.23 & 0.507	&	10.67 & 22923 &22953	&	430	&	10.2	&-0.20	&	-0.06	&	2.95&{\it N}\\
\hline 
\multicolumn{16}{l}{Column (1): Sequence ID number.}\\
\multicolumn{16}{l}{Columns (2): SDSS names; ($^b$) PGC names \citep{Paturel89}.  } \\
\multicolumn{16}{l}{Columns (3) and (4): RA, Dec coordinates from NED. } \\
\multicolumn{16}{l}{Column (5): The morphological type from NED, except ($^*$) taken from HyperLEDA.}  \\
\multicolumn{16}{l}{Columns (6) and (7): The $g$ magnitude and ($g-r$) 
color index estimated in this work from CFHT/PanSTARRS images.}\\ 
\multicolumn{16}{l}{Column (8): Stellar mass given by log(M$_{*}$/\msolar).}\\
\multicolumn{16}{l}{Column (9): Optical velocity from our
redshift membership catalog (see Sect.\,\ref{obs}).}   \\
\multicolumn{16}{l}{Column (10): Heliocentric \hi\, 
velocity. The uncertainty in taken as one-half the 
velocity resolution, $i.e.$ \aprox24 \kms\, (see Table\,\ref{tab_cubes}).}  \\
\multicolumn{16}{l}{Column (11): The galaxy velocity, the uncertainty is \aprox48\,\kms (see Table\,\ref{tab_cubes}).}  \\
\multicolumn{16}{l}{Column (12): The \hi-Mass following \cite{Haynes-Giovanelli84}; uncertainties are lower than 10\%; see the text for more details.}  \\
\multicolumn{16}{l}{Columns (13) and (14): The upper and lower limits for \defhi, following \cite{Haynes-Giovanelli84}.}\\
\multicolumn{16}{l}{Column (15): The projected cluster-centric distance considering the position of the corresponding BCG. }  \\
\multicolumn{16}{l}{Column (16): Qualitative code indicating 
the \hi\, appearance, either normal or perturbed. See the text 
for more details.} 
\end{tabular}
}%end resizebox
\end{table}
%\end{landscape}

%\begin{landscape}
% \begin{table}
%    \centering
%    \contcaption{}
%    %The \hi-detected galaxies in Abell\,2670 
    %\label{tab_A2670_det}
%    \begin{tabular}{clccccccccrcccc}
%    \hline 
%\end{tabular}
%\end{table}
%\end{landscape}

%\newpage

%\appendix
%\onecolumn

\section{Spirals not detected in HI, A85, A496, A2670}
\label{app_hi_no_det}

\onecolumn

\begin{table*}
    \centering
    \caption{Catalog of bright spirals (B-Sp) not detected in \hi\ in Abell\,85}
    \label{tab_A85_no_det}
    %\resizebox{\textwidth}{!}{%
    \begin{tabular}{clccccllrc}
    \hline 
ID & Name & RA & Dec & Morph. &  Opt-vel  & $g$ & color & M$_{*}$& Dist.  \\
 & & (2000)   & (2000)   & type  & \kms  &   mag &($g-r$)  & & Mpc   \\
(1)  & (2)      & (3)      & (4)   & (5)   & (6)  & (7)  &(8) & (9) & (10)\\
\hline
1	&	044	&	00 39 47.6	&	-09 48 11	&	-	&	15848	&	18.34	&	0.493	&	9.34	&	2.77	\\
2	&	046	&	00 39 48.4	&	-09 14 37	&	S?$^*$	&	17338	&	17.78	&	0.596	&	9.67	&	1.97	\\
3	&	052	&	00 39 59.1	&	-09 26 04	&	S?$^*$	&	16939	&	18.07	&	0.434	&	9.44	&	1.85	\\
4	&	920$^a$	&	00 40 01.9	&	-10 06 05	&	-	&	17458	&	18.53	&	0.433	&	9.22	&	3.56	\\
5	&	93148$^b$	&	00 40 13.6	&	-08 58 01	&	-	&	17935	&	18.10	&	0.311	&	9.34	&	2.03	\\
6	&	070	&	00 40 18.7	&	-08 52 57	&	S0/a	&	17923	&	15.99	&	0.695	&	10.56	&	2.20	\\
7	&	087	&	00 40 40.9	&	-08 48 59	&	-	&	15652	&	18.15	&	0.393	&	9.37	&	2.20	\\
8	&	099	&	00 40 51.6	&	-09 20 28	&	-	&	16506	&	18.82	&	0.536	&	9.15	&	0.95	\\
9	&	127	&	00 41 10.4	&	-09 15 01	&	-	&	18254	&	18.45	&	0.480	&	9.29	&	0.67	\\
10	&	128	&	00 41 11.1	&	-09 55 30	&	S?$^*$	&	17000	&	18.40	&	0.556	&	9.36	&	2.50	\\
11	&	133	&	00 41 12.8	&	-09 32 04	&	S?$^*$	&	17139	&	16.40	&	0.584	&	10.30	&	1.08	\\
12	&	139	&	00 41 14.2	&	-08 55 54	&	S?$^*$	&	15157	&	17.83	&	0.324	&	9.48	&	1.56	\\
13	&	145	&	00 41 19.0	&	-09 23 24	&	S?$^*$	&	14935	&	17.94	&	0.442	&	9.50	&	0.61	\\
14	&	151	&	00 41 21.3	&	-08 43 30	&	S?$^*$	&	16050	&	18.02	&	0.524	&	9.51	&	2.30	\\
15	&	167	&	00 41 27.1	&	-09 13 43	&	S?$^*$	&	14167	&	16.46	&	0.517	&	10.23	&	0.47	\\
16	&	170	&	00 41 27.9	&	-09 23 30	&	S?$^*$	&	14836	&	18.18	&	0.555	&	9.46	&	0.50	\\
17	&	1645$^a$	&	00 41 27.9	&	-09 13 47	&	-	&	16258	&	17.36	&	0.555	&	9.84	&	0.46	\\
18	&	201	&	00 41 36.2	&	-08 59 35	&	E?$^*$	&	17935	&	16.89	&	0.338	&	9.92	&	1.23	\\
19	&	218	&	00 41 42.5	&	-09 21 26	&	-	&	14737	&	18.56	&	0.576	&	9.29	&	0.25	\\
20	&	1875$^a$	&	00 41 44.4	&	-09 26 25	&	-	&	17457	&	18.16	&	0.527	&	9.45	&	0.54	\\
21	&	226	&	00 41 45.5	&	-09 40 33	&	S?$^*$	&	17072	&	17.61	&	0.487	&	9.68	&	1.45	\\
22	&	1940$^a$	&	00 41 49.0	&	-09 57 05	&	S?$^*$	&	15537	&	16.58	&	0.486	&	10.16	&	2.52	\\
23	&	246	&	00 41 50.8	&	-09 13 46	&	-	&	18178	&	18.37	&	0.321	&	9.23	&	0.29	\\
24	&	255	&	00 41 53.3	&	-09 29 31	&	E?$^*$	&	15732	&	16.38	&	0.342	&	10.16	&	0.74	\\
25	&	256	&	00 41 53.4	&	-09 38 46	&	-	&	15924	&	18.81	&	0.397	&	9.07	&	1.34	\\
26	&	267	&	00 41 57.4	&	-09 35 24	&	S?$^*$	&	15918	&	17.22	&	0.551	&	9.90	&	1.12	\\
27	&	273	&	00 41 59.8	&	-09 42 32	&	-	&	15506	&	17.98	&	0.375	&	9.44	&	1.59	\\
28	&	2146$^a$	&	00 41 59.9	&	-09 39 10	&	S0	&	15780	&	16.42	&	0.455	&	10.21	&	1.37	\\
29	&	276	&	00 42 00.7	&	-09 50 04	&	S 	&	15627	&	16.21	&	0.679	&	10.45	&	2.07	\\
30	&	286	&	00 42 05.0	&	-09 32 04	&	Sc	&	15852	&	15.93	&	0.532	&	10.49	&	0.93	\\
31	&	292	&	00 42 06.4	&	-09 33 35	&	-	&	16097	&	18.71	&	0.460	&	9.15	&	1.03	\\
32	&	2260$^a$	&	00 42 08.4	&	-09 31 05	&	-	&	16963	&	18.69	&	0.570	&	9.23	&	0.88	\\
33	&	2276$^a$	&	00 42 09.8	&	-09 28 52	&	-	&	16739	&	18.19	&	0.273	&	9.28	&	0.76	\\
34	&	304	&	00 42 11.7	&	-09 58 14	&	S?$^*$	&	16191	&	18.30	&	0.515	&	9.38	&	2.62	\\
35	&	316	&	00 42 17.3	&	-09 28 48	&	S?$^*$	&	15855	&	17.14	&	0.535	&	9.93	&	0.81	\\
36	&	321	&	00 42 18.5	&	-09 39 12	&	-	&	15449	&	17.91	&	0.311	&	9.43	&	1.43	\\
37	&	325	&	00 42 19.9	&	-09 25 28	&	S?$^*$	&	15341	&	18.01	&	0.501	&	9.51	&	0.67	\\
38	&	338	&	00 42 24.2	&	-09 16 17	&	S?$^*$	&	18195	&	17.02	&	0.576	&	10.01	&	0.55	\\
39	&	345	&	00 42 28.4	&	-09 49 38	&	S?$^*$	&	15006	&	18.42	&	0.557	&	9.35	&	2.13	\\
40	&	355	&	00 42 32.9	&	-09 21 44	&	-	&	17064	&	18.94	&	0.356	&	8.98	&	0.72	\\
41	&	356	&	00 42 33.8	&	-08 52 53	&	Sc	&	17722	&	16.79	&	0.697	&	10.19	&	1.78	\\
42	&	358	&	00 42 33.9	&	-09 08 46	&	S?$^*$	&	16677	&	17.16	&	0.600	&	9.96	&	0.92	\\
43	&	362	&	00 42 34.7	&	-09 04 36	&	S?$^*$	&	14745	&	17.48	&	0.451	&	9.72	&	1.13	\\
44	&	366	&	00 42 37.0	&	-09 45 20	&	-	&	17065	&	18,64	&	0.559	&	9.25	&	1.91	\\
45	&	382	&	00 42 43.9	&	-09 44 21	&	Sb	&	15231	&	17.02	&	0.660	&	10.06	&	1.90	\\
46	&	391	&	00 42 48.4	&	-09 34 41	&	S?$^*$	&	17940	&	17.63	&	0.500	&	9.68	&	1.41	\\
47	&	413	&	00 42 58.7	&	-08 55 06	&	-	&	17081	&	18.57	&	0.520	&	9.26	&	1.85	\\
48	&	429	&	00 43 04.2	&	-09 32 43	&	Sc	&	15812	&	17.25	&	0.618	&	9.93	&	1.51	\\
49	&	435	&	00 43 06.0	&	-09 50 15	&	Sb	&	14742	&	16.66	&	0.650	&	10.22	&	2.40	\\
50	&	439	&	00 43 08.2	&	-09 49 37	&	-	&	15203	&	17.39	&	0.690	&	9.91	&	2.39	\\
51	&	2972$^a$	&	00 43 08.1	&	-09 34 44	&	-	&	15024	&	18.13	&	0.403	&	9.39	&	1.64	\\
52	&	2997$^a$	&	00 43 09.8	&	-10 04 43	&	-	&	15960	&	18.09	&	0.395	&	9.40	&	3.27	\\
53	&	449	&	00 43 11.0	&	-09 18 00	&	-	&	15945	&	18.94	&	0.511	&	9.08	&	1.29	\\
54	&	451	&	00 43 11.6	&	-09 38 16	&	Se	&	16253	&	15.84	&	0.586	&	10.57	&	1.84	\\
55	&	483	&	00 43 30.5	&	-09 43 59	&	S?$^*$	&	15125	&	16.80	&	0.516	&	10.08	&	2.31	\\
56	&	514	&	00 44 04.1	&	-09 41 06	&	-	&	14934	&	18.83	&	0.290	&	8.99	&	2.60	\\
 \hline
\multicolumn{10}{l}{Column (1): Sequence ID number.}\\
\multicolumn{10}{l}{Column (2): [DFL98] galaxy name \citep{Durret98}; ($^a$) [SDG98] galaxy name \citep{Slezak98};}\\
\multicolumn{10}{l}{($^b$) PGC names \citep{Paturel89}.  }\\
\multicolumn{10}{l}{Columns (3) and (4): RA, Dec coordinates from NED.}\\
\multicolumn{10}{l}{Column (5): The morphological type from NED, except ($^*$) taken from HyperLEDA. }\\
\multicolumn{10}{l}{Column (6): Optical velocity from our
redshift membership catalog (see Sect.\,\ref{obs}).}\\
\multicolumn{10}{l}{Columns (7) and (8): The $g$ magnitude and ($g-r$) 
color index estimated in this work from}\\
\multicolumn{10}{l}{CFHT/PanSTARRS images.}\\ 
\multicolumn{10}{l}{Column (9): Stellar mass given by log(M$_{*}$/\msolar).}\\
\multicolumn{10}{l}{Column (10): The cluster-centric distance from the BCG position.}\\
\end{tabular}
%}%end resizebox
\end{table*}

\begin{table*}
   % \centering
    \caption{Catalog of bright spirals (B-Sp) not detected in 
    \hi\ in Abell\,496}
    \label{tab_A496_no_det}
    \begin{tabular}{clccccllrc}
    \hline 
ID & Name & RA & Dec & Morph. &  Opt-vel  & $g$ & color & M$_{*}$& Dist.  \\
 & & (2000)   & (2000)   & type  & \kms  &   mag &($g-r$)  &  & Mpc   \\
(1)  & (2)      & (3)      & (4)   & (5)   & (6)  & (7)  &(8) & (9) & (10)\\
\hline
1	&	576	&	04 30 45.1	&	-12 37 39	&	S0-a$^*$	&	9325	&	15.04	&	0.793	&	10.47	&	2.26	\\
2	&	671	&	04 31 03.1	&	-12 58 26	&	SBb$^*$	&	10524	&	14.40	&	0.438	&	10.60	&	1.65	\\
3	&	718	&	04 31 10.7	&	-12 39 20	&	-	&	9115	&	16.51	&	0.588	&	9.69	&	2.04	\\
4	&	799	&	04 31 23.7	&	-13 04 13	&	E?$^*$	&	9587	&	15.80	&	0.508	&	9.98	&	1.38	\\
5	&	924	&	04 31 42.9	&	-12 32 38	&	-	&	9519	&	17.75	&	0.568	&	9.12	&	2.05	\\
6	&	B042925.77-124531.4	&	04 31 45.6	&	-12 39 10	&	-	&	8656	&	17.70	&	0.119	&	8.93	&	1.82	\\
7	&	1080	&	04 32 06.6	&	-13 04 59	&	Sc	&	10840	&	17.12	&	0.737	&	9.49	&	0.98	\\
8	&	1193	&	04 32 24.0	&	-12 46 48	&	SBa	&	10140	&	15.89	&	0.735	&	10.06	&	1.36	\\
9	&	1216	&	04 32 27.7	&	-12 38 43	&	-	&	10339	&	16.46	&	0.620	&	9.70	&	1.62	\\
10	&	1443	&	04 33 08.9	&	-13 02 36	&	Sbc	&	10197	&	17.18	&	0.651	&	9.42	&	0.59	\\
11	&	1520	&	04 33 20.2	&	-13 26 22	&	SBb	&	9937	&	16.59	&	0.707	&	9.72	&	0.46	\\
12	&	1660	&	04 33 38.7	&	-12 32 04	&	E?$^*$	&	9812	&	15.20	&	0.519	&	10.27	&	1.74	\\
13	&	2193	&	04 35 03.0	&	-12 53 51	&	SBab	&	9591	&	15.90	&	0.734	&	10.05	&	1.20	\\
14	&	2195	&	04 35 03.7	&	-13 39 20	&	Sa	&	9577	&	16.97	&	0.893	&	9.62	&	1.26	\\
15	&	2210	&	04 35 05.9	&	-13 10 54	&	-	&	10230	&	16.86	&	0.409	&	9.46	&	0.88	\\
16	&	2253	&	04 35 11.1	&	-13 14 40	&	SBab	&	10575	&	15.27	&	0.744	&	10.35	&	0.91	\\
17	&	2374	&	04 35 32.4	&	-13 33 23	&	SBaa	&	10121	&	16.46	&	0.928	&	9.89	&	1.31	\\
\hline 
\multicolumn{10}{l}{Column (1): Sequence ID number.}\\
\multicolumn{10}{l}{Column (2): [SDG99]-SRC galaxy name \citep{Slezak99}.}\\
\multicolumn{10}{l}{Columns (3) and (4): RA, Dec coordinates from NED.}\\
\multicolumn{10}{l}{Column (5): The morphological type from NED, except ($^*$) taken from HyperLEDA. }\\
\multicolumn{10}{l}{Column (6): Optical velocity from our
redshift membership catalog (see Sect.\,\ref{obs}).}\\
\multicolumn{10}{l}{Columns (7) and (8): The $g$ magnitude and ($g-r$) 
color index estimated in this work
%}\\
%\multicolumn{10}{l}{
from CFHT/PanSTARRS images.}\\ 
\multicolumn{10}{l}{Column (9): Stellar mass given by log(M$_{*}$/\msolar).}\\
\multicolumn{10}{l}{Column (10): The cluster-centric distance from the BCG position.}  \\
\end{tabular}
\end{table*}

\begin{table*}
    \centering
    \caption{Catalog of bright spirals (B-Sp) not detected in 
    \hi\ in Abell\,2670}
    \label{tab_A2670_no_det}
    \begin{tabular}{clccccllrc}
    \hline 
ID & Name & RA & Dec & Morph. &  Opt-vel  & $g$ & color & M$_{*}$ & Dist.  \\
 & & (2000)   & (2000)   & type  & \kms  &   mag &($g-r$)  &  &Mpc   \\
(1)  & (2)      & (3)      & (4)   & (5)   & (6)  & (7)  &(8) & (9) &(10)\\
\hline
1	&	J235258.26-105430.3	&	23 52 58.3	&	-10 54 30	&	S?$^*$	&	23588	&	18.08	&	0.493	&	9.81	&	3.06	\\
2	&	J235258.58-104111.7	&	23 52 58.6	&	-10 41 12	&	S?$^*$	&	21817	&	16.70	&	0.627	&	10.54	&	2.16	\\
3	&	J235303.21-101748.0	&	23 53 03.2	&	-10 17 48	&	Sbc	&	22296	&	17.53	&	0.522	&	10.08	&	1.66	\\
4	&	J235311.63-101835.4	&	23 53 11.6	&	-10 18 35	&	S?$^*$	&	21754	&	17.98	&	0.539	&	9.89	&	1.46	\\
5	&	J235326.80-102442.1	&	23 53 26.8	&	-10 24 42	&	S?$^*$	&	21684	&	17.86	&	0.360	&	9.82	&	1.02	\\
6	&	J235344.11-102200.7	&	23 53 44.1	&	-10 22 01	&	-	&	24283	&	18.92	&	0.101	&	9.20	&	0.70	\\
7	&	J235356.81-100930.2	&	23 53 56.8	&	-10 09 30	&	-	&	23204	&	19.05	&	0.387	&	9.29	&	1.43	\\
8	&	J235357.29-102550.0	&	23 53 57.3	&	-10 25 50	&	-	&	22168	&	19.29	&	0.319	&	9.13	&	0.36	\\
9	&	J235358.85-104130.7	&	23 53 58.9	&	-10 41 31	&	S?$^*$	&	22207	&	17.51	&	0.334	&	9.97	&	1.48	\\
10	&	J235405.85-102624.5	&	23 54 05.9	&	-10 26 25	&	-	&	21388	&	19.44	&	0.513	&	9.19	&	0.20	\\
11	&	141266$^b$	&	23 54 08.9	&	-10 27 49	&	-	&	22108	&	18.97	&	0.564	&	9.45	&	0.26	\\
12	&	J235414.84-102448.8	&	23 54 14.8	&	-10 24 49	&	-	&	21178	&	17.31	&	0.728	&	10.32	&	0.04	\\
13	&	J235416.65-101454.6	&	23 54 16.7	&	-10 14 55	&	Sd	&	22630	&	16.42	&	0.598	&	10.65	&	0.90	\\
14	&	J235420.58-100813.5	&	23 54 20.6	&	-10 08 14	&	-	&	24458	&	19.06	&	0.332	&	9.25	&	1.50	\\
15	&	J235422.17-102022.4	&	23 54 22.2	&	-10 20 22	&	-	&	21208	&	19.26	&	0.458	&	9.23	&	0.46	\\
16	&	J235425.88-095923.7	&	23 54 25.9	&	-09 59 24	&	E?$^*$	&	22779	&	17.93	&	0.639	&	9.98	&	2.29	\\
17	&	J235431.44-100210.2	&	23 54 31.5	&	-10 02 10	&	-	&	22283	&	19.01	&	0.379	&	9.30	&	2.06	\\
18	&	J235434.57-095415.5	&	23 54 34.6	&	-09 54 16	&	-	&	23284	&	19.22	&	0.484	&	9.27	&	2.76	\\
19	&	J235436.64-095802.9	&	23 54 36.7	&	-09 58 03	&	-	&	22294	&	18.87	&	0.455	&	9.42	&	2.44	\\
20	&	J235436.83-101758.1	&	23 54 36.8	&	-10 17 58	&	S0/a	&	23244	&	18.04	&	0.508	&	9.84	&	0.81	\\
21	&	J235439.40-100006.0	&	23 54 39.4	&	-10 00 06	&	-	&	21916	&	18.22	&	0.457	&	9.72	&	2.28	\\
22	&	J235445.81-101445.5	&	23 54 45.8	&	-10 14 46	&	Sd	&	23371	&	17.52	&	0.449	&	10.04	&	1.15	\\
23	&	J235506.02-103351.3	&	23 55 06.0	&	-10 33 51	&	-	&	23359	&	18.23	&	0.391	&	9.67	&	1.37	\\
24	&	J235509.50-100353.1	&	23 55 09.5	&	-10 03 53	&	-	&	21885	&	18.19	&	0.434	&	9.72	&	2.23	\\
25	&	J235514.23-103611.3	&	23 55 14.2	&	-10 36 11	&	-	&	23375	&	19.09	&	0.393	&	9.27	&	1.63	\\
26	&	J235514.70-102947.5	&	23 55 14.7	&	-10 29 48	&	S0/a	&	24533	&	17.76	&	0.645	&	10.06	&	1.39	\\
27	&	J235517.83-095359.8	&	23 55 17.8	&	-09 53 60	&	-	&	22762	&	18.57	&	0.453	&	9.55	&	3.08	\\
28	&	J235526.62-100843.4	&	23 55 26.6	&	-10 08 43	&	-	&	21724	&	18.75	&	0.371	&	9.41	&	2.14	\\
\hline 
\multicolumn{10}{l}{Column (1): Sequence ID number.} \\
\multicolumn{10}{l}{Column (2): SDSS names; ($^b$) PGC names \citep{Paturel89}.}\\
\multicolumn{10}{l}{Columns (3) and (4): RA, Dec coordinates from NED.}\\
\multicolumn{10}{l}{Column (5): The morphological type from NED, except ($^*$) taken from HyperLEDA. }\\
\multicolumn{10}{l}{Column (6): Optical velocity from our
redshift membership catalog (see Sect.\,\ref{obs}).}\\
\multicolumn{10}{l}{Columns (7) and (8): The $g$ magnitude and ($g-r$) 
color index estimated in this work
%}\\
%\multicolumn{10}{l}{
from CFHT/PanSTARRS images.}\\ 
\multicolumn{10}{l}{Column (9): Stellar mass given by log(M$_{*}$/\msolar).}\\
\multicolumn{10}{l}{Column (10): The cluster-centric distance from the BCG position.}  \\
\end{tabular}
\end{table*}

\newpage

%  FAKE  END-OF-DOCUMENT
\end{document}